\documentclass[preprint,aps]{revtex4}
\usepackage{amssymb,amsmath}
\usepackage{graphics}% Include figure files
\usepackage{dcolumn}% Align table columns on decimal point
\usepackage{bm}% bold math
\begin{document}
\preprint{~}

\title{
The quark-gluon-plasma phase transition diagram, 
Hagedorn matter and quark-gluon liquid
}

\author{Ismail Zakout$^{1,2}$ and Carsten Greiner$^{2}$}
\affiliation{
$\,^{1}$Frankfurt Institute for Advanced Studies
and
$\,^{2}$Institut f\"ur Theoretische Physik, 
Johann Wolfgang Goethe Universit\"at, 
Frankfurt am Main, Germany
}

%%%%%%%%%%%%%%%%%%%%%%%%%%%%%%%%%%%%%%%%%%%%%%%%%%%%%%%%%%%%%%%%%%%%%%%%%%%%%
%%%%%%%%%%%%%%%%%%%%%%%%%%%%%%%%%%%%%%%%%%%%%%%%%%%%%%%%%%%%%%%%%%%%%%%%%%%%

%%%%%%%%%%%%%%%%%%%%%%%%%%%%%%%%%%%%%%%%%%%%%%%%%%%%%%%%%%%%%%%%%%%%%%%%%%%%%
%%%%%%%%%%%%%%%%%%%%%%%%%%%%%%%%%%%%%%%%%%%%%%%%%%%%%%%%%%%%%%%%%%%%%%%%%%%%
\date{\today} % It is always \today, today,
%  but any date may be explicitly specified

\begin{abstract}
In order to study the nuclear matter in the relativistic 
heavy ion collisions and the compact stars, 
we need the hadronic density of states 
for the entire ($\mu_B-T$) phase transition diagram. 
We present a model for the continuous high-lying mass 
(and volume) 
spectrum density of states that fits the Hagedorn mass spectrum. 
%%%%%
This model explains the origin of the tri-critical point 
besides various phenomena such as the quarkyonic matter 
and the quark-gluon liquid. 
The Hagedorn mass spectrum is derived for 
the color-singlet quark-gluon bag with various internal structures 
such as the unimodular unitary, orthogonal 
and color-flavor locked symplectic symmetry groups.
%%%%%
The continuous high-lying hadronic mass spectrum
is populated at first by the unitary Hagedorn states. 
Then the spectrum turns to be dominated by the colorless 
orthogonal states as the dilute system is heated up.
Subsequently, the liquid/gas of orthogonal Hagedorn states 
undergoes higher order deconfinement phase transition 
to quark-gluon plasma. 
Under the deconfinement phase transition process,  
the color-singlet states is broken badly to form 
the colored $SU(N_c)$ symmetry group. 
%%%%%%
On the other hand, when the hadronic matter 
is compressed to larger $\mu_{B}$ and heated up, 
the colorless unitary states (Hagedorn states) 
undergoes first order phase transition to explosive 
quark-gluon plasma at intermediate baryonic density.
The tri-critical point emerges as a change in the
characteristic behaviour of the matter and as an intersection 
among various phases with different internal symmetries. 
When the saturated hadronic matter is cooled down
and compressed to higher density, it turns 
to be dominated by the colorless symplectic states. 
This matter exhibits the first order phase transition 
to quark-gluon plasma when it is heated up 
to higher temperature.
Furthermore, when the Hagedorn states freeze out, 
they evaporate to the low-lying mass spectrum particles 
through Gross-Witten phase transition.
The role of chiral phase transition is also discussed.
\end{abstract}
\maketitle
%%%%%%%%%%%%%%%%%%%%%%%%%%%%%%%%%%%%%%%%%%%%%%%%%%%%%%%%%%%%%%%%%%%%
%%%%%%%%%%%%%%%%%%%%%%%%%%%%%%%%%%%%%%%%%%%%%%%%%%%%%%%%%%%%%%%%%%%%

\section{Introduction}
%%%%%%%%%%%%%%%%%%%%%%%
The phase transition diagram in the $(\mu_B-T)$ has attracted 
much attention. 
The crucial pieces in 
the phase transition diagram are the existence 
of the tri-critical point and the discovery 
of quark-gluon fluid besides
the other predicted phenomena 
such as the color-conductivity and CFL phase. 
The phase transition diagram persists 
to be proved non-trivial and very rich. 
The ingredient element of any investigation 
is the equation of state and its thermodynamics.
The density of states for the hadronic mass spectrum 
is essential for the thermodynamic description 
of strongly interacting hadronic matter 
as well as the deconfinement phase transition 
to the quark-gluon plasma. 
The deconfined quark-gluon plasma is represented 
by the colored $SU(N_c)$ symmetry group where quarks 
and gluons are liberated 
and attain the color degrees of freedom.
They can also form colored quark-gluon bubbles
It is recently argued that the hadronic matter that is populated 
by the discrete low-lying hadronic mass spectrum exhibits 
Gross-Witten transition to another hadronic matter 
that is dominated by the continuous high-lying hadronic mass 
spectrum when the nuclear matter is compressed 
or heated up ~\cite{Zakout:2007nb,Zakout:2006zj}. 
The broken chiral symmetry generates a discrete mass spectrum. 
When the chiral symmetry is restored under the extreme 
conditions, we expect a continuous mass spectrum to emerge in the system. 
It appears that both the color and flavor 
sectors support the discrete and the continuous mass spectrum 
hadronic phase transition scenario.
The discrete low-lying hadronic mass spectrum consists all 
the known mesons, baryons, resonances and the exotic hadronic 
particles which are found in the data book~\cite{databook:2008} 
such as $\pi, \omega \cdots$ and 
$N, \Lambda, \Sigma, \Xi \cdots$ etc.  
For instance, the non-strange discrete mass spectrum consists 
76 mesons and 64 baryons.
The continuous high-lying hadronic mass spectrum 
is the mass spectrum of the Hagedorn states 
and these states appear as gas/liquid of hadronic fireballs. 
The asymptotic continuous high-lying mass spectrum 
can be calculated using the canonical ensemble 
construction~\cite{Kapusta:1981a,Kapusta:1982a}. 
These predicted states are fireballs or composite bags 
consisting of
quarks and gluons 
with specific internal color and/or flavor symmetry.
In order to ensure the confinement of 
the bag's constituent quarks and gluons 
and subsequently the hadronic state, 
the composite bag is projected 
to the color-singlet state 
(i.e. the colorless state).  
Hence, in QCD, it is naturally 
to assume that 
the Hagedorn states is the mass spectrum of a composite bag 
that is projected into color-singlet state regardless 
of the bag's internal symmetry. 
There nothing to prevent us 
to extrapolate the mass spectrum of 
the colorless $SU(N_c)$ state 
of composite bag to the colorless states 
of other color-flavor symmetries 
such as $U(1)^{N_c}$, orthogonal $O_{(S)}(N_c)$ 
and symplectic $Sp(N_c)$ symmetry groups.
These symmetry groups are restricted 
to an additional unimodular-like constraint. 
These symmetry groups are related to each other
either by the decomposition or by the reduction.
In each symmetry group, the color confinement 
(colorless) is guaranteed 
by projecting only the color-singlet-state wave-function 
though every symmetry group represents the composite 
color and flavor degrees of freedom in a different way.
The unimodular-like constraint is imposed   
in order to be consistent with QCD.
This assumption of the bag's internal symmetry modification
takes us to consider the possibility of the phase transition 
from a specific symmetry group to another one.
For instance, the hadronic matter which is populated 
by the colorless $SU(N)$ state bags can transmute
to another matter that is dominated either 
by the colorless orthogonal $O_{(S)}(N)$ state bags 
or by the colorless symplectic $Sp(N)$ state bags 
or even by colorless $U(1)^{N_c}$ state bags. 
%%%%%%%%%%%%%%%%%%%%%%%%%%%%%%%%%%%%%%%%%%%%%%%%%%%
%%%%%%%%%%%%%%%%%%%%%%%%%%%%%%%%%%%%%%%%%%%%%%%%%%%

The effective Coulomb Vandermonde potential is induced 
by the symmetry group constraint that projects 
the bag's color-singlet state. 
It has been shown that the effective 
Coulomb Vandermonde potential is regulated 
in a nontrivial way in the extreme hot 
and dense nuclear bath. 
The characteristic modification 
of the Vandermonde potential within $SU(N_c)$ symmetry
causes the third order Gross-Witten 
hadronic (Hagedorn) transition~\cite{Gross:1980he}.
The physics around Gross-Witten point 
neighbourhood is very rich. 
The emergence of new class of hadronic matter 
(i.e. Hagedorn states)
is relevant to the tri-critical 
point~\cite{Andronic:2009gj}.
The authors of Refs.~\cite{Lang:1980ws,Azakov:1986pn} 
have shown the existence of a critical chemical 
potential $\mu_{c}$ such that for $T>0$,
the physical properties for the low-lying spectrum 
are unaffected
by the chemical potential $|\mu|<\mu_{c}$. 
The Gross-Witten hadronic (Hagedorn) transition 
has received much attention 
in vast fields such as the weak-strong phase transition 
in AdS/CFT~\cite{Aharony:2003a} 
and the search for quark-gluon 
plasma (see for 
example~\cite{Dumitru:2003hp,Dumitru:2004gd,Dumitru:2005ng}). 
%%%%%%%%%%%%%%%%%%%%%%%%%%%%%%%%%%%%%%%%%%%%%%%%%%%%%%%%%%%%%%% 
%%%%%%%%%%%%%%%%%%%%%%%%
%%%%%%%%%%%%%%%%%%%%%%%%
%%%%%%%%%%%%%%%%%%%%%%%%
%%%%%%%%%%%%%%%%%%%%%%%%

It is reasonable to expand the quark-gluon bag's 
internal structure 
to incorporate the color, flavor 
and angular-momentum degrees of freedom 
$\left(N_{c}, N_{f}, L\right)\rightarrow (N,\cdots)$
in an appropriate configuration 
in order to maintain 
the system's internal symmetry invariance in particular
under the extreme hot and dense conditions.
The value $N\,(\equiv N_c)$ is the number of 
the symmetry group Hamiltonian invariant charges 
(see for example~\cite{Mehta1967}). 
The unimodular-like constraint is essential in QCD-like theory.
It is considered in all the symmetry groups 
which are considered in the present work. 
The symmetry group is defined 
by the $N_{fun}=N$ fundamental charges
and $N_{adj}$ of adjoint charges.
The unimodular constraint reduces $N_{fun}$ and $N_{adj}$
to $N_{fun}=N_{fun}-1$ and $N_{adj}=N_{adj}-1$.
The $N_{adj}$ adjoint charges are invariance 
only over the $N_{fun}$  fundamental charges.
The $\left(N_{adj}-N_{fun}\right)$ extra real independent 
parameters do not appear in the Hamiltonian 
and they are integrated (washed) out.
%
%%%%%%%%%%%%%%%%%%%%%%%%%%%%%%%%%%%%%%%%%%%%%%%%%%%%%%%%%%%%%%
%%%%%%%%%%%%%%%%%%%%%%%%%%%%%%%%%%%%%%%%%%%%%%%%%%%%%%%%%%%%%%
The unitary ensemble $U(N)$ has $N_{fun}=N$ 
fundamental eigenvalues and $N_{adj}=N^2$ parameters. 
The number of extra real independent parameters 
is $N_{adj}-N_{fun}=(N^2-N)$.
This means that the unitary symmetry group 
has $N$ fundamental chemical potentials and $N^2$ adjoint parameters.
The  $N^2$ adjoint parameters are transformed (diagonalized) 
to depend basically only on the $N$ fundamental chemical potentials.
The redundant $(N^2-N)$ parameters are, subsequently, integrated out
and they disappear from the resultant ensemble.
The orthogonal ensemble has $N_{fun}=N$ fundamental eigenvalues 
and $N_{adj}=\frac{1}{2}N(N+1)$ adjoint parameters.
The real symmetric $N\times N$ matrix has 
$N_{adj}-N_{fun}=\frac{1}{2}N(N-1)$ 
redundant real parameters and these parameters 
are washed away by an appropriate transformation. 
The unitary symmetry group can be broken and decomposed 
to an orthogonal symmetry group with the same number 
of conserved charges $N_{fun}=N$ (or $N_{fun}=N-1$ 
when the unimodular constraint is embedded).
The symmetry decomposition (breaking) from $U(N)$ 
to orthogonal $O_{(S)}(N)$ leads to $N(N-1)/2$ 
Goldstone bosons emerge as glueballs (or gluon jets) in the medium.
They escape from the Hagedorn bags and enrich 
the medium with gluonic contents and jets. 
The $N(N-1)/2$ Goldstone bosons are identified 
as free colorless gluon degrees of freedom 
while the remaining 
$\frac{1}{2}N(N+1)$ gluons remain 
as the exchange interacting gluons
for the $O_{(S)}(N)$ symmetry group. 
The definitions of $O_{(S)}(N)$ and other symmetry groups 
such as the symplectic $Sp(N)$ symmetry group 
shall be reviewed below in Sec. III.
On the other hand, the symplectic ensemble 
has $N_{fun}=N$ eigenvalues of  
$N\times N$ quaternion-real matrix 
and $N_{adj}= N\left(2 N - 1\right)$.
The number of redundant parameters 
in the symplectic symmetry group 
is $2N\left(N-1\right)$.
The (unimodular) unitary color $U(N_c)$ symmetry group 
may merge with other degree of freedom such 
as flavor symmetry $U(N_{f})$ and 
the symmetry of the resultant composite transmutes 
to a symplectic (quaternion) symmetry group 
through the symmetry modification mechanism 
that is given by 
$U(N_c)\times U(N_{f})\rightarrow U(N_c+N_{f}) 
\rightarrow O(2N) \rightarrow Sp(N)$.
Hence, under the preceding assumption, 
there is an indication that 
the eventual deconfined quark and gluon matter 
can be reached after 
a chain of multi-phase processes 
and this argument is not that simple.

%%%%%%%%%%%%%%%%%%%%%%%%%%%%%%%%%%%%%%%%%%%%%%%%%%%%%%%%%%%%
%%%%%%%%%%%%%%%%%%%%%%%%%%%%%%%%%%%%%%%%%%%%%%%%%%%%%%%%%%%%
%%%%%%%%%%%%%%%%%%%%%%%%%%%%%%%%%%%%%%%%%%%%%%%%%%%%%%%%%%%%
%
%  Bootstrap model
%
%%%%%%%%%%%%%%
%  SEE SEE EXTERNAL PART CALLED BOOTSTRAP MODEL------------
%%%%%%%%%%%%%%%%%%%%%%%%%%%%%%%%%%%%%%%%%%%%%%%%%%%%%%%%%%%%%%%%%%%%%%%%%
%%%%%%%%%%%%%%%%%%%%%%%%%%%%%%%%%%%%%%%%%%%%%%%%%%%%%%%%%%%%%%%%%%%%%%%%%

The thermodynamics of quark jets with an internal color structure 
has been considered using one (or two) dimensional gas
~\cite{Nambu:1982fb,Bambah:1983px%
,Gross:1982hz,Jaimungal:1997hu,Gattringer:1996fi}.
This model is an alternate approximation to deal 
the color-singlet state of quark gas. 
The quarks are treated as classical particles 
but their non-Abelian interactions are introduced 
by the exact Coulomb gauge potential.  
This model has been considered to study the order 
of the deconfinement phase transition. 
The effective potential in that model is dominated 
by a linear potential and it differs from the effective 
Vandermonde potential that emerges from 
the symmetry group invariance Haar measure.  
Furthermore, the density of states for the classical quark gas 
is not given in these studies.

%%%%%%%%%%%%%%%%%%%%%%%%%%%%%%%%%%%%%%%%%%%%%%%%%%%%%%%%%%%%%%%%%%%%%%%%%
%%%%%%%%%%%%%%%%%%%%%%%%%%%%%%%%%%%%%%%%%%%%%%%%%%%%%%%%%%%%%%%%%%%%%%%%%
The hadronic phase transition has been studied using 
the bootstrap 
model~\cite{Hagedorn:1965a,Frautschi:1971a}. 
The continuous hadronic mass spectrum is an exponentially
increasing mass spectrum and it resembles 
the bootstrap mass density, namely, 
$\rho(m)\,\sim\,c\,m^{-\alpha}\,e^{b\,m}$
~\cite{new-ref1:Cabibbo1975,new-ref2:Dashen1969,
new-ref3:Huang-Weinberg1970,new-ref4:Fubini-Veneziano1969}.
The grand potential density is reduced to
\begin{eqnarray}
\frac{\Omega}{V}&=& -\frac{\partial}{\partial\,V}\left( T\ln Z\right), 
\nonumber\\
&=&
- T\,\int^{\infty}_{m_0}\, {d m}\, 
\rho(m)\, 
\frac{\partial}{\partial\,V}
\,\left[\ln Z(m,\beta,\cdots)\right].
\end{eqnarray}
The simple and reasonable approximation 
is the Boltzmann gas with the canonical ensemble 
that is given by,
\begin{eqnarray}
\ln Z&=& \,\int\,{d^3\vec{r}}\,\int \frac{d^3\vec{k}}{(2\pi)^3} 
\Lambda e^{-\beta \epsilon(p,m,r)},
\nonumber\\
&\approx&  \,V\,
\left(\frac{m\, T}{2\pi}\right)^{3/2}\,\Lambda\,
\exp\left(-\frac{m}{T}\right),
\end{eqnarray}
where $\Lambda$ is the thermodynamic fugacity and 
$\epsilon(p,m,r)$ is the energy of constituent particle. 
For the simplicity, 
the free energy $\epsilon(p,m)=\sqrt{\vec{p}^2+m^2}$ 
is usually adopted.
When the critical point of the deconfinement phase transition 
is reached,
the Hagedorn mass spectrum leads to the following results: 
The grand potential density diverges 
for the exponent $\alpha\le\,5/2$. 
This means that the instant phase transition 
to explosive quark-gluon plasma 
does not exist for this class of mass spectrum.
The deconfinement phase transition turns 
to be smooth cross-over one.
On the other hand, the system likely undergoes 
first order deconfinement phase transition
when the exponent becomes $\alpha>\,7/2$ 
because the grand potential density 
and its derivative 
are finite at the critical point making 
the grand potential density continuous and
making its first derivative with respect to 
the thermodynamic ensemble discontinuous. 
When the exponent $\alpha$ which appears 
in the mass spectral density
runs over $5/2<\alpha\le 7/2$, 
the grand potential density becomes finite and continuous 
over the critical temperature and chemical potential 
while the order of phase transition becomes 
rather more difficult to be determined 
but not really abstruse
and the system undergoes higher order 
deconfinement phase transition.
It is evident that the value of 
the mass/volume power exponent $\alpha$ plays 
a decisive role 
in determining the order and shape of 
the phase transition diagram. 
Every choice for the exponent $\alpha$
leads to a different partition function 
and another distinctive physical
behaviour.
%
%
%%%%%%%%%%%%%%%%%%%%%%%%%%%%%%%%%
%%%%%%%%%%%%%%%%%%%%%%%%%%%%%%%%%

The smooth phase transition has been studied using 
phenomenological exponents of the hadron's mass-volume 
spectral density which are set by inspired 
models~\cite{new-ref1:Cabibbo1975,new-ref2:Dashen1969,
new-ref3:Huang-Weinberg1970,new-ref4:Fubini-Veneziano1969,
Begun:2009an,Gorenstein:2005rc,Gorenstein:1998am}. 
The mass and volume bag spectral density 
can be reduced simply to either mass spectral 
density or volume spectral density for a model 
with an adequate approximation such as the MIT bag 
model with the assumption of the sharp boundary surface.
The dependence of the deconfinement phase transition on 
the mass spectral exponent $\alpha$ which appears
in the mass spectral density, namely, 
$\rho=c\, m^{-\alpha}\, e^{b\,m}$
for the bootstrap-like model have  been 
studied~\cite{Ferroni:2008ej}.
The colorless multi-quark cluster has been considered 
to hint some of the puzzling results in RHIC 
experiments~\cite{Abir:2009sh,Pal:2010es}.
The Hagedorn matter below the deconfinement phase transition
is found to fit  the experimental and theoretical 
observations~\cite{NoronhaHostler:2009cf,
NoronhaHostler:2007jf,NoronhaHostler:2008ju,
NoronhaHostler:2009tz}.
The projection of the color-singlet state
resembles Polyakov Loop approach
~\cite{Dumitru:2003hp,Dumitru:2004gd,Dumitru:2005ng,
Sakai:2008um,Abuki:2008nm,Fukushima:2008wg,Schaefer:2007pw}
where the Polyakov effective potential stems from the
Vandermonde potential of a specific symmetry group 
and interaction.
The variation of the phenomenological parameters in 
the Polyakov effective potential
reveals the internal symmetry modification by altering 
the correspondent Vandermonde potential.
In other words, the modification of Vandermonde 
potential alters the bag's internal symmetry.
%%%%%%%%%%%%%%%%%%%%%%%%%%%%%%%%%%%%%%%%%%%%%
The QCD phase transition is considered 
in the context of the random matrix model
~\cite{Klein:2003fy,Kogut:2000ek,Halasz:1998qr}.
There is a strong indication of a new class 
of hadronic matter above 
the discrete low-lying hadronic matter 
and below the deconfined quark-gluon plasma 
and this matter is identified 
as hadronic matter which is dominated by Hagedorn states 
or the quarkyonic matter~\cite{McLerran:2007qj}.
It is argued that the tri-critical point that appears
in the phase transition diagram is related to the existence 
of the predicted quarkyonic matter~\cite{Andronic:2009gj} 
(or equivalently the Hagedorn matter).  
The role of Hagedorn states in both experimental 
and theoretical observations 
have been studied extensively
~\cite{Bugaev:2009ag,Bugaev:2009ge,Bugaev:2008jj,Bugaev:2008iu,
Bugaev:2008nu,Bugaev:2007ww}.
A recent survey on the importance of quarkyonic matter
(i.e. Hagedorn phase) 
in the phase transition diagram from 
the experimental point of view 
can be found in Ref.~\cite{Stock:2009yg}.
The tri-critical point in the QCD phase transition diagram 
is analysed using the lattice calculations
~\cite{deForcrand:2002ci,Fodor:2009ax}.
The models of clustered quarks are demonstrated
to smoothen the order of phase transition
(for instance see ~\cite{Yukalov1}).
Furthermore, the Hagedorn states might play a significant role
in the formation of new class of compact stars at high densities.
For instance, it has been proposed that
the the hypothetical quark stars are made of very massive quarks
rather than ordinary baryons ~\cite{Itoh-1970}.

%%%%%%%%%%%%%%%%%%%%%%%%%%%%%%%%
%%%%%%%%%%%%%%%%%%%%%%%%%%%%%%%%
For only the sake of simplification, we do not include
the Van der Waals (excluded) volume correction 
in the present work. 
Nonetheless, the extension to include the repulsive excluded 
volume correction is straightforward though 
it requires more tedious numerical calculation. 
This kind of calculation has been considered 
in Ref.~\cite{Zakout:2006zj,Zakout:2007nb}
and the references therein
where the order and the shape of the deconfinement
phase transition to quark-gluon plasma have been studied extensively. 
It has been shown that the order and shape of the  
phase transition depends on the internal structures 
of the quark-gluon bags. 
Different color-flavor configurations 
lead to different exponents $\alpha$ 
for the continuous high-lying mass spectrum 
(Hagedorn states).
The bag's internal structure depends on several 
constraints which are imposed to restrict 
the color and flavor degrees of freedom. 
Since the flavor degree of freedom in the hot nuclear matter 
tends to be invariance even for bags 
with more complicated color-flavor coupling,
the color confinement is maintained basically 
by projecting the color-singlet state.
Hence, the bound state that exists through the 
intermediate processes is a colorless state 
as far the deconfinement phase transition is not reached yet.

%%%%%%%%%%%%%%%%%%%%%%%%%%%%%%%%%%%%%%%%%%%%%%%%%%%
In the present work, we study the color-singlet state canonical ensemble 
for the quark-gluon bag in the context 
of various internal symmetry groups such as unimodular-like 
unitary $U(N_c)$, orthogonal $O_{(S)}(N_c)$ 
and symplectic $Sp(N_c)$ symmetry groups as well as $U(1)^{N_c}$.
The intermediate transmutation from the hadronic matter that 
is dominated by the colorless unitary fireballs 
to another hadronic matter that is populated 
by the colorless orthogonal states 
or the colorless symplectic states is introduced in order 
to demonstrate the multi-nuclear phases in the phase transition diagram 
and the existence of the tri-critical point. 
We discuss possible scenarios 
for the deconfinement phase transition from the colorless states 
of unitary, orthogonal and symplectic hadronic matter
to the colored $SU(N_c)$ quark-gluon plasma. 
These scenarios are relevant to the possible intermediate 
hadronic phases those take place below 
the deconfinement phase transition
and the emergence of the tri-critical point.
%%%%%%%%%%%%%%%%%%%%%%%%%%%%%%%%%%%%%%%%%%%%%%%%%%%%%
 
The outline of the present paper is as follows: 
In Sec.~\ref{sect2}, we derive 
the quark-gluon grand canonical ensemble  
with the color-singlet $SU(N_c)$ state 
in order to be extrapolated to other symmetries. 
The internal orthogonal, unitary and symplectic 
symmetry groups with the unimodular constraint 
are reviewed in Sec. ~\ref{sect3}. 
The equation of state for the nuclear matter 
that is dominated by the Hagedorn states 
is given in Sec. ~\ref{sect4}. 
The continuous Hagedorn mass spectra are derived 
for the unimodular-like unitary, orthogonal 
and symplectic ensembles.
The role of the chiral phase transition is considered 
in Sec.~\ref{sect5}.
Several scenarios for the deconfinement phase transition
are presented in Sec.~\ref{sect-6a}.
Finally, the conclusion is presented in Sec.~\ref{sect-6b}.
%%%%%%%%%%%%%%%%%%%%%%%%%%%%%%%%%%
%%%%%%%%%%%%%%%%%%%%%%%%%%%%%%%%%%%%%%%%%%%%%%%%%%%%%%%%%%%%%%%%%%%%%
%%%%%%%%%%%%%%%%%%%%%%%%%%%%%%%%%%%%%%%%%%%%%%%%%%%%%%%%%%%%%%%%%%%%%
%%%%%%%%%%%%%%%%%%%%%%%%%%%%%%%%%%%%%%%%%%%%%%%%%%%%%%%%%%%%%%%%%%%%%
%%%%%%%%%%%%%%%%%%%%%%%%%%%%%%%%%%%%%%%%%%%%%%%%%%%%%%%%%%%%%%%%%%%%%
%%%%%%%%%%%%%%%%%%%%%%%%%%%%%%%%%%
%%%%%%%%%%%%%%%%%%%%%%%%%%%%%%%%%%
%%%%%%%%%%%%%%%%%%%%%%%%%%%%%%%%%%

\section{\label{sect2}
The nuclear matter for the relativistic 
heavy ion collisions}
%%%%%%%%%%%%%%%%%%%%%%%%
The grand canonical partition function
in the equilibrium reads
\begin{eqnarray}
Z&=&\mbox{tr}\,\,
\exp\left[-\beta\,H\,+i\sum_{n}\vartheta_n\, N_n\right],
\nonumber\\
&=&\mbox{tr}\,\,
\exp\left[-\beta\left(H\,-\,\sum_{n}\mu_n\,N_n\right)\right].
\end{eqnarray}
The thermodynamic quantities are determined 
from the grand canonical partition function 
as follows
\begin{eqnarray}
\begin{array}{l}
\frac{\Omega}{V}
\,=\,
-\frac{\partial}{\partial V}\left(
T\ln Z
\right),
\\
\rho_i
\,=\, \frac{1}{V}\frac{\partial}{\partial \mu_i}\left(
T\ln Z
\right),
\\
s\,=\, \frac{1}{V}\frac{\partial}{\partial T}\left(
T\ln Z
\right),
\\
\frac{E}{V}
\,=\, \frac{\Omega}{V}
+ T s + \mu_i \rho_i,
\end{array}
\end{eqnarray}
for the grand potential density, charge density, entropy 
and energy density, respectively. 
The canonical partition function can be determined by the 
functional integration method. The Lagrangian construction 
is essential.
The QCD Lagrangian for quarks and gluons is given 
by~\cite{Kapusta2006}
\begin{equation}
{\cal L}=-\frac{1}{4} {F_{\mu\nu}^{a}}\,{F^{\mu\nu\,a}}
+
{\bar{\psi}_i }\,
\gamma^{\mu}\,\left( i\,D_{\mu} \right)_{ij}
\,{\psi_j },
\end{equation}
where the covariance derivative 
and field strength, respectively, read
\begin{eqnarray}
D_{\mu}=\partial_{\mu}
+ i\, m
+i\,{g }\,{\bf t}^{a}\,A^{a}_{\mu},
\end{eqnarray}
and
\begin{eqnarray}
F_{\mu\nu}^{a}\,=\,
\partial_{\mu} A^{a}_{\nu} -\partial_{\nu} A^{a}_{\mu}
-g f^{abc}\, A^{b}_{\mu}\,A^{c}_{\nu}.
\end{eqnarray}
The quarks and antiquarks are represented 
by the fundamental matrices
${\bf t}^{a}$, while the gluons are represented by the
adjoint matrices $\left(T^{a}\right)_{bc}=-i\,f^{abc}$. 
The fundamental representation for $SU(N_c)$ 
symmetry group has $N_c-1$ degrees of freedom 
while the adjoint representation has $N^2_c-1$.
The current conservation laws read
\begin{equation}
\begin{array}{c}
j^{\mu}=\bar{\psi}\,\gamma^{\mu}\,\psi,
\\ 
\partial_{\mu}\, j^{\mu}=0.
\end{array}
\end{equation}
The total conserved baryonic charge with 
$U\left(1\right)_{\mbox{B}}$ group representation 
is furnished by
\begin{eqnarray}
Q&=&\int d^{3}x\, \bar{\psi}\gamma_{0}\psi,
\nonumber\\
&=&\frac{1}{\beta}\int^{\beta}_{0} d{\tau }\,
\int d^{3}{x }\, \bar{\psi}\gamma_{0}\psi.
\end{eqnarray}
On the other hand, the color current density
is given by
\begin{equation}
j^{a}_{\mu}\left(\mbox{color}\right)
=
\bar{\psi}\gamma_{\mu}{\bf t}^{a}\psi
\,+\,i\, \left(T^a\right)_{bc}\, F^{b}_{\mu\nu} A^{\nu}_{c},
\end{equation}
where both the fundamental quarks and 
the adjoint gluons 
contribute to the color current.
The conserved color charge generators 
are given by the time-component as follows
\begin{equation}
Q^{a}\left(\mbox{color}\right)=
\frac{1}{\beta}\int^{\beta}_{0} d{\tau }\,
\int d^{3}{x }\, {j^{a}_{0} }\left(\mbox{color}\right),
\end{equation}
where
\begin{equation}
j^{a}_{0}\left(\mbox{color}\right)
=
\bar{\psi}\gamma_{0}{\bf t}^{a}\psi
\,+\,i\, \left(T^a\right)_{bc}\, F^{b}_{0\,\nu} A^{\nu}_{c}.
\end{equation}
The canonical ensemble in the Hilbert space is given by 
the tensor product of the fundamental and 
anti-fundamental particles Fock spaces 
and the adjoint particles Fock space.
The fundamental and anti-fundamental particles are
the quarks and anti-quarks while the adjoint particles 
are the interaction particles which are the gluons.
The partition functions for the gas 
of fundamental particles 
and the gas of adjoint particles
are constructed separately 
and then the tensor product of their 
resultant Fock spaces is taken in the following way,
\begin{eqnarray}
Z_{q\bar{q}g}=Z_{q\bar{q}}\,\times\,Z_{g}.
\end{eqnarray}  
The partition function for quark and anti-quark gas 
is given by
\begin{equation}
Z_{q\overline{q}}=\mbox{tr}\, 
\left[
e^{-\beta\left(H-\mu Q\right)+
i\theta_i Q_{i}}
\right],
\end{equation}
where $Q_i$ are the color charges.
In the functional integral procedure, 
the partition function is reduced to
\begin{eqnarray}
Z_{q\overline{q}}&=&
\prod_{i\,{, }\,\alpha}\int [i d\psi_{i\alpha}^{\dagger}]
[d\psi_{i\alpha}]
\exp\left[
\int^{\beta}_{0} d\tau \int d^{3}x
\sum_{i\,{, }\,\alpha}
\bar{\psi}_{i\alpha}
\right.
\nonumber\\
&~&~~~~~\left.
\times
\left(
-\gamma^{0}\left[\frac{\partial}{\partial \tau}-\mu
-i\frac{\theta_i}{\beta}\right]
+i\vec{\gamma}\cdot \vec{\nabla}-m
\right)\psi_{i\alpha}
\right].
\label{action-qqbar-thermal1}
\end{eqnarray}
By adopting the imaginary time method 
in the thermal field theory, the straightforward 
integration over the Grassmann variables reduces 
the partition function, 
that is given by Eq.(\ref{action-qqbar-thermal1}),
to
\begin{equation}
Z_{q\overline{q}}\left(\beta,V\right)
=\mbox{det}
\left[
-i\beta\left(
\left[-i\omega_n+\mu+i\frac{1}{\beta}\theta_{i}\right]
-\gamma^{0}\vec{\gamma}\cdot\vec{k}-m\gamma^{0}
\right)
\right]^2,
\end{equation}
where the sum over the Matsubara (odd-) frequencies 
$\omega_{n}$ for fermion is performed 
in a proper way over the physical observable 
$T\ln Z_{q\overline{q}}\left(\beta,V\right)$.
The determinant is evaluated 
by considering the following trick
\begin{eqnarray}
Z_{q\overline{q}}\left(\beta,V\right)
&=&\exp \ln \mbox{det}
\left[
-i\beta\left(
\left[-i\omega_n+\mu+i\frac{1}{\beta}\theta_{i}\right]
-\gamma^{0}\vec{\gamma}\cdot\vec{k}-m\gamma^{0}
\right)
\right]^2.
\end{eqnarray}
As usual, the physical observable is determined 
by taking the real part as follows
\begin{eqnarray}
\ln\, Z_{q\overline{q}}\left(\beta,V\right)\rightarrow
{\Re e}\, 
\left(\,\ln\, Z_{q\overline{q}}\left(\beta,V\right)\,\right).
\end{eqnarray}
The partition function's explicit expression reads
\begin{eqnarray}
\ln\, Z_{q\overline{q}}\left(\beta,V\right)&=& 
{\Re e}\,\sum_n\,\left( (2J+1)\,\mbox{tr} \ln  
\left[
\beta^2\left(
\left[\omega_n
+i\left(\mu_{q}+i\frac{1}{\beta}\theta_{i}\right)\right]^2
+\epsilon_{q}^2(\vec{p})
\right)
\right]\right),
\nonumber\\
&=&
{\Re e}\,\left( (2J+1)\,V\int \frac{d^3\vec{p}}{(2\pi)^3}
\sum_n \ln  
\left[
\beta^2\left(
\left[\omega_n
+i\left(\mu_{q}+i\frac{1}{\beta}\theta_{i}\right)\right]^2
+\epsilon_{q}^2(\vec{p})
\right)
\right]\right),
\nonumber\\
\end{eqnarray}
where $\epsilon_{q}(\vec{q})=\sqrt{\vec{p}^2+{m_{q}}^2}$
and $\mu_{q}$ is the flavor chemical potential.
The summation over the number of states is evaluated using 
the standard procedure.  
A possible extension to include 
the effect of the bag's smooth boundary can be 
found in Refs.~\cite{Balian:1970a,Balian:1972a}.
The pre-factor $(2J+1)=2$ comes from the spin degeneracy.
The summation over the Matsubara 
(odd-) frequencies is evaluated 
and the result reads,
\begin{eqnarray}
T\,\ln\, Z_{q\bar{q}} &=& \frac{1}{\beta} \,\ln\, Z_{q\bar{q}},
\nonumber\\
&=&
{\Re e}\,\left( (2J+1) V \int \frac{d^3\vec{p}}{(2\pi)^3}
\left[
\epsilon(\vec{p})
\,+\,\frac{1}{\beta}\,\sum_{i}\ln\left(
1+e^{-\beta
\left(\epsilon(\vec{p})-\mu-i\frac{\theta_i}{\beta}\right)}
\right)\right.\right.
\nonumber\\
&~&\left.\left.
+\,\frac{1}{\beta}\,\sum_{i}\ln\left(
1+e^{-\beta
\left(\epsilon(\vec{p})+\mu+i\frac{\theta_i}{\beta}\right)}
\right)
\right]\right).
\end{eqnarray}
The first term is temperature independent. It leads 
to the divergence at zero temperature 
and can be renormalised in the standard way.
Since, we are interested on the temperature dependent terms,
the first term is trivially dropped.
In the extremely hot medium, the system 
is supposed to take any flavor symmetry
under the unitary representation. 
The mechanism is recognised 
as the flavor invariance in the extreme
hot nuclear matter.
Hence by the assumption that there is no specific
flavor symmetry configuration is preferred, 
we do not need to project specific flavor symmetry. 
It is reasonable to believe that under extreme temperature 
such as the case of hadronic fireballs, 
the system is preferred  to maintain 
the flavor chemical equilibrium 
rather than carries specific internal flavor symmetry. 
Under this assumption a specific flavor structure 
such as mesonic 
or baryonic Hagedorn states becomes unimportant 
and the classical Maxwell-Boltzmann statistics 
becomes satisfactory.
%%%%%%%%%%%%%%%%%%%%%%%%%%%%%%%%%%%%%%%%%%%%%%%%%%%%%

Nonetheless, it is still possible to maintain
the flavor and $SU(N_{f})$ symmetry 
by considering non-strangeness mesonic 
and baryonic Hagedorn states at low baryonic density.
When the nuclear matter is compressed to higher 
baryonic chemical potential, the strange mesonic 
and baryonic Hagedorn states emerge in the system 
and so on.  
Furthermore, it is possible to imagine that
under certain extreme hot and/or dense circumstances
the color and flavor symmetries 
are modified to reproduce either 
the orthogonal or symplectic symmetry. 
%%%
The flavor degree of freedom is introduced trivially 
as follows
\begin{eqnarray}
\ln\, Z_{q\bar{q}} &=&
{\Re e}\,\left( (2J+1)\,V 
\sum^{N_{f}}_{q}\int \frac{d^3\vec{p}}{(2\pi)^3}
\sum_{i}\left[
\ln\left(1+e^{-\beta
\left(\epsilon_{q}(\vec{p})-\mu_{q}-i\frac{\theta_i}{\beta}\right)}
\right)
\right.\right.
\nonumber\\
&+& \left.\left.
\ln\left(1+e^{-\beta
\left(\epsilon_{q}(\vec{p})+\mu_{q}+i\frac{\theta_i}{\beta}\right)}
\right)
\right]\right),
\nonumber\\
\label{real-Z-fund-q-q}
\end{eqnarray}
where
$\epsilon_q(\vec{p})=\sqrt{\vec{p}^2+m^2_{q}}$.
Only the real part is taken in Eq.(\ref{real-Z-fund-q-q}).
The resultant partition function is reduced to
\begin{eqnarray}
\ln Z_{q\overline{q}}\left(\beta,V\right)
&=&
{\Re e}\,\left( (2J+1) V \int \frac{d^3\vec{p}}{(2\pi)^3}
\left[
\sum_{i}\ln\left(
a_{q\bar{q}}\left(\theta_i\right)
+
ib_{q\bar{q}}\left(\theta_i\right)
\right)
\right]\right),
\end{eqnarray}
where
\begin{equation}
\begin{array}{l}
a_{q\bar{q}}\left(\theta_i\right)=
\left(
1\,+\,\cos(\theta_i)
\sum^{N_{f}}_{q}
\left[e^{-\beta\left(\epsilon_{q}(\vec{p})-\mu_{q}\right)}
+e^{-\beta\left(\epsilon_{q}(\vec{p})+\mu_{q}\right)}
\right]
+e^{-2\beta\,\epsilon_{q}(\vec{p})}
\right),
\\
b_{q\bar{q}}\left(\theta_i\right)=
\sin(\theta_i)
\sum^{N_{f}}_{q}
\left[e^{-\beta\left(\epsilon_{q}(\vec{p})-\mu_{q}\right)}
-e^{-\beta\left(\epsilon_{q}(\vec{p})+\mu_{q}\right)}
\right].
\end{array}
\end{equation}
%%%%%%%%%%%%%%%%
The partition function for quark and anti-quark 
in the Fock space becomes
\begin{eqnarray}
Z_{q\overline{q}}\left(\beta,V\right)
&=&\exp\left[
(2J+1) V
\int \frac{d^3\vec{p}}{(2\pi)^3}
\left(
\frac{1}{2}\sum_{i}\ln\left[
a^2_{q\bar{q}}\left(\theta_i\right)
+
b^2_{q\bar{q}}\left(\theta_i\right)
\right]
\right)\right].
\end{eqnarray}
%%%%%%%%%%%%%%%%%%%%%%%%%%%%%%%%%%%%%%%%%%%%%
%%%%%%%%%%%%%%%%%%%%%%%%%%%%%%%%%%%%%%%%%%%%%
The Vandermonde determinant, which appears 
in the invariance Haar measure, 
contributes to the action as an additional effective potential term.
The Vandermonde effective potential term becomes soft 
when the color eigenvalues of the stationary condition 
are distributed uniformly
over the entire circle $|\theta_i|\le \pi$.
The integral of the resultant ensemble 
can be evaluated trivially under such circumstances.
However, this will not be the case 
under the extreme hot and dense conditions in particular 
when the color eigenvalues turn to be distributed
in a narrow interval $|\theta_i|<<\pi$. 
When the saddle points 
congregate around the origin rather than 
are distributed uniformly 
over the entire color circle range $|\theta_i|\le\pi$,
the Vandermonde effective potential 
develops a virtual singularity. 
Subsequently, the action must be
regulated in a proper way in order to remove
the Vandermonde determinant divergence.
This regulation procedure accommodates 
the extreme hot and dense conditions.
It is usually associated with smooth phase transition.
The regulation procedure corresponds 
the Gross-Witten third order Hagedorn phase 
transition in the context of $U(N_c)$ 
in the limit $N_c\rightarrow \infty$.
The same procedure can be extended 
to other symmetry groups such as the orthogonal 
and symplectic ones.
%%%%%%%%%%%%%%%%%%%%%%%%%%%%%%%%%%%%%%%%%%%%%%%%%
The change in the analytic solution corresponds 
higher order phase transition 
from the dilute and relatively cold hadronic matter 
(i.e. discrete low-lying hadronic mass spectrum phase) 
to the highly thermally excited hadronic matter 
(i.e. continuous high-lying hadronic 
mass spectrum phase).
This kind of nuclear phase transition would not mean 
that the deconfinement phase transition is reached.
The highly thermally excited hadronic matter 
is interpreted as an exotic hadronic phase 
that is dominated by the Hagedorn states.
In the lattice theory, this is implying
that the weak and strong coupling
is not described by the same
analytic function.
This strong- and weak-coupling transition
is analogous to the phase transition
from the discrete low-lying mass spectrum particles 
to the highly excited and massive Hagedorn states 
(i.e. continuous high-lying mass spectrum).
The Hagedorn states are given by the mass spectrum 
of the color-singlet state composite (colorless)
where the constituent quarks and gluons are represented 
by the $SU(N_c)$ symmetry group representation
and these states are hadrons
~\cite{Elze:1983du,Elze:1985wv,Gorenstein:1983a,Auberson:1986a}. 
This kind of matter should not be immediately interpreted
as the deconfined quark-gluon plasma. 
The critical Gross-Witten point is the threshold point 
where the Hagedorn states emerge in the system.
The internal color structure is known to be essential 
in the phase transition to the quark-gluon plasma
~\cite{ Elze:1983du,Elze:1984un,Elze:1985wv,Elze:1986db,Elze:1986gz}.
The deconfinement phase transition can either take place
immediately when unstable Hagedorn states are produced
or as a subsequent process when
the metastable Hagedorn phase eventually undergoes 
the phase transition to quark-gluon plasma. 
The Hagedorn phase may undergo multiple 
intermediate transitions before the deconfinement
phase transition is eventually reached.
%%%%%%%%%%%%%%%%%%%%%%%%%%%%%%%%%%%%%%

The effective Vandermonde potential plays 
a crucial role in the intermediate 
phase transition processes from 
the hadronic phase and quark-gluon plasma. 
This potential is back reacted 
to the heat and compression of the nuclear matter.
%%%%%%%
When the invariance Haar measure is regulated, 
another analytical solution with different 
characteristic properties emerges. 
Hence, Vandermonde determinant 
is regulated in a nontrivial way
and the action can be expanded 
over the group fundamental variables 
around the stationary Fourier color points  
up to the quadratic terms.
Fortunately, the saddle 
Fourier color points are convened around the origin
and fortunately this simplifies 
the problem drastically.
%%%%%%%%%%%%%%%%%%%%%%%%%%
Despite the complexity of the action due to the realistic
physical situation that is involved, 
there will be  an easy way to find the quadratic expansion 
around the group saddle points.  
The resultant integral is evaluated using 
the standard Gaussian quadrature 
over the group (i.e. color) variables.
Hence, beyond the Gross-Witten point the partition function 
can approximated by the quadratic Taylor expansion.
over the group variables 
%%%%%%%%%%%%%%%%%%%%%%%%%%%%
The quadratic expansion of the quark 
and anti-quark ensemble 
around the saddle points is reduced 
to the following Gaussian-like function
\begin{eqnarray}
Z_{q\bar{q}}\left(\beta,V\right)=
\exp\left[ 
a^{(0)}_{q\bar{q}}
-\frac{1}{2}a^{(2)}_{q\bar{q}}
\sum^{N_c}_{i=1} \theta^2_i\right].
\end{eqnarray}
This regulated ensemble 
$Z_{q\overline{q}}\left(\beta,V\right)$ 
leads to a continuous high-lying mass spectrum.
The coefficients which appear in the exponential 
are calculated in the following way,
\begin{eqnarray}
a^{(0)}_{q\bar{q}}=
(2J+1) V N_c\,\sum^{N_{f}}_{q}\int \frac{d^3\vec{p}}{(2\pi)^3}
\ln\left[
1+
\left(e^{-\beta\left(\epsilon_{q}(\vec{p})-\mu_{q}\right)}
+e^{-\beta\left(\epsilon_{q}(\vec{p})+\mu_{q}\right)}\right)
+e^{-2\beta\epsilon_{q}(\vec{p})}
\right],
\end{eqnarray}
and
\begin{eqnarray}
a^{(2)}_{q\bar{q}}&=&
(2J+1) V\,\sum^{N_{f}}_{q}\int \frac{d^3\vec{p}}{(2\pi)^3}
\frac{
\left(
e^{-\beta\left(\epsilon_{q}(\vec{p})-\mu_{q}\right)}
+e^{-\beta\left(\epsilon_{q}(\vec{p})+\mu_{q}\right)}
\right)
}
{
\left[
1+
\left(e^{-\beta\left(\epsilon_{q}(\vec{p})-\mu_{q}\right)}
+e^{-\beta\left(\epsilon_{q}(\vec{p})+\mu_{q}\right)}\right)
+e^{-2\beta\epsilon_{q}(\vec{p})}
\right]}
\nonumber\\
&~& -
(2J+1) V\,\sum^{N_{f}}_{q}\int \frac{d^3\vec{p}}{(2\pi)^3}
\frac{
\left(
e^{-\beta\left(\epsilon_{q}(\vec{p})-\mu_{q}\right)}
-e^{-\beta\left(\epsilon_{q}(\vec{p})+\mu_{q}\right)}
\right)^{2}
}
{
\left[
1+
\left(e^{-\beta\left(\epsilon_{q}(\vec{p})-\mu_{q}\right)}
+e^{-\beta\left(\epsilon_{q}(\vec{p})+\mu_{q}\right)}\right)
+e^{-2\beta\epsilon_{q}(\vec{p})}
\right]^{2}}.
\end{eqnarray}
The chemical potential $\mu_{q}$ is decomposed 
to $\mu_{q}=\frac{1}{3}\mu_B+S\mu_S+\cdots$.
From the fugacity definition, 
it is worth to keep in mind that 
$\beta\mu_{q}\rightarrow\frac{1}{T}\mu_{q}$.
%%%%%%%%%%%%%%%%%%%%%%%%%%%%%%%%
In the case of massless flavors, we have
\begin{eqnarray}
a^{(0)}_{q\bar{q}}&=&
(2J+1)
\left(\frac{V}{\beta^3}\right) N_c\,\sum^{massless}_{q=1}\,
\left[
\frac{7\pi^2}{360}+\frac{1}{12}\left(\frac{\mu_{q}}{T}\right)^2
+\frac{1}{24\pi^2}\left(\frac{\mu_{q}}{T}\right)^4
\right]
\nonumber\\
&~&~~~~~~~
+(\mbox{massive flavors}),
\end{eqnarray}
and
\begin{eqnarray}
a^{(2)}_{q\bar{q}}&=&
(2J+1)\left(\frac{V}{\beta^3}\right)
\,\sum^{massless}_{q=1}\,
\frac{1}{6}
\left( 1+\frac{3}{\pi^2}\frac{\mu_{q}^{2}}{T^{2}} \right)
+(\mbox{massive flavors}),
\nonumber\\
&\approx&
(2J+1)
\left(\frac{V}{\beta^3}\right)
\,\sum^{massless}_{q=1}\,
\frac{1}{6}
\left( 1+\frac{3}{\pi^2}\frac{\mu_{q}^{2}}{T^{2}} \right)
+(\cdots),
\end{eqnarray}
where $\left(\cdots\right)$ indicates 
the neglected massive flavors.
%%%%%%%%%%%%%%%%%%%%%%%%%%%%%%%%%%%%%%%%%%%%%%%
%%%%%%%%%%%%%%%%%%%%%%%%%%%%%%%%%%%%%%%%%%%%%%%
%
%
%
%
%%%%%%%%%%%%%%%%%%%%%%%%%%%%%%%%%%%%%%%%%%%%%%%
%%%%%%%%%%%%%%%%%%%%%%%%%%%%%%%%%%%%%%%%%%%%%%%
%
%

The grand canonical ensemble for the adjoint gluons 
can be calculated in a similar way to that one done 
in calculating the fundamental quark 
and antiquark grand canonical ensemble 
in the functional integral representation
(see for instance Eq.(\ref{action-qqbar-thermal1})).
%%%%%%% 
For the sake of simplicity, the axial gauge is considered  
and the result is manifestly gauge invariant
~\cite{Kapusta2006}.
In the axial gauge, the canonical partition function for 
a gas of adjoint gluons reads
\begin{eqnarray}
Z_{g}\left(\beta,V\right)
&=&\int \prod_{a} 
\int [d A^{a}_0]  [d A^{a}_1]  [d A^{a}_2]
\mbox{det}\left(\partial_3\right)
e^{S_{g}},
\end{eqnarray}
where the action is given by
\begin{eqnarray}
S_{g}&=&
\frac{1}{2}\int^{\beta}_0 d\tau\int d^{3} x
(A^{a}_0,A^{a}_1,A^{a}_2){\bf D} (A^{a}_0,A^{a}_1,A^{a}_2)^{T}.
\label{action-gluon-Lagrangian1}
\end{eqnarray}
By constructing the canonical Lagrangian
from the Hamiltonian and then by integrating 
the resultant Lagrangian by parts, the operator 
that appears in Eq.(\ref{action-gluon-Lagrangian1}) 
is reduced to 
\begin{eqnarray}
{\bf D}=\left(
\begin{array}{ccc}
\nabla^{2}& -\partial_1\partial_{\tau} & -\partial_2\partial_{\tau}
\\
-\partial_1\partial_{\tau}&\partial^2_2+\partial^2_3+\partial^2_{\tau}
&-\partial_1\partial_2
\\
-\partial_2\partial_{\tau}& -\partial_1\partial_2&
\partial^2_1+\partial^2_2+\partial^2_{\tau}
\end{array}
\right).
\end{eqnarray}
Following the standard procedure in the thermal quantum field 
theory and the expansion over the Matsubara (even-) frequencies
in the imaginary time formalism, it is possible to invert the 
operators to standard matrices.  
The canonical partition function for a gas of gluons 
is reduced to   
\begin{eqnarray}
Z_{g}\left(\beta,V\right)
&=& 
\mbox{det}(\partial_3)\times 
\frac{1}{\sqrt{\mbox{det}({\bf D})}},
\end{eqnarray}
where the resultant Riemann integrals are evaluated
by the Gaussian integration.
The determinants for the axial gauge constraint 
and the action read, respectively,
\begin{equation}
\mbox{det}(\partial_3)=\prod^{N^2_c-1}_{a}\mbox{det}(\beta p_3),
\end{equation}
and
\begin{equation}
\mbox{det} {\bf D}= 
\prod^{N^2_c-1}_{a=1}\mbox{det} {\bf D}^{a}.
\end{equation}
%%%%%%%%%%%%%%%
By introducing the Fourier transform 
and the Matsubara (even-) frequencies,
the adjoint gluon operator is reduced to
\begin{eqnarray}
{\bf D}^a=\beta^2\left(
\begin{array}{ccc}
\vec{p}^2 & -(\omega_n-i\frac{\phi_a}{\beta}) p_1 &
-(\omega_n-i\frac{\phi_a}{\beta}) p_2
\\
-(\omega_n-i\frac{\phi_a}{\beta}) p_1&
(\omega_n-i\frac{\phi_a}{\beta})^2+p^2_2+p^2_3& -p_1 p_2
\\
-(\omega_n-i\frac{\phi_a}{\beta}) p_2&
-(\omega_n-i\frac{\phi_a}{\beta}) p_1&
(\omega_n-i\frac{\phi_a}{\beta})^2+p^2_1+p^2_3
\end{array}
\right).
\end{eqnarray}
%%%%%%%%%%%%%%%%%%%%%%%%%%%%%%%%%%%
The straightforward calculation
following the standard procedure 
in the thermal field theory
~\cite{Kapusta2006} 
leads to
\begin{eqnarray}
\ln\, Z_{g}\left(\beta,V\right)
&=& 
-(2J+1) V\,\sum_n\,
\int \frac{d^3 p}{(2\pi)^3}\sum^{N^2_c-1}_{a=1}
\ln
\left(\beta^2
\left[
\left(\omega_n -i\frac{\phi_a}{\beta}\right)^2+\vec{p}^2
\right]
\right).
\end{eqnarray}
The sum over the Matsubara (even-) 
frequencies gives the following result
\begin{eqnarray}
\frac{1}{\beta}\,\ln\, Z_{g}\left(\beta,V\right)
= -(2J+1) V
\int \frac{d^3 p}{(2\pi)^3}
\sum^{N^2_c-1}_{a=1}
\left[
\frac{1}{2}\epsilon_{g}(\vec{p})
+
\frac{1}{\beta}
\ln\left(
1-e^{-\left(\beta\epsilon_{g}(\vec{p})-i\phi_a\right)}
\right)
\right],
\label{grand-potential-gluon1}
\end{eqnarray}
where 
$\epsilon_{g}(\vec{p})=\sqrt{\vec{p}^2+m^{2}_{g}}$ 
and $m_{g}=0$.
The first term inside the square bracket 
on the right hand side 
of the grand potential 
$\Omega=-V\frac{\partial}{\partial V}
\frac{1}{\beta}\,\ln\, Z_{g}\left(\beta,V\right)$
that is given by Eq.(\ref{grand-potential-gluon1})
is temperature independent.
This term must be regulated at zero temperature 
and must be dropped from the thermal contribution
because it has no thermal origin.
The grand canonical ensemble for the gluon gas 
is reduced to
\begin{eqnarray}
Z_{g}\left(\beta,V\right)
&=&\exp\left\{ -(2J+1) V \int \frac{d^3 p}{(2\pi)^3}
\sum^{N^2_c-1}_{a=1}
\ln\left[
1-e^{-\left(\beta\epsilon_g(\vec{p})-i\phi_a\right)}
\right]
\right\}.
\end{eqnarray}
The adjoint eigenvalues are calculated 
from the nested commutation relations for  
fundamental eigenvalues in the Lie algebra. 
The adjoint eigenvalues
are related to fundamental eigenvalues by 
the relation
$\phi_a\sim(\theta_i-\theta_j)$. 
This relation diagonalized the adjoint representation
and subsequently commutes with the energy Hamiltonian.
The canonical ensemble for gluon gas becomes
\begin{eqnarray}
\ln Z_{g}\left(\beta,V\right)
&=& -(2J+1) V
\int \frac{d^3 p}{(2\pi)^3}
\left\{
\sum^{N_c}_{1}\sum^{N_c}_{i\ne j}
\ln\left(1-
e^{-\left[\beta\epsilon_{g}
(\vec{p})-i(\theta_i-\theta_j)\right]}
\right)\right\}
\nonumber\\
&~&
-(2J+1) (N_c-1) V
\int \frac{d^3 p}{(2\pi)^3}
\ln\left(
1-e^{-\beta\epsilon_{g}(\vec{p})}
\right).
\end{eqnarray}
The gluon partition function has no imaginary 
and it can be re-written as follows
\begin{eqnarray}
\ln Z_{g}\left(\beta,V\right)
&=& - (2J+1) V
\int \frac{d^3 p}{(2\pi)^3}
\left\{
\sum^{N_c}_{i>j}
\ln\left[
1-
2\cos(\theta_i-\theta_j)\,e^{-\beta\epsilon_{g}(\vec{p})}
+e^{-2\beta\epsilon_{g}(\vec{p})}
\right]\right\}
\nonumber\\
&~&
- (2J+1) \left(N_c-1\right) V
\int \frac{d^3 p}{(2\pi)^3}
\ln\left[
1-e^{-\left(\beta\epsilon_{g}(\vec{p})\right)}
\right].
\label{gluon-partition-1}
\end{eqnarray}
%%%%%%%%%%%%%%%%%%%%%%%%%%%%%%%%%%%%%%%%%%%%%%%%%%%%
Similar to the fundamental quark and antiquark 
canonical ensemble, the grand canonical ensemble for 
the adjoint gluons exhibits 
Gross-Witten Hagedorn transition.
The grand canonical partition function  
above the point of the Gross-Witten Hagedorn transition 
can be approximated 
by the quadratic Taylor expansion around 
the saddle points 
as has been done for the quark and anti-quark 
grand canonical ensemble.
The fact that the saddle points are convened 
at the origin simplifies the calculation remarkably. 
The quadratic Taylor expansion 
of the gluon grand canonical ensemble 
reads
\begin{eqnarray}
\ln\,{Z }_{g}\left(\beta,V\right) &=& a^{(0)}_g 
-\frac{1}{2} a_g \sum^{N_c}_{i=1}\sum^{N_c}_{j=1}
(\theta_i-\theta_j)^2,
\nonumber\\
&=& a^{(0)}_g 
-\frac{1}{2} \frac{\left(2{N_c }\, {a^{(2)}_{g} }\right)}{2N_c}
 \sum^{N_c}_{i=1}\sum^{N_c}_{j=1}
(\theta_i-\theta_j)^2,
\end{eqnarray}
where
\begin{eqnarray}
a^{(0)}_g&=&
- (2J+1) V
N_{adj}
\int \frac{d^3 p}{(2\pi)^3}
\ln\left(
1-e^{-\beta\,\epsilon_{g}(\vec{p})}
\right),
\nonumber\\
&=&
(2J+1)
\left(\frac{V}{\beta^3}\right)\,N_{adj}\,
\left(\frac{\pi^2}{90}\right),
\end{eqnarray}
and
\begin{eqnarray}
a^{(2)}_g&=&
(2J+1) V\int \frac{d^3 p}{(2\pi)^3}
\left[
\frac{e^{-\beta\epsilon_{g}(\vec{p})}}
{
1-2\,e^{-\beta\epsilon_{g}(\vec{p})}
+e^{-2\beta\epsilon_{g}(\vec{p})}
}
\right],
\nonumber\\
&=&
(2J+1)
\left(\frac{V}{\beta^3}\right)\,\left(\frac{1}{6}\right).
\end{eqnarray}
It should be noted that when the color-singlet 
state with the unimodular unitary symmetry group 
representation is projected by integrating 
the partition function over the invariance Haar measure 
$\,\int\, d\mu\, Z_{q\overline{q}g}\,$ in the Hilbert space,
the Fock space component of the adjoint gluon grand canonical 
ensemble is approximated to 
\begin{eqnarray}
\ln\,{Z }_{g}\left(\beta,V\right) 
&=& a^{(0)}_g 
-\frac{1}{2} \frac{\left(2{N_c }
\, {a^{(2)}_{g} }\right)}{2N_c}
 \sum^{N_c}_{i=1}\sum^{N_c}_{j=1}
(\theta_i-\theta_j)^2,
\nonumber\\
&\equiv& a^{(0)}_g 
-\frac{1}{2} \left(2{N_c }\, {a^{(2)}_{g} }\right)
 \sum^{N_c}_{i=1}
\theta_i^2.
\label{adjoint-gluon-quad1}
\end{eqnarray}
The second line in Eq.(\ref{adjoint-gluon-quad1}) 
is obtained under an appropriate transformation.
The above procedure 
for calculating the canonical ensemble
can be extrapolated to other symmetry groups 
such as the orthogonal and symplectic ones 
with the aim to avoid writing the adjoint representation 
for gluons explicitly. 
The same procedure still holds 
when an additional unimodular-like constraint 
is probed in the given symmetry group. 
This extrapolation is essential in order to simplify 
the calculation drastically.
The grand canonical ensemble for a blob 
of quarks and gluons is the tensor product of the Fock spaces 
for the quark and anti-quark canonical partition function
and the gluon canonical partition function. 
Therefore, the quark-gluon partition function just above 
the Gross-Witten Hagedorn transition from
the discrete low-lying mass spectrum 
to the continuous high-lying 
mass spectrum reads
\begin{eqnarray}
Z_{q\bar{q}g}\left(\beta,V\right)
&=&\exp\left(a^{(0)}_{q\bar{q}}+a^{(0)}_g\right)
e^{-\frac{1}{2}
\left(
a^{(2)}_{q\overline{q}}+
2{N_c }\, {a^{(2)}_{g} }\right)
\sum^{N_c}_{i=1}\theta^2_i},
\nonumber\\
&\equiv&
Z_{q\bar{q}g}
\left(\theta_1,\cdots,\theta_{N}\right).
\end{eqnarray}
The confined quark-gluon bag 
must be colorless.
The color-singlet-state (colorless) bags 
correspond the Hagedorn states.
The color-singlet grand canonical partition function 
for the confined quark-gluon bag 
in the $SU(N_c)$ representation reads
\begin{eqnarray}
Z_{H}\left(\beta,V\right)&=&
\int d\,\mu(g) Z_{q\bar{q}g}\left(\beta,V\right),
\nonumber\\
&\sim&
\frac{\exp\left(a^{(0)}_{q\bar{q}g}\right)}{N_c!}
\int\, \left(\prod^{N_c}_{n=1} \frac{d\theta_{n}}{2\pi}\right)\,
2\pi\,\delta\left(\sum^{N_c}_{i=1}\theta_i\right)\,
\left(\prod^{N_c}_{j>i}\left(\theta_{j}-\theta_{i}\right)^{2}\right)
e^{-\frac{1}{2}
a^{(2)}_{q\overline{q}g}
\sum^{N_c}_{i}\theta^2_i},
\nonumber\\
&\sim&
\frac{\exp\left(a^{(0)}_{q\bar{q}g}\right)}{N_c!}
\sqrt{\frac{2\pi a^{(2)}_{q\overline{q}g}}{N_c}}
\int\, \left(\prod^{N_c}_{n=1} \frac{d\theta_{n}}{2\pi}\right)\,
\left(\prod^{N_c}_{j>i}\left(\theta_{j}-\theta_{i}\right)^{2}\right)
e^{-\frac{1}{2}
a^{(2)}_{q\overline{q}g}
\sum^{N_c}_{i}\theta^2_i},
\nonumber\\
\label{hag-canonical1}
\end{eqnarray}
where
\begin{eqnarray}
a^{(0)}_{q\bar{q}g}&=&
a^{(0)}_{q\bar{q}}+a^{(0)}_g,
\nonumber\\
&=&
(2J+1)\left(\frac{V}{\beta^3}\right)
\left(
N_c\sum^{N_{f}}_{q=1}\left[
\frac{7 \pi^2}{360}+\frac{1}{12}\left(\frac{\mu_{q}}{T}\right)^2
+\frac{1}{24 \pi^2}\left(\frac{\mu_{q}}{T}\right)^4
\right]
+N_{adj} \frac{\pi^2}{90}
\right),
\end{eqnarray}
and
\begin{eqnarray}
a^{(2)}_{q\overline{q}g}&=&
a^{(2)}_{q\bar{q}}+2 N_c\, a^{(2)}_g,
\nonumber\\
&=&
(2J+1)\left(\frac{V}{\beta^3}\right)
\left(
\sum^{N_{f}}_{q=1}\frac{1}{6}
\left[1+\frac{3}{\pi^2}\left(\frac{\mu_{q}}{T}\right)^2
\right]
+\frac{2N_c}{6}
\right).
\end{eqnarray}
The partition function that is given by Eq.(\ref{hag-canonical1}) 
leads to the high-lying continuous hadronic mass spectrum 
when it is transformed to the micro-canonical ensemble.

It is useful to mention the following relations.
The canonical ensembles for adjoint particles
in $SU(N_c)$ and $U(N_c)$ symmetry group 
representations are related by 
\begin{eqnarray}
Z_{(Adj)}&=&
\int\, d\mu_{[U(N_c)]} 
e^{-\frac{1}{2} a_{Adj} \sum^{N_c}_{i}\sum^{N_c}_{j}
\left(\theta^{2}_{i}-\theta^{2}_{j}\right)^{2}},
\nonumber\\
&\sim&
N_{c}\,\int\, d\mu_{[SU(N_c)]} 
e^{-\frac{1}{2} a_{Adj} \sum^{N_c}_{i}\sum^{N_c}_{j}
\left(\theta^{2}_{i}-\theta^{2}_{j}\right)^{2}}.
\nonumber\\
&\sim&
N_{c}\,\int\, d\mu_{[SU(N_c)]} 
e^{-\frac{1}{2} [2N_{c}\,a_{Adj}] 
\sum^{N_c}_{i}\theta^{2}_{i}}.
\end{eqnarray}
The canonical ensembles for gas of fundamental 
and adjoint particles 
in $SU(N_c)$ and $U(N_c)$ symmetry group 
representations are related in the following way:
\begin{eqnarray}
Z_{(Fun+Adj)}&=&
\int\, d\mu_{[U(N_c)]} 
e^{-\frac{1}{2}a_{Fun} \sum^{N_c}_{i}\theta^2_{i}
-\frac{1}{2} a_{Adj} \sum^{N_c}_{i}\sum^{N_c}_{j}
\left(\theta^{2}_{i}-\theta^{2}_{j}\right)^{2}},
\nonumber\\
&\sim&
\frac{1}{\sqrt{a_{Fun}}}
\,
\int\, d\mu_{[SU(N_c)]} 
e^{-\frac{1}{2}a_{Fun} \sum^{N_c}_{i}\theta^2_{i}
-\frac{1}{2} a_{Adj} \sum^{N_c}_{i}\sum^{N_c}_{j}
\left(\theta^{2}_{i}-\theta^{2}_{j}\right)^{2}},
\nonumber\\
&\sim&
\frac{1}{\sqrt{a_{Fun}}}
\,
\int\, d\mu_{[SU(N_c)]} 
e^{-\frac{1}{2} \left[a_{Fun}+2\,N_c\,a_{Adj}\right] 
\sum^{N_c}_{i}\theta^{2}_{i}}.
\end{eqnarray}
%%%%%%%%%%%%%%%%%%%%%%%%%%%%%%%%%%%%%%%%%%%%%%%%%%%%%%
%%%%%%%%%%%%%%%%%%%%%%%%%%%%%%%%%%%%%%%%%%%%%%%%%%%%%%%%%%
%%%%%%%%%%%%%%%%%%%%%%%%%%%%%%%%%%%%%%%%%%%%%%%%%%%%%%%%%%
%%%%%%%%%%%%%%%%%%%%%%%%%%%%%%%%%%%%%%%%%%%%%%%%%%%%%%%%%%

\section{\label{sect3} Internal symmetry}

The canonical ensemble for the quark-gluon bag 
with specific internal symmetry is constructed 
in the standard way by defining the group's eigenvalues 
of the conserved color charges 
which are commutated with the Hamiltonian.
The group's eigenvalues are defined 
by an appropriate transformation 
of the group's generators 
to take a diagonal format as follows
\begin{eqnarray}
\sum^{N}_{i=1} a_i\, {\bf t}_i=
\mbox{diag}\left(\theta_1,\cdots,\theta_N\right),
\end{eqnarray}
where ${\bf t}_i$ are the group's generators 
in the fundamental representation. 
The continuous group's parameters 
$\theta_i$ (where $i=1,\cdots N$)
are treated as variables of the conserved color charges 
which control the composite internal structure. 
In the same way, it is possible to write the diagonal form 
of the adjoint representation as follows
\begin{eqnarray}
\sum^{N_{adj}}_{a=1} a_a\, {\bf T}_a=
\mbox{diag}\left(\phi_1,\cdots,\phi_{N_{adj}}\right).
\end{eqnarray} 
The conserved charges in the adjoint representation 
depend essentially on the fundamental charges and 
they are written as functions 
of the fundamental charges: 
\begin{eqnarray} 
\phi_a = f_{a}\left(\theta_1,\cdots,\theta_N\right),
\end{eqnarray} 
where the subscript index, namely, $a$ 
runs over the adjoint index $a=1,\cdots,N_{adj}$. 
For instance with a suitable choice 
of the unitary representation $U(N)$, 
the conserved adjoint charges $\phi_{a}$
are related to the fundamental charges
$\theta_{i}$ 
by $\phi_{\underbrace{(ij)}}=(\theta_i-\theta_j)$. 
The adjoint index $a$ is written in term 
of the double fundamental indexes $i\,j$. 
The group's (energy-) eigenvalues which control 
the composite internal structure degrees of freedom 
correspond the conserved charges. 
The dual role of the group's energy-eigenvalues 
and the conserved charges is the mechanism behind 
the symmetry breaking/restoration 
(i.e. symmetry decomposition/reduction 
from symmetry group to another). 
It is responsible of the phase transition
from a composite with a specific internal structure 
to another state with different symmetry configuration. 
This mechanism becomes obvious in the example 
of the phase transition scenario $SU(N)\rightarrow U(1)^{N-1}$.
The chemical reaction fugacity 
and the set of characteristic chemical potentials 
of the conserved charges are realized by 
the analytic continuation of the group eigenvalues 
$i\theta_i\rightarrow -\beta\mu_i$. 
The group invariance is transformed into system's fugacities
$\Lambda_{i}=e^{i\theta_i}\rightarrow \Lambda_{i}=e^{-\beta\mu_i}$.
%%%%%%%%%%%%%%%%%%%%%%%%%%%%%%%%%%%%%%%%%%%%%%%%%%%%%%%%%%%%%%%%%%%%%
The internal structure of the composite is incorporated 
by writing the canonical ensemble as a function of 
the group energy eigen-variables, namely, 
$Z\left(\theta_1,\cdots,\theta_{N}\right)$.
%%%%%%%%%%%%%%%%%%%%%%%%%%%%%%%%%%%%%%%%%%%%%%%%%%%%%%%%%%%%%%%%
The easiest way to define the canonical ensemble 
of the composite in the Hilbert space
is to write it as a tensor product of 
the Fock space canonical ensembles 
of the constituent quarks and gluons. 
The canonical ensemble for the quark-gluon bag 
is written as follows
\begin{eqnarray}
Z_{q\overline{q}g}\left({\bf R}(g)\right)&=&
\mbox{tr}\left[ {\bf R}_{q\overline{q} g}(g) 
\exp\left(-\beta H_{q\overline{q}g}\right)\right],
\nonumber\\
&=&
\mbox{tr}
\left[ 
{\bf R}_{q}(g) \exp\left(-\beta H_{q}\right)
\right]
\mbox{tr}
\left[ 
{\bf R}_{\overline{q}}(g) \exp\left(-\beta H_{\overline{q}}\right)
\right]
\mbox{tr}
\left[ 
{\bf R}_{g}(g) \exp\left(-\beta H_{g}\right)
\right].
\nonumber\\
\end{eqnarray}
The canonical ensemble for quark gas in the Fock space reads
\begin{eqnarray}
\mbox{tr}
\left[
{\bf R}_{q}(g) \exp\left(-\beta H_{q}\right)
\right]
=
\exp\left[
\sum_{\alpha} \mbox{tr} 
\ln\left(1+ {\bf R}_{fun}(g) e^{-\beta {H_{q}}_{\alpha}}\right)
\right],
\end{eqnarray}
while the canonical ensemble for gluon gas becomes 
\begin{eqnarray}
\mbox{tr}
\left[
{\bf R}_{g}(g) \exp\left(-\beta H_{g}\right)
\right]
=
\exp\left[
-\sum_{\alpha} \mbox{tr} 
\ln\left(1- {\bf R}_{adj}(g) e^{-\beta {H_{g}}_{\alpha}}\right)
\right].
\end{eqnarray}
%%%%%%%%%%%%%%%%%%%%%%%%%%%%%%%%%%%%%%%%%%%%%%%%%%%%%%%%%%%%%%%%%
The internal structure is incorporated by the group eigenvalues 
of the conserved charges. 
The fundamental and adjoint color charges are specified, 
respectively, by
\begin{eqnarray}
{\bf R}_{fun}(g)=
U^{-1}\,
\left(
\exp\left[ i\,\mbox{diag}(\theta_1,\cdots\theta_N)\right]
\right)
\,U,
\end{eqnarray}
and 
\begin{eqnarray}
{\bf R}_{adj}(g)=
{U_{adj}}^{-1}\,
\left(
\exp\left[ i\,\mbox{diag}(\phi_1,\cdots\phi_{N_{adj}})\right]
\right)
\,{U_{adj}},
\end{eqnarray}
where $U$ and $U_{adj}$ are appropriate transformation matrices 
and they are reduced to the identity matrices 
(i.e. $U=1$ and $U_{adj}=1$)
for the unitary group representation.
%%%%%%%%%%%%%%%%%%%%%%%%%%%%%%%%%%%%%%%%%%%%%%%%%%%%%%%%%%%%%%%%%
Finally, the canonical ensemble with a specific internal structure 
is projected using the standard orthogonal expansion procedure.
The specific symmetry state is projected by the integration 
of a specific projection state 
over the symmetry group invariance Haar measure 
in the following way,
\begin{eqnarray}
Z_{\chi}=\int d\mu_{G}\left(\theta_1,\cdots \theta_N\right) 
\left[\chi\times
Z\left(\theta_1,\cdots \theta_N\right)\right],
\end{eqnarray} 
where $\chi=\chi\left(\theta_1,\cdots \theta_N\right)$ 
is an orthogonal basis.
The integral element $d\mu_{G}$ is the symmetry group invariance 
Haar measure for a specific symmetry group.
The invariance Haar measure is essential 
in order to accommodate the composite internal symmetry. 
The color confinement is manifested by maintaining 
the color singlet state. 
Fortunately, the singlet state is the simplest case 
and its basis is given by $\chi_{singlet}=1$.
The prime objective is to confine the quarks 
in a color singlet state regardless of 
the composite internal group symmetry.
%%%%%%%%%%%%%%%%%%%%%%%%%%%%%%%%%%%%%%%%

\subsection{Unitary symmetry group $U(N)$}

In the context of the unitary symmetry group transformation, 
the physical Hamiltonian is subject 
to the Hermitian condition
\begin{eqnarray}
H=H^{\dagger}.
\end{eqnarray}
The unitary symmetry group subjects to the following constraint
\begin{eqnarray}
{U^{\dagger}}\, {U}=1, 
\mbox{hence}\, {U^{-1}}={U^{\dagger}},
\end{eqnarray}
where the matrix $U$ is a complex one 
and 
the Hermitian matrix is defined by 
$U^{\dagger}={U^{*}}^{T}$.
%%%%%%%%%%%%%%%%%%%%%%%%%%%
The unitary matrix $U$ has $2N^2$ real parameters 
where $N^2$ comes from the real part of the unitary matrix 
and $N^2$ comes from the imaginary part. 
The unitary condition imposes $\left[N+2 N(N-1)/2\right]$
constraints on $2N^2$ real parameters 
to give $N_{adj}=N^2$ essential real parameters.
Hence, the unitary symmetry 
has $N_{fun}=N$ and $N_{adj}=N^2$ 
for the fundamental and adjoint charges, respectively.
%%%%%%%%%%%%%%%%%%%%%%%%%%%%%%%%%%%%%%%%%%%%%%%%%%%%%%%%

The Gaussian unitary ensemble multiplied 
by the group volume element is defined in terms 
of Hermitian matrices as follows
\begin{eqnarray}
Z\left(H\right)\, dH \left(\mbox{invariant }\right),
\end{eqnarray}
which is invariant under automorphism unitary transformation.
The invariance Haar measure is given 
by the volume element
\begin{eqnarray}
d H = \prod_{k\le j} d H^{(0)}_{jk}\, 
\prod_{k < j} d H^{(1)}_{kj},
\end{eqnarray}
where $H^{(0)}_{kj}$ and $H^{(1)}_{kj}$ are real and imaginary
parts of $H_{kj}$.
The partition function $Z\left(H\right)$ 
is a function of only the eigenvalues $\theta_i$ 
where $i$ runs over $1,\cdots, N$.
This means that the Hermitian Hamiltonian 
commutes with the group generators 
those correspond the group eigenvalues 
$\theta_i$ where $i$ runs over $1,\cdots, N$.
The invariance Haar measure is reduced to
\begin{eqnarray}
\int d\mu\left(\theta_1\cdots\theta_N\right)
&=&
N_{U}\,\frac{1}{N!}\int^{\pi}_{-\pi} 
\left({\prod^{N}_{i=1} \frac{d\theta_{i}}{2\pi}}
\right)
\,
\prod^{N}_{i< j} 
4\left(e^{i\theta_i}-e^{i\theta_j}\right)
\left(e^{i\theta_i}-e^{i\theta_j}\right)^{*}
\nonumber\\
&=&
N_{U}\,
\frac{1}{N!}\int^{\pi}_{-\pi} 
\left({\prod^{N}_{i=1} \frac{d\theta_{i}}{2\pi}}\right)
\,\prod^{N}_{i< j} 
4\sin^2\left(\frac{\theta_i-\theta_j}{2}\right),
\end{eqnarray}
where $-\pi\le\theta_{i}\le\pi$ 
for the circular ensemble.
The variables $(\theta_1,\cdots,\theta_{N})$ 
are the diagonalized group eigenvalues. 
The number of adjoint charges is reduced to $N_{adj}=N^2$.
The normalisation constant $N_{U}$ is determined by
\begin{eqnarray}
\frac{1}{N_{U}}&=&\frac{1}{N!}\int^{\pi}_{-\pi} 
\left({\prod^{N}_{i=1} \frac{d\theta_{i}}{2\pi}}\right)
\,\prod^{N}_{i< j} 
4\sin^2
\left(\frac{\theta_i-\theta_j}{2}\right).
\end{eqnarray}
The saddle points of the group eigen-variables accumulate 
around the origin for
the hot and dense nuclear matter under extreme conditions.
Subsequently, the canonical ensemble can be written 
as the Gaussian-like ensemble in the following way,
\begin{eqnarray}
\int d\mu\left(\theta_1\cdots\theta_N\right) 
e^{-a\sum^{N_c}_{i} \theta^2_i}.
\end{eqnarray}
Therefore, the invariance Haar measure 
is approximated to
\begin{eqnarray}
\int d\mu\left(\theta_1\cdots\theta_N\right) 
\left(\cdots\right)
&\approx&
N_{U}\,
\frac{1}{N!}\int^{\infty}_{-\infty} 
{\prod^{N}_{i=1} \frac{d\theta_{i}}{2\pi}} 
\,
\prod^{N}_{i< j} 
4\left(\frac{\theta_i-\theta_j}{2}\right)^2
\, \left(\cdots\right),
\nonumber\\
&\approx&
{N}_{U}\,
\frac{1}{N!}\int^{\infty}_{-\infty} 
{\prod^{N}_{i=1} \frac{d\theta_{i}}{2\pi}} 
\,
\prod^{N}_{i< j} 
\left(\theta_i-\theta_j\right)^2
\, \left(\cdots\right).
\end{eqnarray}
The unimodular condition for the special 
unitary symmetry group $SU(N)$
is imposed by 
\begin{eqnarray}
\sum^{N}_{i}\theta_i=0.
\end{eqnarray}
Hence, the invariance Haar measure 
for $SU(N)$ becomes
\begin{eqnarray}
\int d\mu_{SU}\left(\theta_1\cdots\theta_N\right) 
e^{-\frac{1}{2} a\sum^{N}_{i}\theta^2_i}
&=&
N_{U}\,
\frac{1}{N!}\int^{\infty}_{-\infty} 
{\prod^{N}_{i=1} \frac{d\theta_{i}}{2\pi}}
2\pi\delta\left(\sum^{N}_{i}\theta_i\right)
\nonumber\\
&~&\,\,\,\,\times
\,\prod^{N}_{i< j} 
\left(\theta_i-\theta_j\right)^2\, 
e^{-\frac{1}{2} a\sum^{N}_{i}\theta^2_i}.
\end{eqnarray}
By writing the delta function as
\begin{eqnarray}
\delta\left(\sum^{N}_{i}\theta_i\right)=
\int^{\infty}_{-\infty} 
d x\, \exp\left(i\,x\sum^{N}_{i}\theta_i\right),
\end{eqnarray}
the color-singlet state of the canonical ensemble
for the unimodular-like $U(N_c)$ (i.e. $SU(N_c)$) 
reads
\begin{eqnarray}
\int d\mu_{SU}\left(\theta_1\cdots\theta_{N}\right) 
e^{-\frac{1}{2} a\sum^{N}_{i=1}\theta^2_i}
&=&
\sqrt{\frac{2\pi a}{N}}
\int d\mu_{U}\left(\theta_1\cdots\theta_N\right) 
e^{-\frac{1}{2} a\sum^{N}_{i=1}\theta^2_i}.
\nonumber\\
\end{eqnarray}
%%%%%%%%%%%%%%%%%%%%%%%%%%%%%%%%%%%%%%%%%%%%%%%%%%%%%%%%%%%%%%%%%%%%%%%%%
%%%%%%%%%%%%%%%%%%%%%%%%%%%%%%%%%%%%%%%%%%%%%%%%%%%%%%%%%%%%%%%%%%%%%%%%%
%%%%%%%%%%%%%%%%%%%%%%%%%%%%%%%%%%%%%%%%%%%%%%%%%%%%%%%%%%%%%%%%%%%%%%%%%

\subsection{Orthogonal symmetry group $O_{(S)}(N)$}

The unitary Hermitian system may be decomposed 
into symmetric Hermitian and anti-symmetric Hermitian 
real orthogonal components.
The orthogonal symmetry group is symmetric one
and is invariant under the time-reversal transformation.
The unitary Hermitian system can be fully or partially 
projected into (symmetric) orthogonal Hermitian system.  
The projection of the unitary Hermitian system 
into the real symmetric orthogonal system 
is very relevant to the quark-gluon plasma 
under extreme conditions such as the hot nuclear matter.
The transformation matrix, namely, $U$ 
of the orthogonal group is a symmetric one. 
It has $N^{2}$ elements and these elements 
are restricted to $N+N(N-1)/2$ conditions. 
These constraints reduce the number of 
the adjoint variables to 
$N_{adj}=N(N-1)/2$ 
while the number of the fundamental variables 
remain the same $N_{fun}=N$ 
in the orthogonal symmetry group.
The invariance Haar measure for 
the orthogonal symmetry group reads
\begin{eqnarray}
\int d\mu_{O_{(S)}}\left(\theta_1\cdots\theta_N\right)
&=&
N_{O}\,\int^{\pi}_{-\pi} 
{\prod^{N}_{i=1} \frac{d\theta_{i}}{2\pi}}
\,\prod^{N}_{i< j} 
\left|\,\sin\left(\frac{\theta_i-\theta_j}{2}
\right)\,\right|.
\end{eqnarray}
The normalisation constant $N_{U}$ is determined by
\begin{eqnarray}
N_{O}\,\int^{\pi}_{-\pi} 
{\prod^{N}_{i=1} \frac{d\theta_{i}}{2\pi}}
\,\prod^{N}_{i< j} 
\left|\, 2\sin
\left(\frac{\theta_i-\theta_j}{2}\right)\,\right|
=1.
\end{eqnarray}
When the saddle points accumulate around the origin,
the invariance Haar measure becomes
\begin{eqnarray}
\int d\mu_{O_{(S)}}\left(\theta_1\cdots\theta_N\right)
e^{-\frac{1}{2} a\sum^{N}_{i=1}{\theta}^2_i}
&=&
N_{O}\,\int^{\infty}_{-\infty} 
{\prod^{N}_{i=1} \frac{d\theta_{i}}{2\pi}}
\,\prod^{N}_{i< j} 
\left|\,\theta_i-\theta_j\,\right|
e^{-\frac{1}{2} a\sum^{N}_{i=1}{\theta}^2_i}.
\nonumber\\
\end{eqnarray}
%%%%%%%%%%%%%%%%%
The unimodular-like orthogonal $O_{(S)}(N_c)$ 
symmetry group is constructed by imposing 
the unimodular-like constraint 
$\sum^{N_c}_{i=1}\theta_{i}=0$ 
to the standard orthogonal symmetry group. 
The color-singlet Gaussian integral 
for the unimodular-like orthogonal symmetry group 
$S\,O_{(S)}(N)$ 
is related to $O_{(S)}(N)$ in the following way
\begin{eqnarray}
&~&\int d\mu_{O_{(S)}}
\left(\theta_1\cdots\theta_N\right)
e^{
-\frac{1}{2} a\,\sum^{N}_{i=1}{\theta}^2_i
},   ~~(\mbox{with the unimodular-like constraint}),
\nonumber\\
&~& \,~~~~~~~~~~~\,
\sim 
\int d\mu_{O_{(S)}}\left(\theta_1\cdots\theta_N\right)
2\pi\,\delta\left(\sum^{N}_{i=1}\theta_i\right)\,
e^{-\frac{1}{2} a\sum^{N}_{i=1}{\theta}^2_i}
\nonumber\\
&~& \,~~~~~~~~~~~\,
\sim
\sqrt{\frac{2\pi\, a}{N}}
\int d\mu_{O_{(S)}}\left(\theta_1\cdots\theta_N\right)
e^{-\frac{1}{2} a\sum^{N}_{i=1}{\theta}^2_i}.
\end{eqnarray}
%%%%%%%%%%%%%%%%%%%%%%%%%%%%%%%%%%%%%%%%%%%%%%%%%%%%%%%%%%%%%%%%%%%%%%%
%%%%%%%%%%%%%%%%%%%%%%%%%%%%%%%%%%%%%%%%%%%%%%%%%%%%%%%%%%%%%%%%%%%%%%%
%%%%%%%%%%%%%%%%%%%%%%%%%%%%%%%%%%%%%%%%%%%%%%%%%%%%%%%%%%%%%%%%%%%%%%%

\subsection{Symplectic symmetry group 
$Sp(N)$ (N is the number of quaternion)}
%%%%%%%%%%%%%%%%

The symplectic symmetry group is time-reversal invariance. 
The anti-symmetrization in the symmetry group $Sp(N)$ 
leads to odd-spin.
The algebra of the symplectic symmetry group can be expressed 
in terms of quaternions. 
The quaternions are convenient to deal with 
the anti-symmetrization. 
The notation $Sp(N)$ is equivalent to $Sp(2N,R)$ 
where $N$ corresponds to the number of quaternions.
%%%%%%%%%%%%%%%%%%%%%%%%%%%%%%%%%%%%%%%%%%%%%%%%%%%%%%
%
%
%%%%%%%%%%%%%%%%%%%%%%

The transformation matrix $U$ of the $Sp(N)$ group 
is of the size of $(2N\,\times\, 2N)$. 
This transformation matrix is considered 
as cut into $N^2$ blocks
of $\, 2\,\times\, 2\,$ and each $\, 2\,\times\, 2\,$ 
block is expressed in terms 
of the quaternions.
In this sense the number of real 
independent parameters these are defined 
in the $(\, 2N\,\times\, 2N\,)$ 
self-dual Hermitian matrix 
is given by the number of the adjoint variables $N_{adj}$. 
The number of the fundamental (energy-) eigen variables
is $N_{fun}=N$ while the number of the adjoint variables 
is $N_{adj}=N\,(2N-1)$.
%%%%%%%%%%%%%%%%%%%%%%%%%
%%%%%%%%%%%%%%%%%%%%%%%%%
%%%%%%%%%%%%%%%%%%%%%%%%%
The invariance Haar measure 
for the symplectic symmetry group 
$Sp(N)$ with the circular canonical 
ensemble is given by 
\begin{eqnarray}
\int d\mu_{Sp}\left(\theta_1\cdots\theta_N\right)
=
N_{Sp}\,\int^{\pi}_{-\pi} 
{\prod^{N}_{i=1} \frac{d\theta_{i}}{2\pi}}\,
\prod_{i< j} 
\left[4\left(e^{i\theta_i}-e^{i\theta_j}\right)
\left(e^{i\theta_i}-e^{i\theta_j}\right)^{*}
\right]^2.
\end{eqnarray}
The normalisation constant $N_{Sp}$ is determined by
\begin{eqnarray}
\frac{1}{N_{Sp}}&=&
\int^{\pi}_{-\pi}{\prod^{N}_{i=1} \frac{d\theta_{i}}{2\pi}}
\,\prod^{N}_{i< j} 
16\sin^4
\left(\frac{\theta_i-\theta_j}{2}\right).
\end{eqnarray}
Under extreme conditions, 
the saddle points 
of the group continuous variables 
accumulate around the origin.
Therefore, the canonical ensemble is 
approximated to the Gaussian-like ensemble
in the following way
\begin{eqnarray}
\int d\mu_{Sp}\left(\theta_1\cdots\theta_N\right) 
e^{
-a\sum^{N_c}_{i} 
\theta^2_i +b\sum^{N_c}_{i} \theta_i+c
},
\end{eqnarray}
where $a$ is real and positive 
and $b$ and $c$ are real. 
The invariance Haar measure 
for $Sp(N)$ is approximated to
\begin{eqnarray}
\int d\mu_{Sp}\left(\theta_1\cdots\theta_N\right) 
\left(Z\right) 
\approx
{N}_{Sp}\,\int^{\infty}_{-\infty} 
{\prod^{N}_{i=1} \frac{d\theta_{i}}{2\pi}} \,
\prod^{N}_{i< j} 
\left(\theta_i-\theta_j\right)^4 
\left(Z\right)\,.
\end{eqnarray}
%%%%%%%%%%%%%%%%%

The result for the canonical partition function
with the symplectic symmetry group 
can be extrapolated around the saddle points 
to take the following general formula
\begin{eqnarray}
Z\left(\mbox{I.S.}\right)&=&
\int^{\pi}_{-\pi} 
d\,\mu_{Sp}\left(\theta_1,\theta_2,\cdots,\theta_{N}\right)
Z\left(\theta_1,\theta_{2},\cdots,\theta_{N}\right),
\nonumber\\
&=&
Z_0\int^{\infty}_{-\infty} 
d\,\mu_{Sp}\left(\theta_1,\theta_2,\cdots,\theta_{N}\right)
e^{-\frac{1}{2}\, a\,\sum^{N}_{i=1} \theta^2_i},
\end{eqnarray}
where 
$\,Z\left(\theta_1,\theta_{2},\cdots,\theta_{N}\right)
\approx
Z_0\left(\{\theta_{i}\approx 0\}\right)
\,e^{-\frac{1}{2}\, a\,\sum^{N}_{i=1} \theta^2_i}\,$.
The abbreviation (I.S.) means 
that the internal structure and 
the group variables satisfy the constraint  
$(\theta_1-\pi)\le\theta_2
\le\cdots\le\theta_{2N}\le (\theta_1+\pi)$.
%%%%%%%%%%%%%%%%%%%%%%%%%%%%%%%%%%%%%%%%%%%%%
When the unimodular-like constraint is imposed
(i.e. $\sum^{N}_{i}\theta_{i}=0$), 
the color-singlet state of 
the (unimodular-like) symplectic canonical ensemble 
is reduced to
\begin{eqnarray}
Z\left(\mbox{I.S.}\right)&\sim&
Z_{0}\,\times
\,{N}_{Sp}\,
\sqrt{\frac{2\pi a}{N}}\,
\int^{\infty}_{-\infty} 
{\prod^{N}_{i=1} \frac{d\theta_{i}}{2\pi}} \,
\prod^{N}_{i< j} 
\left(\theta_i-\theta_j\right)^4 
\left(
e^{-\frac{1}{2}\, a\,\sum^{N}_{i=1} \theta^2_i}
\right).
\end{eqnarray}
%%%%%%%%%%%%%%%%%%%%%%%%%%%%%%%%%%%%%%%%%%%%%%%%%%%%
%%%%%%%%%%%%%%%%%%%%%%%%%%%%%%%%%%%%%%%%%%%%%%%%%%%%
%%%%%%%%%%%%%%%%%%%%%%%%%%%%%%%%%%%%%%%%%%%%%%%%%%%%

\section{\label{sect4} Equation of State 
for composite nuclear matter}
%%%%%%%%%%%%%%%%%%%%%%%%%%%%%

The color-singlet-state unitary (mixed-grand) 
canonical ensemble 
can be extrapolated to another color-singlet-state
ensemble with different symmetry groups. 
The most important issue in the extrapolation 
from one symmetry group to another one 
is that these symmetry groups have 
the same number $N_c$ 
of fundamental Fourier color charges  
(i.e. group eigen-variables). 
The difference among various symmetry groups comes
from the difference in the numbers 
of the conserved adjoint Fourier charges 
with the same number of fundamental Fourier charges. 
The second difference comes from 
the different invariance Haar measure 
for every kind of symmetry group. 
The invariance Haar measure for each symmetry group 
is essential in order to project 
the color-singlet-state 
of the quark and gluon bag.
Therefore, the effective Vandermonde 
potential is modified drastically 
when the internal symmetry is transferred 
from one symmetry group to another one.
%%%%%%%%%%%%%%%%%%%%%%%%%%%%%%%%%%%%%%%%

The canonical ensemble with specific internal symmetry group 
and that is projected to a certain quantum wavefuntion 
is given by
\begin{eqnarray}
Z_{state}\left(\beta,V\right)&=&
\int \, d\mu_{G}(g) \,\left[ \chi_{state}\,
\times\, 
Z_{q\bar{q}g}\left(\beta,V;\theta_1,\cdots,\theta_{N}\right)
\right],
\nonumber\\
&\sim&
{\cal N}\, Z_{0}\left(\beta,V\right)\,
\int\, \left(\prod^{N_c}_{n=1} d\theta_{n}\right)\,
\prod_{j>i}\left|\theta_{j}-\theta_{i}\right|^{G}
\,
\left(\chi_{state}\right)\,
e^{-\frac{1}{2}
a^{(2)}_{q\overline{q}g}
\sum^{N_c}_{i}\theta^2_i},
\end{eqnarray}
where 
$Z_0\left(\beta,V\right)=\exp\left(a^{(0)}_{q\bar{q}g}\right)$.
%%%
The symmetry group structure is defined 
by $G=1$, $2$, $4$  
for the orthogonal, unitary and symplectic ensemble, 
respectively.
The color-singlet-state quantum wave-function is given 
by $\chi_{singlet}=1$ regardless of the symmetry group.
In general the physical state 
in QCD is assumed to satisfy the condition 
$\mbox{det}(G)=1$ or equivalently 
it is restricted to the unimodular-like
constraint, namely, $\sum^{N_c}_{i=1}\theta_i=0$.
The unimodular-like constraint modifies 
the extrapolated
color-singlet-state canonical Gaussian ensemble
with a specific group 
such as unitary, orthogonal and symplectic 
symmetry groups 
to the following quantity
\begin{eqnarray}
Z_{singlet}\left(\beta,V\right)
&\sim&
{\cal N}\, Z_{0}\left(\beta,V\right)
\,
\sqrt{\frac{2 \pi\,a^{(2)}_{q\overline{q}g}}{N_c}}
\int\, \left(\prod^{N_c}_{n=1} d\theta_{n}\right)\,
\prod^{N_c}_{j>i}\left|\theta_{j}-\theta_{i}\right|^{G}
\nonumber\\
&~&
~~~~\times\,
e^{-\frac{1}{2}
a^{(2)}_{q\overline{q}g}
\sum^{N_c}_{i=1}\theta^2_i},
\label{group-extrapolation-1}
\end{eqnarray} 
where ${\cal N}=N_{U}$, $N_{O}$ and $N_{Sp}$ 
for the unitary, orthogonal and symplectic symmetry groups, 
respectively. 
Eq.(\ref{group-extrapolation-1}) is derived by imposing 
the unimodular-like constraint 
using the following trick
\begin{eqnarray}
\delta\left(\sum^{N}_{i=1}\theta_i\right)=
\int^{\infty}_{-\infty}\, d x 
\, e^{i x\sum^{N}_{1}\theta_i}.
\end{eqnarray}
The integral part in 
Eq.(\ref{group-extrapolation-1}) 
is evaluated as follows
\begin{eqnarray}
I&=&
\int\, \left(\prod^{N_c}_{n=1} d\theta_{n}\right)\,
\prod^{N_c}_{j>i}\left|\theta_{j}-\theta_{i}\right|^{G}
\,e^{-\frac{1}{2}
a^{(2)}_{q\overline{q}g}
\sum^{N_c}_{i=1}\theta^2_i},
\nonumber\\
&=&
\left(
\frac{G}{ a^{(2)}_{q\overline{q}g} }
\right)^{G\,(N^2_c-N_c)/4+N_c/2}
\int^{\infty}_{-\infty}\, \left(\prod^{N_c}_{n=1} dx_{n}\right)\,
\prod_{j>i}\left|x_{j}-x_{i}\right|^{G}
e^{-\frac{1}{2}\,G\,
\sum^{N_c}_{i}\, x^2_i},
\nonumber\\
&=&
{\cal N}_{N_c\,G}
\left(
\frac{G}{ a^{(2)}_{q\overline{q}g} }
\right)^{\alpha}.
\end{eqnarray}
The exponent $\alpha$ is determined by
\begin{eqnarray}
\alpha=G\,(N^2_c-N_c)/4+N_c/2.
\label{mass-exponent-1}
\end{eqnarray}
In QCD where the number of colors becomes $N_c=3$, 
the exponent $\alpha$ is reduced 
to $3$, $9/2$ and $15/2$
for the orthogonal, unitary and 
symplectic color-singlet ensembles, respectively, 
and these groups are fallen 
under an additional unimodular-like constraint.
The normalisation constant is determined by
\begin{eqnarray}
{\cal N}_{N_c\,G}&=&
(2\pi)^{\frac{N_c}{2}}
G^{-\alpha}
{\prod^{N_c}_{n=1}\Gamma\left(1+\frac{1}{2}\,G\, n\right)}/
{\left[\Gamma\left(1+\frac{1}{2}\,G\,\right)\right]^{N_c}},
\end{eqnarray}
where $G=1,2$ and $4$ for the orthogonal, unitary 
and symplectic symmetry groups, respectively.
It should be noted that 
$\Gamma(n)$ is the gamma function.
%%%%%%%%%%%%%%%%%%%%%%%%%%%%%%%%%%%
The color-singlet-state canonical ensemble 
with a certain internal symmetry 
that is fallen under the additional 
unimodular-like condition is reduced 
to the following general formula,
\begin{eqnarray}
Z_{singlet}\left(\beta,V\right)&\sim&\sqrt{2\pi/N_c}
{\cal N}\, {\cal N}_{N_c\,G}\, G^{\alpha} 
\left(
a^{(2)}_{q\overline{q}g}
\right)^{-(\alpha-1/2)}
\exp\left(a^{(0)}_{q\bar{q}g}\right).
\label{singlet-grand-canonical-general1}
\end{eqnarray}
When the term $\left(V/\beta^{3}\right)$ is parameterised,
the color-singlet-state (mixed-grand) canonical ensemble 
which is given in 
Eq.(\ref{singlet-grand-canonical-general1})
becomes 
\begin{eqnarray}
Z_{q\overline{q}g}\left(\beta,V\right)
&=&
C_{N_c\,G}
\,
\left(
\frac{\beta^3/V}{\tilde{a}^{(2)}_{q\overline{q}g} }
\right)^{\alpha-\frac{1}{2}}
\,
\exp
\left(\frac{V}{\beta^3}\,\tilde{a}^{(0)}_{q\bar{q}g}
\right).
\end{eqnarray}
where
\begin{eqnarray}
\begin{array}{l}
\tilde{a}^{(0)}_{q\bar{q}g}=
a^{(0)}_{q\bar{q}g}/
\left(\frac{V}{\beta^3}\right),
\\
\tilde{a}^{(2)}_{q\bar{q}g}=
a^{(2)}_{q\bar{q}g}/
\left(\frac{V}{\beta^3}\right).
\end{array}
\end{eqnarray}
The normalisation constant $C_{N_c\,G}$ for short reads
\begin{eqnarray}
C_{N_c\,G}=
\sqrt{2\pi/N_c}
{\cal N}\, {\cal N}_{N_c\,G}\, G^{\alpha}.
\end{eqnarray}
We can relax the definition of the normalisation constant,
namely, $C_{N_c\,G}$.
This modified constant $C_{N_c\,G}$ 
will be determined from the physical constraint 
in order to satisfy the thermodynamic ensemble 
under realistic situations. 
The thermodynamic ensemble is concave up with respect 
to the  variable $x=V/\beta^3$~\cite{Zakout:2007nb}.
%%%%%%%%%%%%%%%%%%%%%%%%%%
It is evident that the physical threshold 
must start from bottom point of the concave up curve. 
The line on the left hand side of the threshold 
point is unphysical and must be excluded. 
This behaviour furnishes an additional physical constraint 
that mentioned above. 
This provides a strong signature 
that analytical solution has been changed somewhere 
on the right hand side of the threshold point. 
The change in the analytical solution behaviour 
indicates that the canonical ensemble transmutes 
from one state to another but the deconfinement 
phase transition is still not reached yet.
%%%%%%%%%%%%%%%%%%%%%%%%%%%
Furthermore, the value of the thermodynamic ensemble 
above the threshold of the Gross-Witten point is restricted 
to fall in a physical region (i.e. non-negative) 
as follows, 
\begin{eqnarray}
\ln\left[\,Z_{q\overline{q}g}(\mbox{threshold})\,\right]
\,\ge\,0.
\end{eqnarray}
The critical (threshold) point 
$x_c\,\equiv\,x_{min}$ 
is calculated by extremizing the logarithm 
of the canonical ensemble.
The canonical ensemble is written as follows
\begin{eqnarray}
Z_{q\overline{q}g}\left(x\right)&\sim& 
C\,\left(\tilde{a}^{(2)}_{q\overline{q}g}\, x\right)^{-(\alpha-1/2)}
\, \exp\left(\tilde{a}^{(0)}_{q\overline{q}g}\,x
\right),
\nonumber\\
\ln Z_{q\overline{q}g}\left(x\right) &\sim&
-(\alpha-1/2)\ln(x) + \tilde{a}^{(0)}_{q\overline{q}g}\,x
+
\left[
\ln C-(\alpha-1/2)\ln \tilde{a}^{(2)}_{q\overline{q}g}
\right].
\end{eqnarray}
The value of the relaxation constant $C=C_{N_c G}$
is fixed in order to regulate the solution and 
match some physical requirements~\cite{Zakout:2007nb}.
The threshold point $x_{th}$ is determined by 
calculating the minimum point of the concave 
up curve as follows 
\begin{eqnarray}
x_{th}&=&x_{min},
\nonumber\\
&=&(\alpha-1/2)/\tilde{a}^{(0)}_{q\overline{q}g}.
\end{eqnarray}
The limit of the normalisation 
constant $C$ is determined by
\begin{eqnarray}
\ln Z\left(x_{th}\right)\ge \ln Z_{0}\ge 0.
\end{eqnarray}
The energy threshold is found 
\begin{eqnarray}
\left.\left(\frac{V}{\beta^3}\right)\right|_{th}&=&x_{th},
\nonumber\\
&=&(\alpha-1/2)/\tilde{a}^{(0)}_{q\overline{q}g}.
\end{eqnarray}
The threshold Hagedorn mass 
is estimated from 
$\left.\left(
\frac{V}{\beta^3}
\right)\right|_{threshold}$.
As it will be demonstrated below 
when the canonical ensemble is transformed 
to the micro-canonical ensemble using
Eq.(\ref{Laplace-ensemble1}), 
the saddle point method leads 
to the following dual connection
\begin{eqnarray}
\left(\frac{V}{\beta^3}\right)\rightarrow 
\frac{1}{3 \tilde{a}^{(0)}_{q\overline{q}g}}
\left(
3 \tilde{a}^{(0)}_{q\overline{q}g}\,VW^{3}
\right)^{1/4}.
\end{eqnarray}
In the context of standard MIT bag model, 
the above relation is reduced to
\begin{eqnarray}
\left(\frac{V}{\beta^3}\right)&\rightarrow& 
\frac{m}{
4\,(\tilde{a}^{(0)}_{q\overline{q}g})^{3/4}
\,B^{1/4}},
\nonumber\\
&\ge& 
\left(\alpha-\frac{1}{2}\right)\left(\frac{1}
{\tilde{a}^{(0)}_{q\overline{q}g}}\right),
\end{eqnarray}
where $B$ is the MIT bag constant. 
The bag constant is assigned 
to be approximately $B^{1/4}\,\approx\,200$ MeV.
The constraint of Hagedorn mass threshold 
becomes
\begin{eqnarray}
m_{threshold}\ge 4\left(\alpha-\frac{1}{2}\right)
\left(
\frac{B}{\tilde{a}^{(0)}_{q\overline{q}g}}
\right)^{1/4}.
\end{eqnarray}
%%%%%%%%%%%%%%%%%%%%%%%%%%%%%%%%%%%%%%%%%%%%%%%%%%%% 
%%%%%%%%%%%%%%%%%%%%%%%%%%%%%%%%%%%%%%%%%%%%%%%%%%%%
The asymptotic density of states 
for the micro-canonical ensemble 
with large blob energy $W$ is calculated 
by taking the inverse Laplace transform 
for the grand canonical ensemble. 
The inverse Laplace transform for large $W$ 
is evaluated using the steepest descent method,
\begin{eqnarray} 
Z_{q\overline{q}g}
\left(W,V\right)&=& 
\frac{1}{2\pi i}
\int^{\beta_0+i\infty}_{\beta_0-i\infty}
d\beta e^{\beta W}  
Z_{q\overline{q}g}\left(\beta,V\right)
\nonumber\\
&=&
\frac{1}{2\pi i} 
\int^{\beta_0+i\infty}_{\beta_0-i\infty}
d\beta\,\left[ C\, 
\left(
\tilde{a}^{(2)}_{q\bar{q}g} \frac{V}{\beta^3}
\right)^{-\alpha+\frac{1}{2}}\,  
e^{ \beta W+\frac{V}{\beta^3} \tilde{a}^{(0)}_{q\bar{q}g}}
\right].
\label{Laplace-ensemble1} 
\end{eqnarray}
The straightforward calculation leads 
to the following micro-canonical ensemble,
\begin{eqnarray}
Z_{q\overline{q}g}\left(W,V\right)
&=&
C^{\star} 
\frac{1}{W}\left[
\left(3 \tilde{a}^{(0)}_{q\bar{q}g}
\,V W^{3}\right)^{1/4}
\right]^{-(\alpha-1)}
\,\exp\left[\frac{4}{3}
\left(
3 \tilde{a}^{(0)}_{q\bar{q}g} V W^{3}\right)^{1/4}
\right],
\label{micro-volume-energy}
\end{eqnarray}
where
\begin{eqnarray}
C^{\star}=\frac{1}{2\sqrt{2\pi}}\,
C\,\left(
\frac{3 \tilde{a}^{(0)}_{q\bar{q}g}}
{\tilde{a}^{(2)}_{q\bar{q}g}}
\right)^{\alpha-1/2}.
\end{eqnarray}
The saddle point of the Laplace transform 
is found
\begin{eqnarray}
\beta=\frac{1}{W}\left[
3 \tilde{a}^{(0)}_{q\bar{q}g} V W^3
\right]^{1/4}.
\label{beta-saddle-point1}
\end{eqnarray}
In the standard MIT bag model, 
the bag's mass and volume are given 
by the following relations
\begin{eqnarray}
\begin{array}{l}
m=W+BV,\\
m=4BV,\\
W=\frac{3}{4} m,\\
V=\frac{m}{4B},\\
\left(3 \tilde{a}^{(0)}_{q\bar{q}g} VW^{3}\right)^{1/4}=
\frac{3}{4}
\left(\frac{\tilde{a}^{(0)}_{q\bar{q}g}}{B}\right)^{1/4}
\, m.
\end{array}
\label{standard-MIT-1}
\end{eqnarray}
The density of states for the continuous 
high-lying mass spectrum is given by calculating 
the canonical ensemble and its transformation 
to the micro-canonical ensemble. 
The continuous high-lying mass spectral densities
for the color-singlet-state orthogonal, unitary 
and symplectic symmetry groups 
with an additional unimodular-like constraint
are reduced to the following general 
form
\begin{eqnarray}
\rho_{high}\left(m\right)&=& 
\frac{4}{3}\, C^{\star}
\left[
\frac{3}{4}
\left(\frac{\tilde{a}^{(0)}_{q\bar{q}g}}{B}\right)^{1/4}
\right]^{-\alpha+1}
m^{-\alpha}\, 
\exp\left[
\left(\frac{\tilde{a}^{(0)}_{q\bar{q}g}}{B}\right)^{1/4}
\, m\right].
\end{eqnarray}
The value of the exponent $\alpha=G(N^2_c-N_c)/4+N_c/2$ 
depends on the bag's internal symmetry 
and again $G=1,2$ and $4$ correspond 
the orthogonal, unitary and symplectic symmetry groups, 
respectively. 
The unimodular-like constraint is assumed 
implicitly in all symmetry groups 
those are considered in the present work. 
The normalisation constant $C^{\star}$ 
depends on the group kind.
Moreover, the continuous high-lying mass 
spectral density is simplified to
\begin{eqnarray}
\left.\begin{array}{c}
\rho_{(II)}(m)=\rho_{high}\left(m\right),
\\
\rho_{high}\left(m\right)
= c\, m^{-\alpha}\, e^{b\, m}, 
\\ 
b=\left(
\frac{\tilde{a}^{(0)}_{q\bar{q}g}}{B}
\right)^{1/4}, 
\\
c\sim\frac{4}{3}\, C^{\star}
\left[\frac{3}{4}
\left(\frac{\tilde{a}^{(0)}_{q\bar{q}g}}{B}\right)^{1/4}
\right]^{-\alpha+1}
\end{array}\right\}.
\end{eqnarray}
The mass spectral exponent $\alpha$ 
in the mass spectral density $\rho_{II}(m)$ 
is reduced to 
$\alpha=\frac{N^2_c}{4}+\frac{N_c}{4}$,
$\alpha=\frac{N^2_c}{2}$ 
and
$\alpha=N^2_c-\frac{N_c}{2}$ 
for the unimodular-like orthogonal, unitary 
and symplectic Hagedorn states, respectively.
%%%%%%%%%%%%%%%%%%%%%%%%%%%%%%%%%%%%%%%%%%%%%%%%%%%%%
%%%%%%%%%%%%%%%%%%%%%%%%%%%%%%%%%%%%%%%%%%%%%%%%%%%%%
Furthermore, the density of states for the colored bags 
with color $SU(N_c)$ symmetry group 
(i.e. non-singlet color states)
can be written in similar manner 
but the mass spectral exponent $\alpha$ takes 
the minimum value $\alpha=\frac{1}{2}$ 
regardless the number of colors.
%%%%%%%%%%%%%%%%%%%%%%%%%%%%%%%%%

The scenario of the phase transition for 
the dilute and hot nuclear matter 
(i.e. $\mu_B\approx 0$ and 
$T\rightarrow T_{critical}$)
seems to pass multi-process phase transitions.
The color-singlet-state $SU(N_c)$ symmetry group 
(Hagedorn states)
is broken to the color-singlet-state
unimodular-like $U(1)^{N_c}$ symmetry group 
(equivalent to $U(1)^{N_c-1}$) 
and, subsequently, the number 
of adjoint color degrees of freedom
is reduced from $N^2_c-1$ to $N_c-1$.
This can be interpreted that $N^{2}_c-N_{c}$ 
gluon degrees of freedom are released from 
the unitary Hagedorn states when they 
are mutated to $U(1)^{N_c}$ Hagedorn states.
Furthermore, the symmetry transmutation 
from the $SU(N_c)$ color-singlet state 
to the $U(1)^{N_c}$ color-singlet state 
likely takes place through the intermediate 
the $O_{S}(N_c)$ color-singlet state.

%%%%%%%%%%%%%%%%%%%%%%%%%%%%%%%%%%%%%%%%%%%%
The crucial phenomenon of the model is that 
the quarks and anti-quarks tends 
to be confined in the $U(1)^{N_c}$ 
Hagedorn states (i.e. colorless states).
These states can be interpreted 
as quark liquid bubbles. 
The construction of the color-singlet-state 
for (unimodular-like) $U(1)^{N_c}$ quark-bag
is different from those for the color-singlet-state 
bags with the unitary, orthogonal 
and symplectic internal symmetry groups 
with the unimodular-like constraint.
The $U(1)^{N_c}$ symmetry group maintains 
the minimal gluonic content in the quark-bag
(i.e. the minimal interaction among 
the quarks and antiquarks since 
$N_{fun}=N_{c}$ and $N_{adj}=N_c$).
The $U(1)^{N_c}$ color-singlet-state  
canonical ensemble for quark and antiquark 
bag is constructed in the following way
\begin{eqnarray}
Z_{qbag}(\beta,V)&=&
\left[
\int\, d\mu_{U(1)}(\theta)\, Z_{q\overline{q} {g}}
(\beta,V,\theta)
\right]^{N_c-1},
\nonumber\\
&=&
\left[\frac{1}{2\pi}
\int^{\pi}_{-\pi}\, d\theta\, Z_{q\overline{q} {g}}
(\beta,V;\theta)
\right]^{N_c-1}.
\end{eqnarray}
Instead of considering the $U(1)^{N_c}$ symmetry group, 
the $U(1)^{N_c}$ symmetry group 
with the unimodular-like constraint 
$\left(\sum^{N_c}_{i}\theta_{i}=0\right)$ 
is adopted in order to be consistent 
with unitary, orthogonal and symplectic 
symmetry groups those are restricted 
to the unimodular-like condition and have 
the same number of the fundamental color charges.
%%%%%%%%%%%%%%%%%%%%%%%%%%%%%%%%%%
Hence, the canonical ensemble for 
the $U(1)^{N_c}$ color-singlet state 
of the quark-bag is constructed as follows
\begin{eqnarray}
Z_{qbag}(\beta,V)&=&
\int\,\prod_{i}\,d\mu_{U(1)}(\theta_i)
\,2\pi\delta\left(\sum^{N_c}_{i}\theta_i\right)\,
\prod^{N_c}_{i}
Z_{q\overline{q} {g}}(\beta,V,\theta_i),
\nonumber\\
&=&
\int\, \prod^{N_c}_{i}\,\frac{d\theta_i}{2\pi}
\,2\pi\delta\left(\sum^{N_c}_{j}\theta_j\right)\,
\prod^{N_c}_{k}
Z_{q\overline{q} {g}}(\beta,V,\theta_k),
\end{eqnarray}
where it is restricted 
to the unimodular-like constraint.
At extreme condition, it is reduced to
\begin{eqnarray}
Z_{qbag}(\beta,V)&=&
\sqrt{
\frac{2\pi}{N_c} 
\frac{V}{\beta^{3}}
\tilde{u}^{(2)}_{q\overline{q} {g}}
}\,\exp\left(
\frac{V}{\beta^{3}}
\tilde{u}^{(0)}_{q\overline{q} {g}}
\right)
\int \prod^{N_c}_{i}\frac{d\theta_i}{2\pi}
\exp\left(
-\frac{1}{2} \frac{V}{\beta^{3}}
\tilde{u}^{(2)}_{q\overline{q} {g}}\,
\sum^{N_c}_{i}\theta_{i}^{2}
\right),
\nonumber\\
&=&
\frac{1}{\sqrt{N_c}}
\left(2\pi\right)^{-N_c/2+1/2}
\left(
\frac{V}{\beta^{3}}
\tilde{u}^{(2)}_{q\overline{q} {g}}
\right)^{-N_c/2+1/2}\,
\exp\left(
\frac{V}{\beta^{3}}
\tilde{u}^{(0)}_{q\overline{q} {g}}
\right),
\end{eqnarray}
where
\begin{eqnarray}
\tilde{u}^{(0)}_{q\overline{q} {g}}&=&
(2J+1)N_{c}\sum^{N_{f}}_{f=1}
\left[
\frac{7\pi^2}{360}
+\frac{1}{12}
\left(\frac{\mu_f}{T}\right)^2
+\frac{1}{24\pi^2}
\left(\frac{\mu_f}{T}\right)^4
\right]
\nonumber\\
&~& 
+ 
(2J+1) N_{c} \left(\frac{\pi^2}{90}\right),
\end{eqnarray}
and
\begin{eqnarray}
\tilde{u}^{(2)}_{q\overline{q} {g}}&=&
(2J+1)\sum^{N_{f}}_{f=1}
\frac{1}{6}\left[
1+\frac{3}{\pi^2}\left(\frac{\mu_f}{T}\right)^2
\right]
+ 
(2J+1)\left(\frac{1}{6}\right).
\end{eqnarray}
The color-singlet-state micro-canonical ensemble 
for the unimodular-like $U(1)^{N_c}$ symmetry 
reads
\begin{eqnarray}
Z_{qbag}(W,V)&=&
C_{qb}\,
\frac{1}{W}
\left[\left(
3\tilde{u}^{(0)}_{q\overline{q} {g}}\,VW^{3}\right)^{1/4}
\right]^{-(\frac{N_c}{2}-1)}
\,\exp\left[
\frac{4}{3}\left(
3\tilde{u}^{(0)}_{q\overline{q} {g}}\,VW^{3} 
\right)^{\frac{1}{4}}
\right],
\end{eqnarray}
where 
\begin{eqnarray}
C_{qb}&=&
\frac{1}{2\sqrt{N_c}}(2\pi)^{-N_c/2}
\left(
\frac{3\tilde{u}^{(0)}_{q\overline{q} {g}}}
{\tilde{u}^{(2)}_{q\overline{q} {g}}}
\right)^{\frac{N_c}{2}-\frac{1}{2}}.
\end{eqnarray}
%%%%%%%%%%%%%%%
%%%%%%%%%%%%%%%
Furthermore, using the standard MIT bag 
model approximation, 
the mass spectral density is reduced to
\begin{eqnarray}
\rho_{(II)}&=&\rho_{qbag}(m),
\nonumber\\
&=&
\frac{4}{3}
C_{qb}\,
\left[
\frac{3}{4}
\left(\frac{\tilde{u}^{(0)}_{q\overline{q} {g}}}
{B}
\right)^{\frac{1}{4}}
\right]^{-\frac{N_c}{2}+1}
\,
m^{-\frac{N_c}{2}}\,
\exp\left[
\left(\frac{\tilde{u}^{(0)}_{q\overline{q} {g}}}
{B}
\right)^{\frac{1}{4}}\,m
\right].
\end{eqnarray}
The high-lying mass spectral density
for the color-singlet-state 
(unimodular-like) 
$U(1)^{N_c}$ symmetry quark-bag reads,
\begin{eqnarray}
\rho_{(II)}(m)&=&\rho_{qbag}(m),
\nonumber\\
&=&
{c}_{qb}\,m^{-{\alpha}_{qb}}\, e^{{b}_{qb}\,\,m},
\nonumber\\
&\equiv&
{c}\,m^{-{\alpha}}\, e^{{b}\,\,m},
\end{eqnarray}
where $\alpha=\alpha_{qb}=N_c/2$ 
and
\begin{eqnarray}
{c}_{qb}&=&
\frac{4}{3}\,
C_{qb}\,
\left[
\frac{3}{4}
\left(\frac{\tilde{u}^{(0)}_{q\overline{q} {g}}}
{B}
\right)^{\frac{1}{4}}
\right]^{-\frac{N_c}{2}+1},
\nonumber\\
{b}_{qb}&=&
\left(\frac{\tilde{u}^{(0)}_{q\overline{q} {g}}}
{B}
\right)^{\frac{1}{4}}.
\end{eqnarray}
Generally speaking, 
it is possible to write the high lying mass spectral density 
as a linear combination of the Hagedorn states 
with different internal symmetries 
in the following way, 
\begin{eqnarray}
\rho_{(II)}(m)=\sum^{symmetries}_{i} a_{(i)} \rho^{(i)}_{(II)}(m),
\end{eqnarray}
where the index $i$ run over the unimodular-like
orthogonal $O_{S}(N_c)$, unitary $U(N_c)$, symplectic $Sp(N_c)$ and
$U(1)^{N_c}$ Hagedorn states, respectively.  
%%%%%%%%%%%%%%%%%%%%%%%%%%%%%%%%%%%%%%%%%%%%%%%%%%%%%
%%%%%%%%%%%%%%%%%%%%%%%%%%%%%%%%%%%%%%%%%%%%%%%%%%%%%

The hadronic matter consists all the known discrete low-lying 
mass spectrum particles those are found 
in the data book~\cite{databook:2008}
and the continuous high-lying mass spectrum for 
the Hagedorn states with various internal structures.
For instance the the data book~\cite{databook:2008} 
consists of 76 non-strange mesons and 64 non-strange baryons
besides strange and other flavor hadrons.
The density of states can be written as follows
\begin{eqnarray}
\rho\left(T,\Lambda,m\right)=
\rho_{(I)}\left(T,\Lambda,m\right)
+\rho_{(II)}\left(T,\Lambda,m\right).
\label{total-density1}
\end{eqnarray}
The first term on the right hand side of Eq.(\ref{total-density1})
is the discrete low-lying hadron mass spectrum 
\begin{eqnarray}
\rho_{(I)}\left(T,\Lambda,m\right)
&=&
\sum^{mesons}_{M}\delta\left(m-m_{M}\right)
\delta(\Lambda-\Lambda_{M})
+
\sum^{baryons}_{B}\delta\left(m-m_{B}\right)
\delta(\Lambda-\Lambda_{B})
\nonumber\\
&+&
\sum^{exotic}_{E}\delta\left(m-m_{E}\right)
\delta(\Lambda-\Lambda_{E})
\end{eqnarray}
where 
$m_{M}$ and $\Lambda_{M}$, respectively, 
are the meson's mass and fugacity
and
$m_{B}$ and $\Lambda_{B}$, 
respectively, are the baryon's mass 
and fugacity while $m_{E}$ and $\Lambda_{E}$, 
respectively, are exotic particle's mass and fugacity.
The baryon and meson mass spectra are satisfying 
the Fermi-Dirac and Bose-Einstein statistics, 
respectively. 
The exotic particle satisfies either Fermi-Dirac 
or Bose-Einstein statistics depends on the number 
of its constituent quarks and antiquarks.
Since the discrete low-lying mass spectrum 
is limited for a relatively light hadron mass 
$m\le m_{critical}\approx 2$ GeV for non-strange 
hadrons, 
the  Fermi-Dirac and Bose-Einstein are distinctive 
and they are taken exactly:
\begin{eqnarray}
I&=&Z_{(I)}\left(T,V;\Lambda\right),
\nonumber\\
&=&\exp\left(\int^{m_0}_{0}\,dm\,
\rho_{(I)}\left(T,\Lambda,m\right)
\,\left[\ln z_{(Stats)}\left(T,m,\Lambda\right)
+\ln z_{(Stats)}
\left(T,m,\Lambda^{-1}\right)\right]\right),
\nonumber\\
&=&
Z_{(I)}\left(T,V;\Lambda,\mbox{Mesons}\right)
\,\times\,
Z_{(I)}\left(T,V;\Lambda,\mbox{Baryons}\right)
\nonumber\\
&\,&\times\,
Z_{(I)}\left(T,V;\Lambda,\mbox{Exotics}\right),
\nonumber\\
\end{eqnarray}
where the subscript notation, namely, 
$(Stats)$ indicates either Fermi-Dirac 
or Bose-Einstein statistics for baryons 
or mesons, respectively.
The low-lying mass spectrum ensemble 
is decomposed into mesonic, baryonic 
and exotic ensembles as follows
\begin{eqnarray}
Z_{(I)}\left(T,V;\Lambda,\mbox{Mesons}\right)
&=&
\prod^{Mesons}_{M}
\,\left[z_{BE}\left(T,m_{M},\Lambda_{M}\right)
\,\times\,
z_{BE}\left(T,m_{M},\Lambda_{M}^{-1}\right)
\right],
\end{eqnarray}
and
\begin{eqnarray}
Z_{(I)}\left(T,V;\Lambda,\mbox{Baryons}\right)&=&
\prod^{Baryons}_{B}
\,\left[z_{FD}\left(T,m_{B},\Lambda_{B}\right)
\,\times\, z_{FD}\left(T,m_{B},\Lambda_{B}^{-1}\right)\right],
\end{eqnarray}
as well as
\begin{eqnarray}
Z_{(I)}\left(T,V;\Lambda,\mbox{Exotics}\right)
&=&
\prod^{Exotics}_{E}
\,\left[ z_{BE}\left(T,m_{E},\Lambda_{E}\right)
\,\times\, z_{BE}\left(T,m_{E},\Lambda_{E}^{-1}\right)\right]
\nonumber\\
&~&~\times\prod^{Exotics}_{E}
\,\left[ z_{FD}\left(T,m_{E},\Lambda_{E}\right)
\,\times\, z_{FD}\left(T,m_{E},\Lambda_{E}^{-1}\right)\right],
\nonumber
\\
&\approx& 
\exp\left[\mbox{negligible}\right]
\approx 1.
\end{eqnarray}
The exotic hadrons are suppressed because of 
their relative large masses in comparison 
to the ordinary mesons and baryons.
Hence its contribution to the discrete 
hadronic mass spectrum is negligible.
%%%%%%%%%%%%%%%%%%%%%%%%%%%%%%%%%%%%%%
%%%%%%%%%%%%%%%%%%%%%%%%%%%%%%%%%%%%%%

The Fermi-Dirac and Bose-Einstein 
statistics, respectively, read
\begin{eqnarray}
z_{FD}\left(T,m,\Lambda_{B}\right)=
\exp\left[
(2J_{B}+1) V\int \frac{d^3 k}{(2\pi)^3} 
\ln\left(1+\Lambda_{B}e^{-\sqrt{(k^2+m_{B}^2)}/T}
\right)\right],
\label{FD-ensemble}
\end{eqnarray}
and
\begin{eqnarray}
z_{BE}\left(T,m,\Lambda_{M}\right)=
\exp\left[
-(2J_{M}+1) V\int \frac{d^3 k}{(2\pi)^3} 
\ln\left(1-\Lambda_{M}
e^{-\sqrt{(k^2+m_{M}^2)}/T}\right)\right].
\label{BE-ensemble}
\end{eqnarray}
%%%%%%%%%%%%%%%%%%%%%%%
The mesonic and baryonic fugacities 
are given by 
$\Lambda_{M}=e^{(S\,\mu_S)/T}$ 
and 
$\Lambda_{B}=e^{(\mu_B+S\,\mu_S)/T}$, 
respectively, where $S=0,-1$ etc 
is the strangeness quantum number.
%%%%%%%%%%%%%%%%%%%%%%%%%%%%%%%%%
%%%%%%%%%%%%%%%%%%%%%%%%%%%%%%%%%
%%%%%%%%%%%%%%%%%%%%%%%%%%%%%%%%%

The second term 
$\rho_{(II)}\left(T,\Lambda,m\right)$ 
on the right hand side 
of Eq.(\ref{total-density1})
is the continuous high-lying 
mass spectrum and corresponds 
the Hagedorn's density of states. 
These states correspond the hadronic 
bubbles with relatively large hadronic 
masses and they appear as fireballs 
just above the highest known 
hadronic particles those are represented 
by the discrete low-lying mass spectrum. 
The Hagedorn threshold is estimated 
to be $m\ge 2$ GeV 
for the non-strange hadrons.
%%%%%%%%%%%%%%%%%%%%%%%%%%%%%%%%%%%%%%%%%%
In the simplest approximation, 
the mass of the composite hadron 
is considered relatively large 
and the flavor degree of freedom 
is assumed to maintain 
the maximal invariance. 
Such exotic hadrons (Hagedorns) have 
no reason to prefer the Fermi-Dirac 
or Bose-Einstein Statistics. 
These Hagedorn states are assumed to obey
the classical Maxwell-Boltzmann statistics 
for the simplicity due 
to their relatively large masses.
Nonetheless, in any realistic physical situation, 
the Hagedorn states violate the flavor invariance 
and this violation becomes negligible 
when the medium tends to be extreme hot 
and the chiral symmetry tends to be restored 
(i.e. the is no sufficient time 
for the flavor violation). 
Furthermore, the Maxwell-Boltzmann statistics 
is not adequate one for the highly compressed 
the cold 
(or slight warm $T\approx\mbox{few tens MeV}$) 
nuclear matter. 
In the case of highly compressed nuclear matter 
where $\mu_{B}$ becomes large and $T$ is small,
the quantum statistics such as Fermi-Dirac 
or Bose-Einstein Statistics becomes important.
It is remarkable to note that 
the Bose-Einstein statistics is likely 
to lead to the Hagedorn condensation
when the baryon chemical potential 
$\mu_{B}$ becomes very large at relatively 
low temperature $T$.
In this sense the quantum statistics 
turns to be important.
%%%%%%%%%%%%%%%%%%%%%%%%%%%%%%%%%%%%%%%%%%%%%%
%%%%%%%%%%%%%%%%%%%%%%%%%%%%%%%%%%%%%%%%%%%%%%

The grand canonical partition function 
for the Hagedorn states is given by
\begin{eqnarray}
I&=&Z_{(II)}\left(T,V,\Lambda\right),
\nonumber\\
&=&\exp\left(\int^{\infty}_{m_0}
\,dm\,\rho_{(II)}\left(T,\Lambda,m\right)
\,\left[\ln z_{(Stats)}\left(T,m,\Lambda\right)
+\ln z_{(Stats)}\left(T,m,\Lambda^{-1}\right)\right]\right),
\nonumber\\
&=&
Z_{(II)}\left(T,V,\Lambda;\mbox{Mesonic}\right)
\,\times\,
Z_{(II)}\left(T,V,\Lambda;\mbox{Baryonic}\right)
\,\times\,
Z_{(II)}\left(T,V,\Lambda;\mbox{Exotic}\right)
\nonumber\\
&~&
\,\times\,\left(\cdots\right)\,.
\end{eqnarray}
The continuous high-lying mass spectrum 
can be decomposed to mesonic ($B=0$)
and baryonic ($B=1$) Hagedorn states and
the exotic Hagedorn states with baryonic
quantum number $B>1$.
The mesonic and baryonic fugacities 
are given by $\Lambda_{M}=e^{(S\,\mu_S)/T}$ 
and $\Lambda_{B}=e^{(B\,\mu_B+S\,\mu_S)/T}$, 
respectively, 
where $S$ is the strangeness quantum number
and $B$ is the baryonic number. 
The (ordinary-) baryonic Hagedorns carries $B=1$. 
The Exotic Hagedorn's states 
can carry various baryonic and strangeness
quantum number $B=0,1,2 \cdots$ etc and $S=0,-1,\cdots$ etc
where the fugacity is reduced to 
$\Lambda_{Exotic}=e^{(B\,\mu_B+S\,\mu_S)/T}$. 
In the case of no strangeness degree 
of freedom, the strangeness 
quantum number is reduced to $S=0$.
The other flavor degrees of freedom such as charmonium 
can be introduced in the same way.
The continuous high-lying density of states
$\rho_{(II)}\left(T,\Lambda,m\right)$
is decomposed to
\begin{eqnarray}
\rho_{(II)}\left(T,\Lambda,m\right)&=&
\sum_{S}
\rho_{(II)}\left(T,\Lambda_{M},m;\mbox{Mesonic}\right)
\nonumber\\
&\,&
+\sum_{S}
\rho_{(II)}\left(T,\Lambda_{B},m;\mbox{Baryonic}\right)
\nonumber\\
&\,&
+\sum_{B}\sum_{S}
\rho_{(II)}\left(T,\Lambda_{Exotic},m;\mbox{Exotic}\right),
\label{Hagedorn-states-Q1}
\end{eqnarray}
where summations are carried over the baryonic and strangeness
quantum number $B$ and $S$, respectively.
The exotic Hagedorn states are assumed negligible
and they are dropped from the calculations.
This approximation  seems to be reasonable for 
the hot and diluted matter. 
Fortunately, the situation is simpler 
in the case of small $\mu_{B}$ and high $T$ 
rather than for large $\mu_{B}$ and low $T$.
%%%%%%%%%%%%%%%%%%%%%%%%%%%%%%%%%%%%%%%%%%%%

The partition functions for mesonic 
and baryonic Hagedorns read
\begin{eqnarray}
\ln\,Z_{(II)}\left(\mbox{Mesonic}\right)
&=&
\ln\,Z_{(II)}
\left(T,V,\Lambda_{M};\mbox{Mesonic}\right),
\nonumber\\
&=&
\int^{\infty}_{m_0}\,dm\,
\rho_{(II)}\left(T,\Lambda_{M},m;\mbox{Mesonic}\right)
\nonumber\\
&\,&\,\times\,\left[\ln z_{BE}\left(T,m,\Lambda_{M}\right)
+\ln z_{BE}\left(T,m,\Lambda_{M}^{-1}\right)\right],
\nonumber\\
\end{eqnarray}
and
\begin{eqnarray}
\ln\,Z_{(II)}\left(\mbox{Baryonic}\right)&=&
\ln\,Z_{(II)}\left(T,V,\Lambda_{B};\mbox{Baryonic}\right),
\nonumber\\
&=&
\int^{\infty}_{m_0}\,dm\,\rho_{(II)}
\left(T,\Lambda_{B},m;\mbox{Baryonic}\right)
\nonumber\\
&\,&\,\times\,
\left[\ln z_{FD}\left(T,m,\Lambda_{B}\right)
+\ln z_{FD}\left(T,m,\Lambda_{B}^{-1}\right)\right],
\nonumber\\
\end{eqnarray}
respectively, where $m_0$ is determined 
from the threshold mass $m_{0}=m_{th}$.
The Fermi-Dirac and Boson-Einstein ensembles for the baryonic 
and mesonic Hagedorns are given in Eqs.(\ref{FD-ensemble}) 
and (\ref{BE-ensemble}), respectively.
For example, the mesonic and baryonic fugacities 
are given by $\Lambda_{M}=1$ 
in the case of no strange degree of freedom 
is included (i.e. $\mu_S=0$) 
and $\Lambda_{B}=e^{\mu_B/T}$, 
respectively.
%%%%%%%%%%%%%%%%%%%%%%%%%%%%%%%%%%%%%%%%%%%%%%
%%%%%%%%%%%%%%%%%%%%%%%%%%%%%%%%%%%%%%%%%%%%%%
%%%%%%%%%%%%%%%%%%%%%%%%%%%%%%%%%%%%%%%%%%%%%%%%
%%%%%%%%%%%%%%%%%%%%%%%%%%%%%%%%%%%%%%%%%%%%%%%%%%%%%%%%%
%%%%%%%%%%%%%%%%%%%%%%%%%%%%%%%%%%%%%%%%%%%%%%%%%%%%%%%%%
%%%%%%%%%%%%%%%%%%%%%%%%%%%%%%%%%%%%%%%%%%%%%%%%%%%%%%%%%
%%%%%%%%%%%%%%%%%%%%%%%%%%%%%%%%%%%%%%%%%%%%%%
%%%%%%%%%%%%%%%%%%%%%%%%%%%%%%%%%%%%%%%%%%%%%%

The grand potential for the mesonic and baryonic
Hagedorn states are calculated,
respectively, as follow
\begin{eqnarray}
\frac{\Omega_{(II)}}{V}
\left(\mbox{Mesonic Hagedorn gas}\right)
=-T\,\frac{\partial}{\partial\,V}
\ln Z_{(II)}\left(\mbox{Mesonic}\right),
\end{eqnarray}
and
\begin{eqnarray}
\frac{\Omega_{(II)}}{V}
\left(\mbox{Baryonic Hagedorn gas}\right)&=&
-T\,\frac{\partial}{\partial\,V}
\ln Z_{(II)}\left(\mbox{Baryonic}\right).
\end{eqnarray}
%%%%%%%%%%%%%%%%%%%%%%%%%%%%%%%%%%%%%%%%%%%%%
In order to simplify the treatment,
we follow the standard procedure
for heavy Hagedorn states ($m_{0} > 2$ GeV)
those dominates the dilute and hot nuclear matter.
The partition functions for classical mesonic and
baryonic Hagedorn states, respectively, read
\begin{eqnarray}
\ln z_{BE}\left(T,m,\Lambda_{M}\right)
&=&
-(2J_{M}+1)\,V\,\int \frac{d^3 k}{(2\pi)^3}
\ln\left(1-\Lambda_{M}
e^{-\sqrt{(k^2+m^2)}/T}\right),
\nonumber\\
&\approx&
(2J_{M}+1) \Lambda_{M}
\,V\,\int \frac{d^3 k}{(2\pi)^3} e^{-\sqrt{(k^2+m^2)}/T},
\nonumber\\
&\approx&
(2J_{M}+1) \Lambda_{M}
\,V\,\left(\frac{m T}{2\pi}\right)^{3/2} e^{-m/T},
\end{eqnarray}
and
\begin{eqnarray}
\ln z_{FD}\left(T,m,\Lambda_{B}\right)
&=&
(2J_{B}+1) \,V\,\int \frac{d^3 k}{(2\pi)^3}
\ln\left(1+\Lambda_{B}e^{-\sqrt{(k^2+m^2)}/T}
\right),
\nonumber\\
&\approx&
(2J_{B}+1) \Lambda_{B}
\,V\,\int \frac{d^3 k}{(2\pi)^3} e^{-\sqrt{(k^2+m^2)}/T},
\nonumber\\
&\approx&
(2J_{B}+1) \Lambda_{B}
\,V\,\left(\frac{m T}{2\pi}\right)^{3/2} e^{-m/T}.
\end{eqnarray}
The exotic Hagedorn states with the baryonic number $B>1$
are assumed to be suppressed and subsequently
they are dropped from the calculations.
The classical statistics is an adequate approximation
for the flavor invariance
when the dilute nuclear matter is heated.
However, in the case of the cold
and highly compressed nuclear matter,
the classical Maxwell-Boltzmann statistic fails
for the mesonic Hagedorn states
when the meson condensation
(in particular the strangeness)
takes place in the system.
Hence, the exact quantum Bose-Einstein
and Fermi-Dirac statistics become essential
for cold and dense nuclear matter.
The high-lying strangeness
condensation takes place
when the mesonic Hagedorn frequency approaches zero.
This phenomenon takes place when the strange chemical
potential $\mu_S$ reaches a critical value.
Hence, under such a circumstance, the Bose-Einstein statistics
must be taken exactly in the calculation.
Furthermore, the exotic Hagedorn states may turn
to be important when the nuclear matter becomes
extremely dense in order
to soften the equation of state.
Fortunately, in the neighborhood of the tri-critical point,
the condensation is not expected
in the hot and dilute nuclear matter
and the classical Maxwell-Boltzmann statistics
remains a good approximation.
Hence, the Hagedorn's partition function is approximated to 
\begin{eqnarray}
\ln\, Z_{(II)}\left(T,V,\Lambda\right)&\approx& 
\frac{V\,T^{3/2}}{(2\pi)^{3/2}}
\sum_{Q} \left(2J_Q+1\right)\,
\Lambda_{Q}
\,\int^{\infty}_{m}\,
\rho_{(II)}\left(T,\Lambda_{Q},T\right)\,m^{3/2} e^{-m/T}
d\,m,\nonumber\\
&\sim&
\frac{V\,T^{3/2}}{(2\pi)^{3/2}} 
\sum_{Q} \left(2J_Q+1\right)\,
\Lambda_{Q}\, c_{Q}
\,\int^{\infty}_{m}\,      
m^{-\alpha_{Q}+3/2} e^{-\left(\frac{1}{T}-b_{Q}\right)\, m} 
d\,m,
\end{eqnarray}
where $Q\sim S,B,\cdots$ is the Hagedorn's quantum number 
that is indicated by 
Eq.(\ref{Hagedorn-states-Q1}) and the mass spectral density 
of the Hagedorn species, namely, 
$\rho_{Q}\, \sim\,  c_{Q}\, m^{-\alpha_{Q}}\,e^{b_{Q}\, m}$ 
is identified by its specific internal symmetry 
as constructed above.
The mass spectral exponents $\alpha_{Q}\equiv \alpha$
for various symmetries are displayed 
in Table~(\ref{table-1}).
%%%%%%%%%%%%%%%%%%%%%%%%%%%%%%%%%%%%%%%%%%%%%%%%%

The mass spectral density with the conserved
charges of the group symmetry
$U(1)_{B}\times U(1)_{S} \cdots$
(i.e. with a specific flavor symmetry
but not the maximal flavor invariance)
can be evaluated by adopting
the imaginary chemical potentials
as follows
\begin{eqnarray}
\rho_{(II)}
\left(
T, e^{B\,\frac{\mu_{B}}{N_{c}}/T}, e^{S\,\mu_{S}/T}, \cdots, m
\right)
\rightarrow
\rho_{(II)}\left(
e^{i B\,\Theta_{B}/N_{c}}, e^{i S\,\Theta_{S}}, \cdots, m
\right).
\end{eqnarray}
Therefore, the mass spectral density with the conserved baryon
and strange numbers is calculated by finding
the Fourier inverse as follows,
\begin{eqnarray}
\rho_{(II)}\left(B,S, m\right)
&=&
\int^{\pi}_{-\pi}\frac{d\Theta_{B}}{2\pi} e^{-i B\,\Theta_{B}}
\,
\int^{\pi}_{-\pi}\frac{d\Theta_{S}}{2\pi} e^{-i S\,\Theta_{S}}
\,
\rho_{(II)}\left(
e^{i B\,\Theta_{B}/N_{c}}, e^{i S\,\Theta_{S}}, m
\right).
\label{conservb1}
\end{eqnarray}
The conservation of the baryon number
$B$ modifies
the mass spectral exponent slightly
to $\alpha\rightarrow \alpha+\frac{1}{2}$.
Moreover, the strange degree of freedom
(i.e. the conservation of strange number $S$)
modifies the mass spectral exponent by
an additional term
$m^{-\frac{1}{2}}$
and, subsequently,
the resultant mass spectral density becomes
$\rho\sim m^{-(\alpha+1)}\, e^{b(B,S\cdots)\, m}$
where the explicit formula for $b(B,S\cdots)$
can be calculated using the saddle point approximation.
In this case, the partition functions
for the mesonic and baryonic Hagedorns
are reduced to, respectively,
\begin{eqnarray}
\ln Z_{(II)}\left(\mbox{mesonic}\right)&=&
\int^{\infty}_{m_{0}} d m \rho_{(II)}\left(B=0,S,m\right)
\nonumber\\
&\times&
\left[\ln z_{BE}(T,m,e^{S\mu_{S}/T})
+\ln z_{BE}(T,m,e^{-S\mu_{S}/T})\right],
\end{eqnarray}
and
\begin{eqnarray}
\ln Z_{(II)}\left(\mbox{baryonic}\right)&=&
\int^{\infty}_{m_{0}} d m \rho_{(II)}\left(B=1,S,m\right)
\nonumber\\
&\times&
\left[\ln z_{BE}
\left(T,m,e^{\mu_{B}/T+S\mu_{S}/T}\right)
+
\ln z_{BE}
\left(T,m,e^{-\mu_{B}/T-S\mu_{S}/T}\right)
\right].
\end{eqnarray}
The contribution of the liberated gluons due to the modification 
of the Hagedorn state's internal symmetry is given as follows
\begin{eqnarray}
\ln\, Z_{lg}(T)&=& (2J+1)\,V\,T^{3}\,N_{lg}
\,\left(\frac{\pi^{2}}{90}\right), 
~~ \left(\mbox{where}~~ J=\frac{1}{2}\right).
\end{eqnarray}
The liberated gluons due to symmetry reduction is given by $N_{lg}= 0$,
$\frac{1}{2} N_{c}\left(N_{c}+1\right)$ and 
$N_{c}\left(N_{c}-1\right)$
for unitary, orthogonal and $U(1)^{N_c}$ Hagedorn states, respectively.
The total partition function for the Hagedorn matter becomes
\begin{eqnarray}
\ln Z_{H-matter}\left(T,V,\Lambda\right)&=&
\ln\, Z_{(II)}\left(T,V,\Lambda\right)\,+\,\ln\, Z_{lg}(T).
\end{eqnarray}
%%%%%%%%%%%%%%%%%%%%%%%%%%%%%%%%%%%%%%%%%%%%%%%%%%%%%%%%% 
%%%%%%%%%%%%%%%%%%%%%%%%%%%%%%%%%%%%%%%%%%%%%%%%%%%%%%%%%
%%%%%%%%%%%%%%%%%%%%%%%%%%%%%%%%%%%%%%%%%%%%%%%%%%%%%%%%%
%%%%%%%%%%%%%%%%%%%%%%%%%%%%%%%%%%%%%%%%%%%%%%%%%%%%%%%%%
%%%%%%%%%%%%%%%%%%%%%%%%%%%%%%%%%%%%%%%%%%%%%%%%%%%%%%%%%
%%%%%%%%%%%%%%%%%%%%%%%%%%%%%%%%%%%%%%%%%%%%%%%%%%%%%%%%%%%%%%%%%%%
%%%%%%%%%%%%%%%%%%%%%%%%%%%%%%%%%%%%%%%%%%%%%%%%%%%%%%%%%%%%%%%%%%%
%%%%%%%%%%%%%%%%%%%%%%%%%%%%%%%%%%%%%%%%%%%%%%%%%%%%%%%%%%%%%%%%%%%
%%%%%%%%%% newnew

\section{\label{sect5} Chiral phase transition} 

The role of chiral phase transition can be considered explicitly 
in the context of the present model~\cite{Zakout:2007nb}.
The chiral restoration is expected to take place at the Gross-Witten point
or at its neighbourhood but below the point of the color deconfinement. 
The chiral phase transition is likely to be a first order one 
if it takes place at the Gross-Witten point~\cite{Kogut:1981ez}. 
We shall demonstrate in the present section that the chiral phase transition
takes place below the deconfinement phase transition point 
for the orthogonal Hagedorn states.
If this is the case, then the standard treatment given in 
the preceding sections is sufficient. 
However, if the chiral restoration takes place far away above 
the Gross-Witten point in the matter that is dominated by 
the Hagedorn states then the effect of the chiral field 
becomes important.
In this case, the chiral Hagedorn states have to be considered 
explicitly in the calculation. 
The contribution of the discrete low-lying hadron mass spectrum
is negligible since the nuclear matter at extreme conditions 
turns to be dominated by the Hagedorn states 
above the Gross-Witten point.
The partition function for the quark and antiquark given 
by Eq.(\ref{real-Z-fund-q-q}) in the absent of the chiral 
background is modified when the chiral field is considered 
in the calculation. 
In the mean field approximation, Eq.(\ref{real-Z-fund-q-q}) 
is modified to become 
\begin{eqnarray} 
\ln Z_{q\overline{q}\sigma}(\beta,V,\sigma)
&=& 
(2J+1) V \sum_{q}^{N_f}
\int \frac{ d^3\vec{p} }{ {(2\pi)}^{3} }
\,\sum_{i} 
\left[ \ln\left( 1+
e^{-\beta\left(\epsilon_{q}(\vec{p},\sigma)-\mu_{q}-i\frac{\theta_{i}}{\beta}\right)} 
\right) 
\right.
\nonumber\\
&~& ~~~ ~~~ ~~~
\left.
+ 
\ln\left( 1+e^{-\beta\left(\epsilon_{q}(\vec{p},\sigma)+\mu_{q}+i\frac{\theta_{i}}{\beta}\right)} 
\right) \right], 
\end{eqnarray} 
where 
$\epsilon(\vec{p},\sigma)=\sqrt{ \vec{p}^2+ ({m^{*}_{q}}(\sigma))^{2} }$, 
$m^{*}_{q}(\sigma)=m_{q}+g_{\sigma}\,\sigma$
and $\sigma=<\sigma>$ 
is the chiral mean field condensate. 
Hence in the background of the chiral field, 
the quark-gluon partition function, in the Hilbert space, reads 
\begin{eqnarray} 
Z_{q\overline{q}g\,\sigma}(\beta,V,\sigma)= 
Z_{q\overline{q}\sigma}(\beta,V,\sigma)
\,\times\,Z_{g}(\beta,V).
\end{eqnarray}  
The gluon partition function $Z_{g}(\beta,V)$ 
is given by Eq.(\ref{gluon-partition-1}).
The color-singlet state of the mixed-grand canonical ensemble 
for the chiral quark-gluon bag is projected out 
in the following way
\begin{eqnarray} 
Z_{Singlet}(\beta,V,\sigma)&=&\int d\mu_{G}(g)\,\chi_{Singlet}(g)
\times Z_{q\overline{q}g\,\sigma}(\beta,V,\sigma), 
\nonumber\\
&=&\int d\mu_{G}(g)\, Z_{q\overline{q}g\,\sigma}(\beta,V,\sigma),
\end{eqnarray} 
where $\chi_{Singlet}(g)=1$ 
and $G$ refers to the group classification either unitary, orthogonal 
or symplectic one.
The integration over the invariance Haar measure is performed 
in the standard way using the Gaussian integration method 
as that is presented in Sec.~\ref{sect2} 
(see also Refs.~\cite{Zakout:2007nb,Zakout:2006zj}).
Furthermore, the micro-canonical ensemble with large energy 
$W$ is calculated by taking the inverse Laplace transform 
of the mixed-grand canonical ensemble in the following way
\begin{eqnarray} 
Z_{Singlet}\left(W,V,\sigma\right)=\frac{1}{2\pi\, i} 
\int_{\beta_0-i\infty}^{\beta_0+i\infty} \,d\beta\, e^{\beta W} 
\, Z_{Singlet}(\beta,V,\sigma). 
\end{eqnarray} 
The inverse Laplace transform is evaluated 
using the steepest decent method~\cite{Zakout:2007nb,Zakout:2006zj}.
The Laplace stationary point is determined by extremizing the 
exponential with respect to the Laplace variable $\beta$.
Because of the finite value of the quark's chiral mass, 
the solution of the Laplace saddle point becomes 
a transcendental one and can not be written in a closed form. 
In this case, the saddle point and the inverse Laplace transform 
is evaluated by iteration~\cite{Zakout:2007nb}). 
The first iteration is sufficient 
to shed the light on the chiral restoration phase transition.
The standard MIT bag model furnishes the following relations between
the bag's mass and volume:
\begin{eqnarray} 
\begin{array}{l} 
m = W + B V,\\ 
W = \frac{3}{4} m, \\ 
V = \frac{m}{4 B}, 
\end{array} 
\end{eqnarray} 
where $B$, $W$, $V$ and $m$ are the bag's constant, cavity energy, 
volume and mass, respectively. 
Hence, the Hagedorn's density of states for the chiral color-singlet 
quark-gluon bag with a specific internal color symmetry 
is reduced to 
\begin{eqnarray} 
\rho_{H}\left(m,\sigma\right)&\propto& 
\left.
Z_{Singlet}\left(W,V,\sigma\right)
\right|_{(W =\frac{3}{4} m , V =\frac{m}{4 B})}.
\end{eqnarray}
The mixed-grand canonical ensemble for the matter 
of chiral Hagedorn states reads
\begin{eqnarray} 
\ln\, Z_{(II)}(T,\mu,V,\sigma) &=&
2 (2J_{H}+1)\, \cosh\left(\mu_{H}/T\right)
\nonumber\\
&~& \times
V\,\int_{m_0}^{\infty} d m \,\rho_{H}\left(m,\sigma\right)
\,
\left(\frac{m T}{2\pi}\right)^{3/2} \,e^{-m/T}.
\end{eqnarray}
The total grand potential density for 
the chiral Hagedorn matter in the presence 
of the chiral field is given by
\begin{eqnarray}
\frac{1}{V}
\Omega_{\sigma}\left(T,\mu,V,\sigma\right)&=&
- T \,\frac{\partial}{\partial V} 
\ln Z_{(II)}\left(T,\mu,V,\sigma\right) 
+ U(\sigma),
\nonumber\\
&=&
\frac{1}{V} 
\Omega_{H\,\sigma}\left(T,\mu,V,\sigma\right)
+ U(\sigma). 
\label{total-grand-potential-sigma1}
\end{eqnarray}
The chiral Hagedorn's grand potential density 
( i.e. the first term on the right hand side 
of Eq.(\ref{total-grand-potential-sigma1}) ) 
is given as follows,
\begin{eqnarray}
\frac{1}{V}
\Omega_{H\,\sigma}\left(T,\mu,V,\sigma\right)&=& 
- T \,\frac{\partial}{\partial V}
\ln Z_{(II)}\left(T,\mu,V,\sigma\right),
\nonumber\\
&=&
-
2 (2J_{H}+1)\, \cosh\left(\mu_{H}/T\right)
\nonumber\\
&~& \times
T\,\int_{m_0}^{\infty} d m \rho_{H}\left(m,\sigma\right)
\,
\left(\frac{m T}{2\pi}\right)^{3/2} \,e^{-m/T}.
\label{chiral-hagedorn1}
\end{eqnarray}
The chiral potential $U(\sigma)$ 
( i.e. the second term on the right hand side of
Eq.(\ref{total-grand-potential-sigma1}) )
can be introduced by a phenomenological one 
in the following way 
\begin{eqnarray}
U(\sigma)=
\frac{1}{2} m^{2}_{\sigma}\,\sigma^{2}
+\frac{1}{3} c_{2}\, \sigma^{3}
+\frac{1}{4} c_{3}\, \sigma^{4},
\end{eqnarray} 
where the cubic and quartic terms represent 
the scalar self-interactions and they were proposed 
by Boguta and Bodmer~\cite{Boguta:1977xi}. 
In the simplest approximation, the cubic and quartic terms
are neglected by taking $c_{2}=0$ and $c_{3}=0$ 
and this corresponds to the Walecka-like models.
The value of the mean field $\sigma$ is determined 
by calculating the extremum (i.e. minimum) 
of the total grand potential density, namely,
\begin{eqnarray}
\frac{1}{V}
\frac{\partial}{\partial \sigma} 
\Omega_{\sigma}\left(T,\mu,V,\sigma\right)=0.
\end{eqnarray}
In order to calculate the chiral Hagedorn's grand potential
density that is given in Eq.(\ref{chiral-hagedorn1}),
the high-lying mass spectral density for chiral Hagedorn states
is rewritten in the following way
\begin{eqnarray}
\rho_{H}\left(m,\sigma\right)= 
c m^{-\alpha} e^{b\left(\sigma\right)\, m}.
\label{chiral-mass-spectrum-1}
\end{eqnarray}
In the context of standard MIT bag model 
(see for instance Eq.(\ref{standard-MIT-1})), 
the exponential term, namely, $b(\sigma)$ reads
\begin{eqnarray}
b\left(\sigma\right)=\frac{1}{B^{1/4}}\, 
\left[
\tilde{a}^{(0)}_{q\overline{q}g}\left(\sigma\right)
\right]^{1/4}.
\end{eqnarray}
Furthermore, the function 
$\tilde{a}^{(0)}_{q\overline{q}g}\left(\sigma\right)$
with the chiral field background is determined as follows
\begin{eqnarray}
\tilde{a}^{(0)}_{q\overline{q}g}\left(\sigma\right)=
\tilde{a}^{(0)}_{q\overline{q}}\left(\sigma\right)+
\tilde{a}^{(0)}_{g},
\end{eqnarray}
where the chiral quark-antiquark term reads
\begin{eqnarray}
\tilde{a}^{(0)}_{q\overline{q}}\left(\sigma\right)&=&
(2J+1) N_{c}\frac{1}{2\pi^2} \sum^{N_f}_{q} \int^{\infty}_{0} dx\, x^{2}
\nonumber\\
&~& ~~\times
\ln\left[
1+2\cosh\left(\frac{\mu_{q}}{T}\right) 
e^{ -\sqrt{x^2+\left(\overline{\beta} {m^{*}_ {q}}({\sigma})\right)^2} }
+  e^{ -2\sqrt{x^2+\left(\overline{\beta} {m^{*}_ {q}}({\sigma})\right)^2 } }
\right],
\nonumber\\
&=&
(2J+1) N_{c}\frac{1}{3\pi^2} \sum^{N_f}_{q} 
\int^{\infty}_{0} dx\,
\frac{x^{4}}
{\sqrt{x^2+\left(\overline{\beta} {m^{*}_ {q}}({\sigma})\right)^2} }
\nonumber\\
&~& ~~~~~
\times\frac{
e^{ -\sqrt{x^2+\left(\overline{\beta} {m^{*}_ {q}}({\sigma})\right)^2} }
\left[
e^{ -\sqrt{x^2+\left(\overline{\beta} {m^{*}_ {q}}({\sigma})\right)^2} }
+
\cosh\left(\frac{\mu_{q}}{T}\right)
\right]
}
{
\left[
1+2\cosh\left(\frac{\mu_{q}}{T}\right)
e^{ -\sqrt{x^2+\left(\overline{\beta} {m^{*}_ {q}}({\sigma})\right)^2} }
+  e^{ -2\sqrt{x^2+\left(\overline{\beta} {m^{*}_ {q}}({\sigma})\right)^2 } }
\right]
}.
\end{eqnarray}
The gluon term remains intact and is given by
$\tilde{a}^{(0)}_{g}=(2J+1) N_{adj} \left(\frac{\pi^2}{90}\right)$.
The term 
$\left.\tilde{a}^{(0)}_{q\overline{q}}\left(\sigma\right)\right|_{\sigma=0}$
with zero chiral field background is reduced to 
\begin{eqnarray}
\left.\tilde{a}^{(0)}_{q\overline{q}}\left(\sigma\right)\right|_{\sigma=0}&=& 
\tilde{a}^{(0)}_{q\overline{q}}\nonumber\\
&=&
\left(2J+1\right) N_{c}
\sum^{N_f}_{q}\left[
\frac{7\pi^2}{360}+\frac{1}{12}\left(\frac{\mu_{q}}{T}\right)^{2}
+\frac{1}{24 \pi^{2}}\left(\frac{\mu_{q}}{T}\right)^{4}
\right].
\end{eqnarray}
The saddle point $\overline{\beta}$ is calculated by iteration 
with respect to the chiral mean field $\sigma$. 
The first iteration in the context of the MIT bag model 
leads to ( see also Eq.(\ref{beta-saddle-point1}) ),
\begin{eqnarray}
\overline{\beta}&=& \frac{1}{W}\left[
3\tilde{a}^{(0)}_{q\overline{q}g} V W^3 
\right]^{\frac{1}{4}},
\nonumber\\
&=& \left[\tilde{a}^{(0)}_{q\overline{q}g}/B\right]^{\frac{1}{4}}.
\end{eqnarray}
The mass spectral exponent $\alpha$ that appears in the mass spectral density 
that is given in Eq.(\ref{chiral-mass-spectrum-1}) is determined by 
Eq.(\ref{mass-exponent-1}).
It is reduced to $\alpha=\,3,\, \frac{9}{2}$ and $\frac{15}{2}$
for the orthogonal, unitary and symplectic Hagedorn states, 
respectively.
It becomes $\alpha=\frac{3}{2}$ for $U(1)^{N_{c}=3}$ Hagedorn states. 
The color-singlet projection with unimodular-like constraint 
is imposed in the all preceding Hagedorn states. 
It is reduced to $\alpha=\frac{1}{2}$ for color non-singlet 
$SU(3)$ quark-gluon bags.
The values of the mass spectral exponent $\alpha$
with various symmetries are displayed in Table~(\ref{table-1}). 
The variation of the chiral Hagedorn's grand potential density 
with respect to $\sigma$ is reduced to
\begin{eqnarray}
\frac{1}{V} \frac{\partial}{\partial \sigma}
\Omega_{H\sigma}\left(T,\mu,V,\sigma\right)
&=&
-2(2J+1)\,\frac{1}{(2\pi)^{3/2}}\, T^{5/2}\, c\,
\left(
\frac{\partial\, b(\sigma)}{\partial \sigma} 
\right)
\nonumber\\
&~& 
~~~~~~~~~ \times\,
\int^{\infty}_{m_0}\,d m\, m^{-\alpha+5/2} e^{\left(b(\sigma)-T\right)\, m},
\nonumber\\
&\sim& -\left(\dots\right)\,
\int^{\infty}_{m_0}\,d m\, m^{-\alpha+5/2} e^{\left(b(\sigma)-T\right)\, m}.
\end{eqnarray}
The scalar mean field, namely, $\sigma = <\sigma>$ 
is found by extremizing 
$\Omega_{\sigma}\left(T,\mu,V,\sigma\right)$
which is presented in Eq.(\ref{total-grand-potential-sigma1}).
The value of mean field $\sigma=\sigma_{0}$ 
can be approximated to
\begin{eqnarray}
m^{2}_{\sigma}\,\sigma \propto 
\left(
\frac{\partial\, b(\sigma)}{\partial \sigma}
\right)
\times
\int^{\infty}_{m_0}\, m^{-\alpha+5/2} e^{\left(b(\sigma)-T\right)\, m}\,d\, m.
\label{sigma-sol1}
\end{eqnarray}
The consequence of Eq.(\ref{sigma-sol1}) when the temperature reaches
the deconfinement one is important. 
The integral term on the right hand side of Eq.(\ref{sigma-sol1}) 
diverges for the mass spectral exponent $\alpha\le \frac{5}{2}+1$. 
It implies that the only possible solution 
is that $\frac{\partial\, b(\sigma)}{\partial \sigma}=0$ and $\sigma=0$. 
This means that the chiral field is certainly restored 
before the deconfinement phase transition point is reached.
Hence, the chiral symmetry is restored far away below 
the deconfinement point for nuclear matter that 
is dominated by the orthogonal Hagedorn states 
(or the Hagedorn states with mass spectral exponent 
$\alpha\le \frac{5}{2}+1$).
%%%%%%%%%%%%%%%%%%%%%%%%%%%%%%%%%%%%%%%%%%%%%%%%%%%%
%
%%%%%%%%%%%%%%%%%%%%%%%%%%%%%%%%%%%%%%%%%%%%%%%%%%%
On the other hand, it is possible to have a finite value solution 
for $\sigma$ for Hagedorn states with mass spectral exponent 
$\alpha>\frac{5}{2}+1$. 
Furthermore, when the exponent exceeds the limit $\alpha>\frac{7}{2}+1$, 
then it is possible to have the first order chiral phase transition 
simultaneously with the deconfinement phase transition point. 
This support the existence of the tri-critical point at finite 
$\mu_{B}$ and $T$ where the nuclear matter 
at moderate baryonic densities is supposed 
to be dominated by the unitary Hagedorn states 
prior the color deconfinement phase transition.
%%%%%%%%%%%%%%%%%%%%%%%%%%%%%%%%%%%%%%%%%%%%%%%%%%%%%%%%%%%%%%%%%%%
%%%%%%%%%%%%%%%%%%%%%%%%%%%%%%%%%%%%%%%%%%%%%%%%%%%%%%%%%%%%%%%%%%%
%%%%%%%%%%%%%%%%%%%%%%%%%%%%%%%%%%%%%%%%%%%%%%%%%%%%%%%%%%%%%%%%%%%
%%%%%%%%%%%%%%%%%%%%%%%%%%%%%%%%%%%%%%%%%%%%%%%%%%%%%%%%% 
%%%%%%%%%%%%%%%%%%%%%%%%%%%%%%%%%%%%%%%%%%%%%%%%%%%%%%%%%
%%%%%%%%%%%%%%%%%%%%%%%%%%%%%%%%%%%%%%%%%%%%%%%%%%%%%%%%%
%%%%%%%%%%%%%%%%%%%%%%%%%%%%%%%%%%%%%%%%%%%%%%%%%%%%%%%%%
%%%%%%%%%%%%%%%%%%%%%%%%%%%%%%%%%%%%%%%%%%%%%%%%%%%%%%%%%
%%%%%%%%%%%%%%%%%%%%%%%%%%%%%%%%%%%%%%%%%%%%%%%%%%%%%%%%%%%%%%%%%%%
%%%%%%%%%%%%%%%%%%%%%%%%%%%%%%%%%%%%%%%%%%%%%%%%%%%%%%%%%%%%%%%%%%%
%%%%%%%%%%%%%%%%%%%%%%%%%%%%%%%%%%%%%%%%%%%%%%%%%%%%%%%%%%%%%%%%%%%
\section{\label{sect-6a}
Discussions}

In order to study the deconfinement phase transition diagram in QCD, 
the hadronic density of states must be known for 
the entire energy domain below the borderline of 
the deconfinement phase transition 
to the true deconfined quark-gluon plasma.
The hadronic density of states for the discrete low-lying mass spectrum
is found from the available experimental data 
that is listed in the particle data group 
book~\cite{databook:2008}.
The discrete low-lying mass spectrum 
is extrapolated to the continuous 
high-lying one using reasonable 
theoretical models.
The standard theoretical procedure 
to find the hadronic density of states 
is carried out by computing 
the micro-canonical ensemble 
for the quark and gluon bag 
with specific internal symmetry 
that is consistent with the experimental 
observation and other theoretical models. 
The confined quark and gluon state 
is guaranteed by projecting the color-singlet state 
(i.e. the colorless state). 
The structure of the underlying internal color symmetry 
of a composite bag has attracted much attention
in order to understand the confinement/deconfinement 
in QCD. 
The Hagedorn model seems to be very useful in studying
the shear viscosity of QGP
and even investigating the extended gauge (conformal) 
field theories such as AdS/CFT.
The bag's constituent quarks and gluons are considered 
in the context of various underlying symmetry groups
such as the orthogonal, unitary and symplectic 
symmetry groups as well as $U(1)^{N_{c}}$ symmetry group.
%%%%%%%%%%%%%%%%%%%%%%%%%%%%%%%%%%%%%%%%%%%%%%%%%%%%%%%%%
In order to be consistent with QCD, the unimodular-like
constraint is imposed in the all symmetry groups 
those are under the present investigation. 
The unimodular-like constraint 
is an additional constraint for 
the orthogonal and symplectic symmetry groups 
and its absence does not affect 
the general discussion of the present work.
Usually, the natural choice in QCD is the unitary symmetry group.
Hence, the transmutation from the unitary symmetry group 
to orthogonal or symplectic symmetry group 
is associated with a new phenomenology.
The major argument is that: is it possible under
certain scenario the the Hagedorn matter
that is dominated by the colorless unitary states 
to be altered (by transmutation) to another one dominated 
by either colorless orthogonal states 
or colorless symplectic ones and vice versa?
It is important to mention here that 
the Hagedorns are colorless regardless 
of their internal color symmetries. 
The colorless is guaranteed by projecting 
the color-singlet-state.
The answer is that: this scenario seems to be 
reasonable in order to explain the elusive behaviour 
of the deconfinement quark gluon plasma 
in the phase transition diagram.

Furthermore, the relevant point to the hadronic 
phase transition is the Gross-Witten point.
This point is associated by modifications 
in the hadronic matter properties.
The Gross-Witten point is strongly believed
to be not the deconfinement point and to be merely 
a transition from hadronic matter that is known 
to be dominated by the low-lying mass spectrum 
to another class of  hadronic matter. 
The Gross-Witten point is interpreted to be the point
where the phase transition from the discrete  
low-lying mass spectrum to the continuous 
high-lying mass spectrum is taken place in the hadronic matter. 
The low-lying mass spectrum is defined by the known hadrons 
those are available and known experimentally 
while  the high-lying mass spectrum is defined by the Hagedorn states.
The phase transition from the low-lying 
to the high-lying hadronic mass spectrum 
is caused by the modification of  the analytical solution 
for the color-singlet-state of the underlying group structure 
of a composite bag. 
The analytical solution for the color-singlet-state 
is modified drastically when the Vandermonde effective 
potential develops a virtual singularity. 
When the solution is regulated, 
another analytical solution emerges and the hadronic phase transition 
from the discrete low-lying mass spectrum 
to the Hagedorn matter takes place in the system.
Evidently, the Gross-Witten point differs 
from the deconfinement phase transition point  and it is located 
in the hadronic phase below the deconfinement's threshold.
Therefore, the Gross-Witten point seems to be necessary 
for the existence of the tri-critical point and 
the multi-phase transition processes below the deconfined 
quark-gluon plasma.
%%%%%%%%%%%%%%%%%%%%%%%

The hadronic phase transition from 
the hadronic matter which is dominated 
by the discrete low-lying hadronic mass spectrum 
(consists baryons such as $N,\Lambda,\Sigma, \Xi$, 
$\cdots$ and mesons such as pions, kaons, $\cdots$)
(i.e. all the hadrons that are found in particle data group 
book~\cite{databook:2008})
to another hadronic matter that is dominated 
by the continuous high-lying hadronic mass spectrum 
(i.e. Hagedorn states) 
takes place at the Gross-Witten point.
It is argued that the hadronic phase transition 
is of the higher order and it is typically of 
a third order one. 
Since the discrete low-lying hadronic mass spectrum 
below the Gross-Witten point is presumed to maintain 
the original internal unitary $SU(N_c)$ structure,
it is natural to assume that the initial Hagedorn states 
which emerge just above the Gross-Witten point 
is the mass spectrum  of a composite bag 
with colorless $SU(N_c)$ states. 
Moreover, it is reasonable under certain conditions 
to imagine that it is easier for the Hagedorn states 
to modify their internal symmetry from 
the unitary $SU(N)$ representation to either 
an orthogonal $O(N)$ symmetry group 
or a symplectic $Sp(N)$ symmetry group. 
The modification of the composite Hagedorn 
internal symmetry 
takes place by either the partial symmetry breaking 
or the partial symmetry restoration
rather than breaking the symmetry group entirely 
from the colorless state to the colored 
$SU(N_{c})$ state (i.e. color-non-singlet state).
Since the discrete low-lying mass spectrum maintains
$SU(N_c)$ symmetry, the continuous high-lying mass 
spectrum threshold likely emerges at first as 
the colorless unitary $SU(N_{c})$ states.
The high thermal excitations modify the internal symmetry 
of the unitary Hagedorn states. 
Therefore, the continuous high-lying mass spectrum 
seems to have the feature to modify 
its own internal symmetry structure and 
subsequently the mass spectral exponent $\alpha$. 
The difficulty to catch a clean signature of the explosive 
quark-gluon plasma can be traced 
to the multi-modifications (or transmutations)
of the hadronic phase just below 
the deconfinement phase transition 
but above the Gross-Witten transition. 
Furthermore, the excited exotic matter that can be formed
aftermath the proton-proton collisions at LHC 
can be traced to Gross-Witten phase transition from the low-lying mass spectrum
to the Hagedorn states and the subsequent multiple phase transition processes 
at low baryonic density and high temperature. 

%%%%%%%%%%%%%%%%%%%%%%%%%%%%
%%%%%%%%%%%%%%%%%%%%%%%%%%%%
The breaking of the quark-gluon bag's internal symmetry
from the colorless $SU(N_c)$ state (i.e. color-singlet state)  
to the colored $SU(N_c)$ state (i.e. color-non-singlet state)
requires much more energy 
than that is needed for the Hagedorn state 
to alter its internal symmetry from 
the colorless unitary state 
to the colorless orthogonal state 
and freeing gluonic jets to the medium 
due to the reduction in the number of adjoint degrees 
of freedom in the composite Hagedorn bags.
In this sense, the colorless $SU(N_c)$ bags 
transmute to orthogonal ones 
and emit gluonic jets (due to the reduction of the adjoint degrees of freedom)
and these jets decay to other (exotic) particles. 
The resultant colorless orthogonal bags
under extreme hot nuclear matter may undergo a direct 
deconfinement phase transition to the quark-gluon plasma.  
In this case, the colorless orthogonal state 
is fully broken to the colored $SU(N_c)$ symmetry 
where the colored quarks and gluons are liberated
to form deconfined quark-gluon plasma.
%%%%%
%%%%%
Hence, it seems that the resultant Hagedorn phase 
which is dominated by 
colorless orthogonal quark-gluon bags 
explains the fluid properties such as 
the low viscosity for the quark-gluon plasma 
in the RHIC and possibly in the LHC. 
The possible explanation is that the effective 
orthogonal Vandermonde potential 
is weaker than the unitary one 
but it still acts as the Coulomb gas 
or gas of fluid droplets.
%%%%%%%%%%%%%%%%%%%%%%%%%%
On the other hand, it is possible 
to think about the coupling of various 
degrees of freedom such as 
the translation, color and flavor 
degrees of freedom when the nuclear 
matter is highly compressed. 
The translation invariance of 
the flavor quarks might be badly 
broken in the extreme dense nuclear matter.
Hence, it is reasonable to assume that 
the color symmetry group partially 
merges with the translation and/or 
flavor symmetry group and the resulting 
color-flavor (or color-translation)
degrees of freedom form an effective 
color-flavor symmetry group. 
In other words, the colorless unitary 
(or orthogonal)  
state couples with flavor degree of freedom 
$SU_{V}(N_{f})$ (or $SO(N_{f})$) and forms 
an effective colorless symplectic $Sp(N)$ state. 
This new symmetry maintains the maximal 
flavor invariance 
(i.e. conserved $U(1)^{N_{f}-1}$) 
and guarantees the colorless state 
(i.e. color-singlet) in a nontrivial way. 
%%%%%%%%%%%%%%%%%%%%%%%%%%%
Hence, the effective symplectic symmetry is basically 
a coupled color-flavor symmetry. 
Furthermore, the color symmetry remains 
colorless when the symplectic $Sp(N)$ symmetry 
is projected to the color-singlet state 
while the flavor symmetry remains to maintain 
its maximal invariance in the following possible scenario 
\begin{eqnarray}
\left.SU(N_{c})\right|_{singlet}\,\times\,
SU_{L}(N_{f})\,\times\,SU_{R}(N_{f})
&\rightarrow& 
\left.SU(N_{c})\right|_{singlet}\,\times\,
SU_{V}(N_{f})
\nonumber\\
&\rightarrow& 
\left.Sp(N_{c})\right|_{singlet}\,\times\,U(1)^{N_{f}-1}.
\end{eqnarray}
Hence, the extremely dense nuclear matter seems to favor 
the phase transition from the unitary Hagedorn states 
to the symplectic Hagedorn states.
%%%%%%%%%%%%%%%%%%%%%%%%%%
%%%%%%%%%%%%%%%%%%%%%%%%%%

The mass spectral exponents for the Hagedorn states 
with different internal symmetries are displayed 
in Table~(\ref{table1}).

\begin{figure} 
\includegraphics{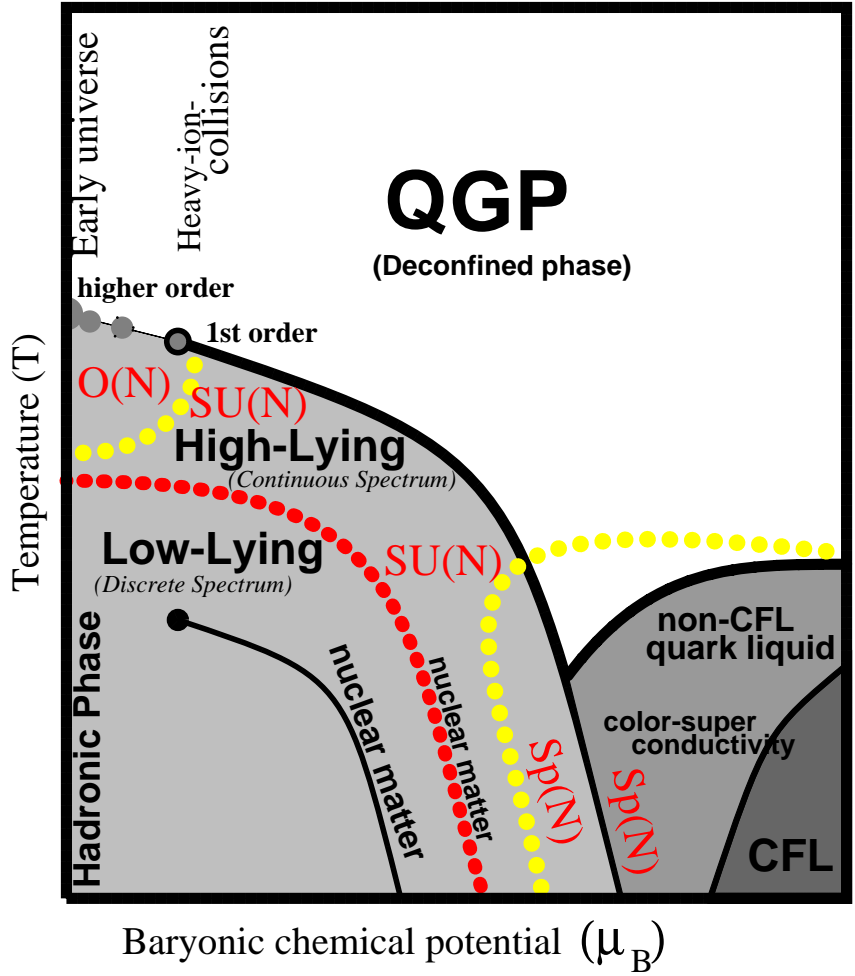}
\caption{\label{phase_sketch1}
(Color online)
The sketch for the order and shape of the phase transition diagram 
($\mu_B$-$T$)
outlining the phase transition between the hadronic matter 
which is dominated by the discrete low-lying mass spectrum 
to another one that is dominated 
by the continuous high-lying mass spectrum 
and the phase transition from the hadronic phase 
to the deconfined quark-gluon-plasma.
%%%%%%%%%%%%%%%%%%%%%%
The contiguity of the the discrete low-lying mass 
spectrum and the continuous high-lying mass spectrum 
is indicated by the lower-left red dotted line.
The Hagedorn phase that is dominated by the  
continuous high-lying mass spectrum states 
splits 
into three individual phases that are dominated 
by $SU(N)$, $O(N)$ and $Sp(N)$ internal color structures.
The conventional phase transition 
diagram for the color superconductivity 
and color-flavor locked phase is depicted. 
The color-superconductivity phase falls under 
the $Sp(N)$ group domain.}
\end{figure}

Fig.~(\ref{phase_sketch1}) depicts the phase transition 
diagram in the $\mu_B\,-\,T$ plane.
It is shown that at low temperature the dilute nuclear 
matter which is dominated 
by the discrete low-lying hadronic 
mass spectrum states,
such as pions, kaons etc and nucleons, Lambdas etc 
and so on, forms a gas of free 
discrete low-lying hadrons.
Under certain conditions, those hadrons
interact with each others and form nuclear matter 
(usually in the simplest approximation, 
the hadrons are treated as ideal gas). 
When the nuclear matter is heated up
and/or compressed, it undergoes higher order 
Gross-Witten Hagedorn phase transition 
from the nuclear matter that is dominated 
by the discrete low-lying mass spectrum hadrons 
to another nuclear matter that is dominated 
by the continuous high-lying mass spectrum hadronic 
(Hagedorn) states. 
The Gross-Witten line appears as the dotted line (red online)
in the lower-left of the phase transition diagram.
It is shown that the nuclear matter which is dominated 
by the discrete low-lying hadrons passes 
higher order Gross-Witten Hagedorn transition 
to another nuclear matter that is dominated 
by the continuous high-lying mass spectrum 
of the colorless unitary $SU(N_c)$ quark-gluon bag. 
The order of the phase transition is likely 
to be a third order one.
The intermediate Hagedorn matter 
is shown above the Gross-Witten line 
which appears as the dotted line (left) and
below the deconfinement phase transition line
which appears as the upper thick solid line.
It is evident that the continuous high-lying 
mass spectrum just above the Gross-Witten point 
(i.e. a continuous line in the $\mu_{B}-T$ plane)
is likely to be dominated by the unitary Hagedorns 
(i.e. colorless unitary $SU(N_c)$ states).
When the dilute $SU(N_{c})$ Hagedorn matter is heated up, 
the Hagedorn states mutate their internal symmetry 
from the colorless $SU(N_c)$ symmetry
to the colorless $O(N_c)$ symmetry.
Nevertheless, the flavor symmetry maintains 
its optimal invariance in the hot nuclear bath.
Hence, the unitary Hagedorn matter undergoes 
phase transition to another Hagedorn matter 
that is dominated 
by the colorless orthogonal $O(N_c)$ states 
and becomes orthogonal Hagedorn matter. 
The phase transition from unitary Hagedorn matter 
to orthogonal Hagedorn matter is shown by 
the light dotted line (yellow online)
on the top left of the $\mu_{B}-T$ phase transition diagram. 
At higher temperature, the Hagedorn matter that 
is dominated by the colorless orthogonal states 
undergoes deconfinement phase transition
to the quark-gluon plasma where 
the color-singlet constraint is broken in order 
to form colored $SU(N_c)$ states. 
When the color singlet constraint is broken badly,
the colored quarks and gluons are liberated
and/or the quark-gluon bags become colored ones. 
The order of the deconfinement phase transition 
is supposed to be varied and it is 
shown by several gray circles on the top and left 
of the $\mu_{B}\,-\,T$ phase transition diagram. 
Two major factors likely modify the order of phase transition:
The first factor comes mainly from 
the Hagedorn internal symmetry modification 
to either the colorless $SU(N_c)$, colorless $O(N_c)$ 
or colorless $Sp(N_c)$ states.
It is also possible to assume that the heat  can 
release most of  the gluonic contents of the Hagedorn states 
and mutates them to the colorless $U(1)^{N_c}$ quark bags 
in the case of the dilute nuclear matter where $\mu_{B}\approx 0$
(i.e. further reduction in the adjoint degrees of freedom). 
Eventually, the heat can break the color-singlet constraint and 
the metastable colored (color-non-singlet-sate) $SU(N_c)$ 
quark-gluon bags (with initial total zero color charges: 
i.e. color-neutral bags) emerge in the system.
The second factor is actually originated from the deformation 
of the Hagedorn cavity boundary surface.
The situation is completely different 
when the nuclear system is compressed 
and cooled to lower temperatures. 
This is marked by the light dotted line (yellow online) 
on the lower right side of the $\mu_{B}\,-\,T$ diagram 
in particular when the baryonic chemical potential 
$\mu_{B}$ becomes relatively large.
The flavor degree of freedom becomes
more important albeit the system tends somehow 
to maintain the optimal flavor invariance 
and restores the chiral symmetry.  
Furthermore, the strangeness degree of freedom is essential, 
though it is not mentioned explicitly throughout the present work. 
Furthermore, along the $\mu_B$ direction, the nuclear matter 
also passes the Gross-Witten Hagedorn phase transition from 
the hadronic matter which is dominated by the discrete low-lying mass spectrum hadrons
to another one that is dominated by the continuous high-lying mass 
spectrum Hagedorns.
The continuous high-lying mass (and volume) 
spectrum phase is initially dominated by the colorless $SU(N_c)$ states 
in particular just above the mete of the Gross-Witten point. 
Evidently the system undergoes phase transition to another 
Hagedorn matter that is populated by colorless $Sp(N)$ states 
when the color and flavor degrees of freedom are coupled 
to form $Sp(N)$ symmetry group.
The phase domain of the new colorless $Sp(N_c)$ states 
is immured by the light dotted line (online green) 
on the lower-right side of the $\mu_B-T$ diagram. 
It sounds that the nuclear matter which 
is dominated by the colorless $Sp(N)$ states 
overlaps and matches the color-superconductivity, 
non-CFL quark liquid and CFL alike phases. 
The complexity of the color-flavor coupling 
and the nontrivial $Sp(N)$ structure increases 
significantly when the system is compressed 
to form denser nuclear matter.
%%%%%%%%%%%%%%%%%%%%%%%
%%%%%%%%%%%%%%%%%%%%%%%

\begin{figure}
\includegraphics{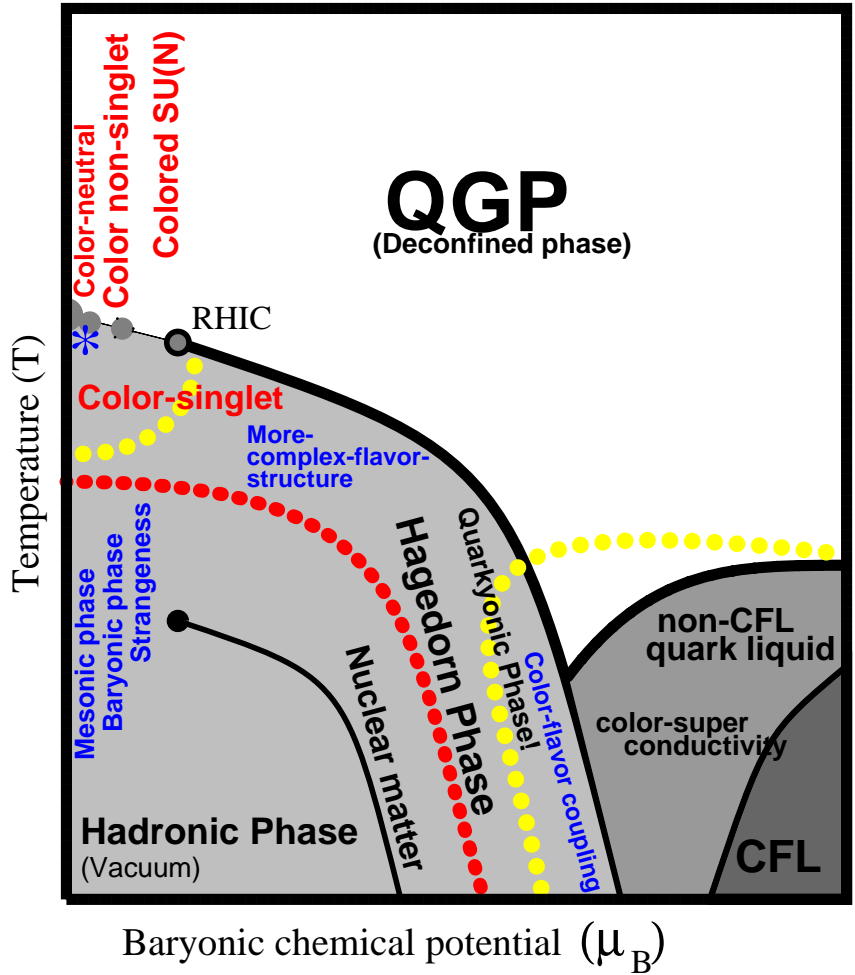}%
\caption{\label{phase_sketch2}
(Color online)
Same as Fig.~(\ref{phase_sketch1}).
The hadronic phases those are dominated 
by the mesonic, baryonic and strangeness phases 
are depicted explicitly.
It is shown that the complexity of the 
Hagedorn state internal color-flavor structure 
increases with respect to $\mu_{B}$. 
The mesonic Hagedorn states are dominated 
at small $\mu_B$ 
while the baryonic and excited Hagedorn states 
appear abundantly as the $\mu_B$ increases 
and when $\mu_B$ becomes significantly large
the Hagedorn states
with strangeness and more other complicated
degrees of freedom emerge in the system.
The $*$ (blue online) indicates the color-singlet 
$U(1)^{N_c}$ states (Hagedorns)
with the unimodular-like constraint.} 
\end{figure}

Fig.~(\ref{phase_sketch2}) focuses on 
the flavor symmetry besides the color symmetry. 
It is shown that the Hagedorn matter is initially dominated 
by the mesonic Hagedorn states for small $\mu_B$. 
When $\mu_B$ increases, the baryonic Hagedorn states 
appear in the system besides 
the existing mesonic Hagedorn states.
It follows from Eq.(\ref{conservb1}) that mass spectral exponent
increases by $\alpha\rightarrow\alpha+\frac{1}{2}$ when the Hagedorn states 
attain specific baryonic number and it is also modified by 
$\alpha\rightarrow\alpha+\frac{1}{2}$ when the strangeness number 
is included.
The strangeness turns to be important 
when the system is compressed 
to larger $\mu_{B}$ and is heated up to higher temperature. 
Furthermore, the Hagedorn matter tends 
to maintain the optimal flavor invariance in the dilute hot nuclear matter 
keeping in mind that the Hagedorn states are colorless regardless
of the type of the internal color structure symmetry 
or if the color degrees of freedom are coupled to other degrees of freedom.  
When the the maximum flavor invariance 
is maintained at high temperature, 
the classical Maxwell-Boltzmann statistics 
becomes a good approximation under certain circumstances. 
The classical statistics approximation also suits 
the exotic Hagedorn states.
It is likely that the hadronic matter which is dominated 
by the colorless orthogonal states undergoes smooth phase transition 
to the colored (non-color-singlet) $SU(N_c)$ quark-gluon matter 
with initial neutral color charges (i.e. color-neutral)
for $\mu_B\approx 0$ at the critical temperature $T_c$.
The alternate scenario is that the colorless $U(1)^{N_c}$ quark bags 
may emerge above the orthogonal Hagedorn states 
and below the deconfinement phase transition 
where the color-singlet constraint of the colorless states 
is broken and the colored (i.e. color-non-singlet) states 
appear abundantly in the system.
This kind of intermediate process smoothen 
the order of the phase-transition remarkably 
at $\mu_{B}\approx 0$.
This process is depicted as the symbol 
$*$ in Fig.~(\ref{phase_sketch2}).
Nonetheless, it is possible for the colorless 
orthogonal (or unitary) states to pass 
the phase transition to metastable colored bags 
via the following reaction:
\begin{eqnarray} 
\mbox{color-singlet}~~ SU(N_{c})~
\,\mbox{or}\,~O(N_{c})~\mbox{Hagedorns}~
\,\mbox{etc}~\,
\rightarrow ~\mbox{colored}~SU(N_{c})~\mbox{plasma}. 
\end{eqnarray}
%%%%%%%%%%%%%%%%%%%%%%%%%%%%%%%%
The overall color charge of each bag remains 
a color-neutral one
(i.e. neutral color charge for $N_{fun}$ 
fundamental quarks and $N_{adj}$ adjoint gluons)
before the bags expand and become colored ones  
and, eventually, form colored quark-gluon plasma.
This kind of phase transition is likely supposed to be of higher order 
and takes place on the left hand side of the 
tri-critical point which appears in the phase transition diagram. 
When $\mu_B$ increases, the flavor structure becomes more rich 
and complex. 
The flavor symmetry invariance turns to be broken and involved. 
Therefore, the complexity of the color-flavor structure increases 
as the system is compressed and cooled down to lower temperature.
The color-flavor internal symmetry is modified 
in an optimal way minimizing the flavor dependence 
and maximizing the preservation of the colorless state. 
This mechanism explain why the quarks avoid to propagate freely 
in the medium.
When $\mu_{B}$ becomes significantly large, the complexity 
of the color-flavor symmetry increases and the $SU(N_c)$ 
symmetry couples with the $SU(N_{f})\times SU(N_{f})$ symmetry.
The internal symmetry of the quark and gluon bag turns 
to be $Sp(N_c)$ color-flavor symmetry.
At high temperature, both the flavor invariance 
and the color-singlet constraint (i.e. colorless state) 
modify the quark-gluon bag's underlying 
internal symmetry from $SU(N_c)$ to $O(N_c)$.
The quarkyonic matter is conjectured to be dominated 
by the continuous high-lying hadronic states (Hagedorns) 
with varied symmetries such as the orthogonal, unitary 
and symplectic ones. 
%%%%%%%%%%%%%%%%%%%%%%%%%%%
%%%%%%%%%%%%%%%%%%%%%%%%%%%

\begin{figure}
\includegraphics{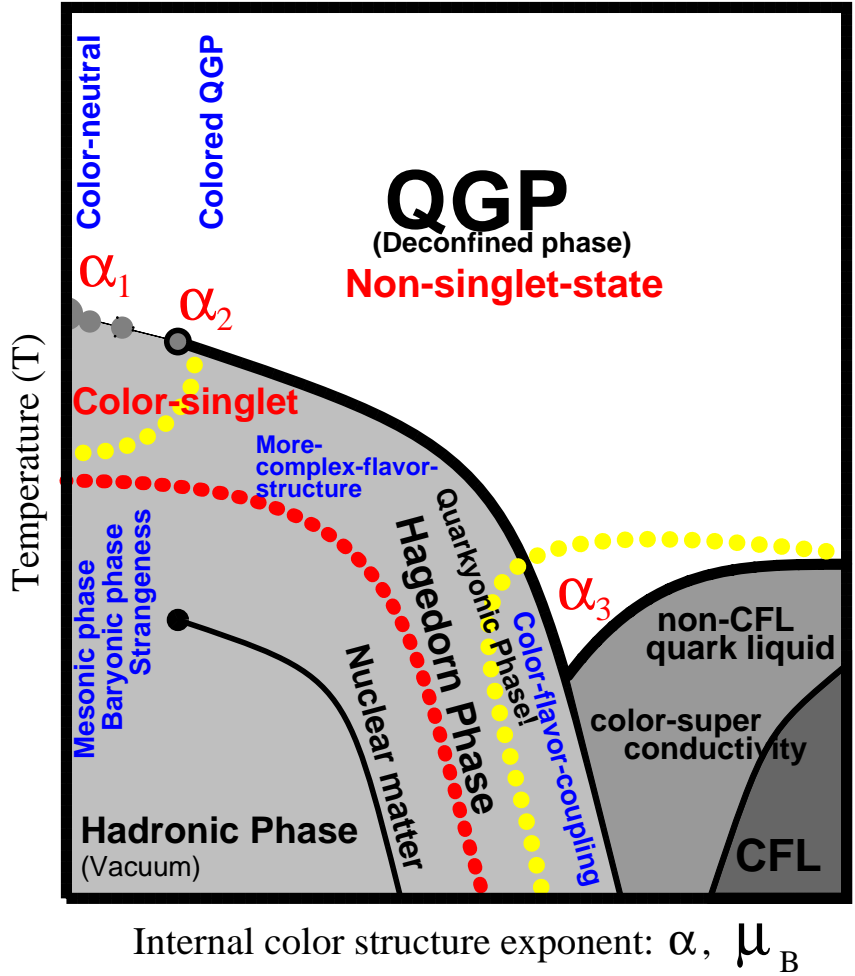}%
\caption{\label{phase_sketch3}
(Color online)
Same as Fig.~(\ref{phase_sketch1}). 
The order of the phase transition 
from the hadronic phase to the quark-gluon-plasma
is shown to depend on 
the Hagedorn mass spectral exponent $\alpha$.
The exponent $\alpha$ is modified 
and increases with respect to $\mu_B$. 
The order of the phase transition is likely 
to be the first order, second order 
and higher order for the exponents 
$\alpha_1$, $\alpha_2$ and $\alpha_3$ 
respectively.
The quark-gluon plasma is likely 
to maintain the color neutrality one
with the higher order phase transition 
at small $\mu_B$ and high $T\sim T_{c}$ 
before it switches to the true deconfined 
colored plasma when $\mu_{B}$ increases.
The exponent $\alpha$ depends on
the Hagedorn's internal color, 
flavor and configuration symmetries.}
\end{figure}

It is shown in Fig.~(\ref{phase_sketch3}) that 
the mass spectral exponent $\alpha$ 
is modified with respect to $\mu_{B}$ and $T$ 
below the brink of the deconfinement phase transition 
borderline to the quark-gluon plasma.
The Hagedorn's mass spectral exponent $\alpha$ 
is small for the dilute and hot hadronic matter. 
The exponent $\alpha$ is modified as the hadronic matter 
is compressed and/or heated. 
The exponent varies and takes the values 
$\alpha_{1}\,\le\,\alpha_{2}\,\le\,\alpha_{3}$. 
The exponents are reduced to 
$\alpha_{1}=3$, $\alpha_{2}=9/2$ 
and $\alpha_{3}=15/2$ 
for the colorless orthogonal, unitary and symplectic
Hagedorn states, respectively, when $N_c=3$.
The mass spectral exponent for 
the colorless $U(1)^{N_c}$ composite states (Hagedorns) 
is reduced to $\alpha=3/2$ for $N_c=3$.
Furthermore, the mass spectral exponent for the finite metastable 
quark-gluon bag with the colored $SU(N_c)$ state is reduced 
to $\alpha_{non}=1/2$ (i.e. simply colored $SU(N_c)$ states). 
The metastable colored bags are expected to emerge before 
they expand smoothly and overlap each other and eventually 
form quark-gluon plasma.  
%%%%%%%%%%%%%%%%%%%%%%%%%%%%%%%%%%%%%%%%%%%%%%%%%%%

Nonetheless, the fuzzy and center of mass corrections modify 
the mass spectral exponent 
$\alpha$ significantly~\cite{Zakout:2006zj}. 
The center of mass correction has been introduced 
by Kapusta~\cite{Kapusta:1982a}.
Therefore, it is expected that when the fuzzy and center 
of mass corrections are considered,
the exponents $\alpha$ for the orthogonal, 
unitary and symplectic Hagedorn states overlap 
with each other under certain conditions 
in the hot and compressed nuclear matter. 
This leads to a smooth transmutation 
from certain composite bag's internal symmetry to another one.
%%%%%%%%%%%%%%%%%%%%%%%%%%%%%%%%%%%%%%%
%%%%%%%%%%%%%%%%%%%%%%%%%%%%%%%%%%%%%%%
%%%%%%%%%%%%%%%%%%%%%%%%%%%%%%%%%%%%%%%

\begin{figure}
\includegraphics{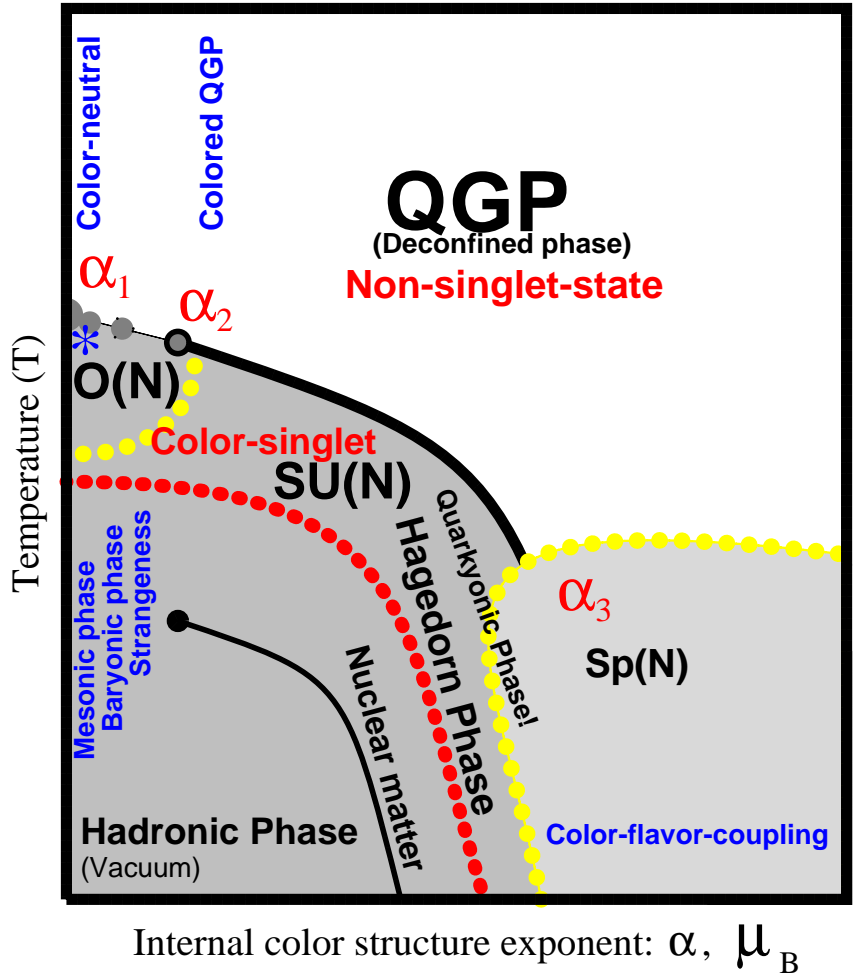}%
\caption{\label{phase_sketch4} 
(Color online)
Same as Fig.~(\ref{phase_sketch1}).
The complexity of the Hagedorn bag's 
internal structure is proportional to $\mu_B$ 
and is inverse proportional to $T$. 
When $\mu_B$ becomes sufficient large, 
the color and flavor degrees of freedom 
are coupled to each other and 
form $Sp(N)$ symmetry group.
The complexity of the color-flavor symmetry 
and space configuration symmetry 
increases as the system is cooled down and 
extremely compressed to large $\mu_B$.}
\end{figure}

The complexity of the color-flavor coupling 
structure for the large  $\mu_{B}$ 
is illustrated in Fig.~(\ref{phase_sketch4}).
The color-flavor coupling is simplified drastically 
by correlating the bag's color and flavor symmetries 
to form the $Sp(N)$ symmetry group.
The mass spectrum for the color superconductivity 
physics is simplified drastically 
by considering the colorless $Sp(N)$ state.  
%%%%%%%%%%%%%%%%%%%%%%%%%%%
It is evident that the mass spectral exponent  $\alpha$
for the continuous high-lying mass spectrum 
$\rho_{(II)}=c\,m^{-\alpha}\,e^{b\,m}$
depends essentially on $\mu_B$ and $T$. 
The order of phase transition is sensitive 
to the exponent $\alpha$. 
This sensitivity indicates the existence 
of the tri-critical point in the phase transition diagram.
The order of deconfinement phase transition
is found first order for the Hagedorn states 
with the exponent $\alpha>\frac{7}{2}$. 
The second and third order phase 
transitions are found for Hagedorn states with exponents
$\frac{7}{2}\,\ge\alpha>\,3$ 
and $3\,\ge\alpha>\,\frac{17}{6}$, respectively.
Moreover, the $n^{th}$ order phase transition 
takes place for 
$\frac{5}{2}+\frac{1}{n-1}\,
\ge\,\alpha\,>\,\frac{5}{2}+\frac{1}{n}$.
There is no direct (explosive) deconfinement phase transition 
to quark-gluon plasma for the hadronic matter that is populated 
by Hagedorn states with the mass spectral exponent 
$\alpha\le\,\frac{5}{2}$.
The phase transition for the Hagedorn states with 
the mass spectral exponent $\alpha\,\le\,\frac{5}{2}$ 
is smooth 
(i.e. smooth cross-over phase transition). 
In this case, the Hagedorn bags expand smoothly and they are gently thermally 
excited  without an abrupt phase transition to explosive quark-gluon plasma.
If the unitary Hagedorns freeze out  they undergo Gross-Witten phase transition 
to the low-lying mass spectrum hadronic matter. 
The means that the Hagedorn bags evaporate 
to the discrete low-lying mass spectrum particles 
and this resembles the black hole evaporation in AdS/CFT duality. 
%%%%%%%%%%%%%%%%%%%

It is reasonable to assume that the thermal excitation 
may change the Hagedorn bag's internal structure.
For instance under certain circumstances the nuclear matter 
that is dominated by the unitary Hagedorn states is altered
to be dominated by the orthogonal Hagedorn states
when the dilute nuclear matter is heated up to higher temperatures.
Since the mass spectral exponent for orthogonal Hagedorns 
(i.e. colorless orthogonal states)
is found to be $\alpha_{1}=3$, 
it is likely that 
the orthogonal Hagedorn matter undergoes 
third order phase transition to quark-gluon plasma.
Furthermore, it is possible that the orthogonal Hagedorn states
are altered to colorless $U(1)^{N_c}$ states 
when the very dilute nuclear matter is further heated up 
to higher temperatures.
The very dilute nuclear matter might be created 
in the $pp$ collisions at LHC besides the heavy ion collisions.
The Hagedorn matter which is dominated by the colorless 
$U(1)^{N_c}$ has the mass spectral exponent $\alpha=3/2$.
Hence, the nuclear matter that is dominated by these states 
does not undergo direct abrupt phase transition 
to quark-gluon plasma but rather smooth cross-over phase transition. 
When the medium is further heated up to higher temperature 
these states 
(i.e. Hagedorn states with the mass spectral exponent $\alpha=3/2$)
may be mutated to metastable colored quark-gluon bags 
with the mass spectral exponent $\alpha=1/2$. 
Since the states with mass spectral exponent $\alpha=1/2$ 
do not pass direct explosive deconfinement phase transition 
to quark-gluon plasma, the colored quark-gluon bags expand smoothly 
and the system undergoes smooth phase transition 
to colored quark-gluon plasma. 
%%%%%%%%%%%%%%%%%%%%%%%%%%%%%%

The orthogonal Hagedorn states 
are mutated to the colorless $U(1)^{N_c}$ 
quark-gluon bags due to the high thermal excitations
in the hot and very dilute nuclear matter 
(i.e. $\mu_{B}\approx 0$).
Since the new nuclear matter turns to be dominated 
by the colorless $U(1)^{N_c}$ quark-gluon bags, 
it does not likely undergo direct phase transition
to explosive quark-gluon plasma. 
But instead, the resultant Hagedorn states are gradually 
altered to metastable colored quark-gluon bubbles.
The metastable colored quark-gluon bags 
expand gradually and overlap each other smoothly  
until the entire space is filled by giant colored (non-singlet) bags. 
The resultant matter have an initial neutral color charge 
aftermath the phase transition.
Therefore, the constraints of the conserved color charges 
must be embedded in the system through the color chemical potentials. 
This kind of (color-non-singlet) matter with the mass spectral exponent 
$\alpha_{non}$ undergoes a smooth cross-over phase transition
to non-explosive quark-gluon plasma. 
The multi-processes mechanism in the phase transition 
from the low-lying hadronic phase to the quark-gluon plasma
strongly indicates the fluid behaviour for the quark-gluon plasma.
The color-singlet states for the quark-gluon bag with an orthogonal color 
representation rather than the unitary one can be interpreted 
as a gas of Coulomb quark-gluon bags 
(or quark-gluon liquid).
Furthermore, the color-singlet states for the quark bag 
with $U(1)^{N_c}$ color symmetry group can be argued
to be quark rich liquid bags and gluon rich medium.  
The phase transition to the colorless orthogonal states
(i.e. orthogonal Hagedorns)
or colorless $U(1)^{N_c}$ states may explain the fluid properties 
and the low viscosity factor for the quark-gluon plasma at the RHIC 
and possibly LHC. 
This class of phase transition enriches the medium with gluonic jets 
besides the Hagedorn states. 
Furthermore, in the case that the orthogonal or  $U(1)^{N_c}$ Hagedorn 
states freeze out at some point before reaching the quark-gluon plasma, 
they evaporate to the (exotic-) low-lying mass spectrum particles
via Gross-Witten like phase transition in an analogous way to the black 
holes evaporation in AdS/CFT.
%%%%%%%%%%%%%%%%%%%%%%%%%%%%%%%%%%

The hot and little compressed hadronic matter ($\mu_{B}\neq 0$)
is likely to be dominated by the orthogonal Hagedorn states 
above the Gross-Witten point.
This kind of nuclear matter tends to undergo higher order phase transition 
( for instance third order with $\alpha$=3 )
to the deconfined quark-gluon plasma.
In this case, the color-singlet state of the orthogonal 
symmetry group is broken smoothly to form the colored 
$SU(N_c)$ state (i.e. color-non-singlet state)
and, subsequently, the quarks and gluons are liberated.
At the intermediate baryonic chemical potential $\mu_{B}$, 
the continuous high-lying unitary Hagedorn 
matter undergoes first order phase transition
to explosive quark-gluon plasma.
The color-singlet state is broken badly to form the colored 
$SU(N_c)$ state. 
The explosive quark-gluon plasma needs 
much thermal energy to be detonated.
Furthermore, it should be noted that when $\mu_{B}$ 
is reduced below a critical value, 
the unitary Hagedorn states tend to 
transmute to the orthogonal Hagedorn states 
prior to the deconfinement phase transition 
to quark-gluon plasma but this is not the case 
for the saturate orthogonal Hagedorn matter
when $\mu_{B}$ increases and exceeds 
a critical value. 
%%%%%%%%%%%%%%%%%%%%%%%%%%%%%%%%%%
%%%%%%%%%%%%%%%%%%%%%%%%%%%%%%%%%%
%%%%%%%%%%%%%%%%%%%%%%%%%%%%%%%%%%
%%%%%%%%%%%%%%%%%%%%%%%%%%%%%%%%%%

On the other hand, the mass spectral exponents 
for the unitary and symplectic Hagedorn states
are reduced to $\alpha_{2}=9/2$ and $\alpha_{3}=15/2$. 
This means that the nuclear matter which is dominated 
by the unitary Hagedorn states likely undergoes 
direct first order phase transition 
to the quark-gluon plasma.
Furthermore, it is obvious that the symplectic nuclear 
matter likely undergoes first order phase transition 
to the explosive quark-gluon plasma when it is heat up.
%%%%%%%%%%%%%%%%%%%%%%%%%%%%%%%%
%%%%%%%%%%%%%%%%%%%%%%%%%%%%%%%%
%%%%%%%%%%%%%%%%%%%%%%%%%%%%%%%%
%%%%%%%%%%%%%%%%%%%%%%%%%%%%%%%%%%%%
%%%%%%%%%%%%%%%%%%%%%%%%%%%%%%%%%%%%

When the hadronic matter is compressed 
to the significantly large $m_{B}$ and cooled down, 
the unitary Hagedorn states are mutated to the symplectic ones due to 
the internal color-flavor symmetry transmutation. 
In this case the saturate unitary Hagedorn matter 
transfers to another nuclear matter that is dominated 
by the symplectic Hagedorn states.
The  mass spectral exponent for the symplectic matter 
is $\alpha_{3}=15/2$.
The physics of the symplectic Hagedorn matter 
is very rich.
In the context of the present model, 
the highly dense neutron star turns to be 
dominated by the symplectic hadronic matter 
rather than the conventional quark matter 
in order to soften the equation of state.
Furthermore, the existence of the color super-conductivity 
phases such as the color-flavor locking phase 
can be interpreted in terms of the the Hagedorn complex 
internal structure and the color-flavor coupling channel.
%%%%%%%%%%%%%%%%%%%%%%%%%%%%%%%%%%%%%%%%%%%%%
%%%%%%%%%%%%%%%%%%%%%%%%%%%%%%%%%%%%%%%%%%%%%
%%%%%%%%%%%%%%%%%%%%%%%%%%%%%%%%%%%%%%%%
%%%%%%%%%%%%%%%%%%%%%%%%%%%%%%%%%%%%%%%%
%%%%%%%%%%%%%%%%%%%%%%%%%%%%%%%%%%%%%%%%
%%%%%%%%%%%%%%%%%%%%%%%%%%%%%%%%%%%%%%%%

%%%%%%%%%%%%%%%%%%%%%%%%%%%%%%%%%%%%%%%%%%%%%
%%%%%%%%%%%%%%%%%%%%%%%%%%%%%%%%%%%%%%%%%%%%%
%%%%%%%%%%%%%%%%%%%%%%%%%%%%%%%%%%%%%%%%
%%%%%%%%%%%%%%%%%%%%%%%%%%%%%%%%%%%%%%%%
%%%%%%%%%%%%%%%%%%%%%%%%%%%%%%%%%%%%%%%%
%%%%%%%%%%%%%%%%%%%%%%%%%%%%%%%%%%%%%%%%

\begin{figure}
\includegraphics{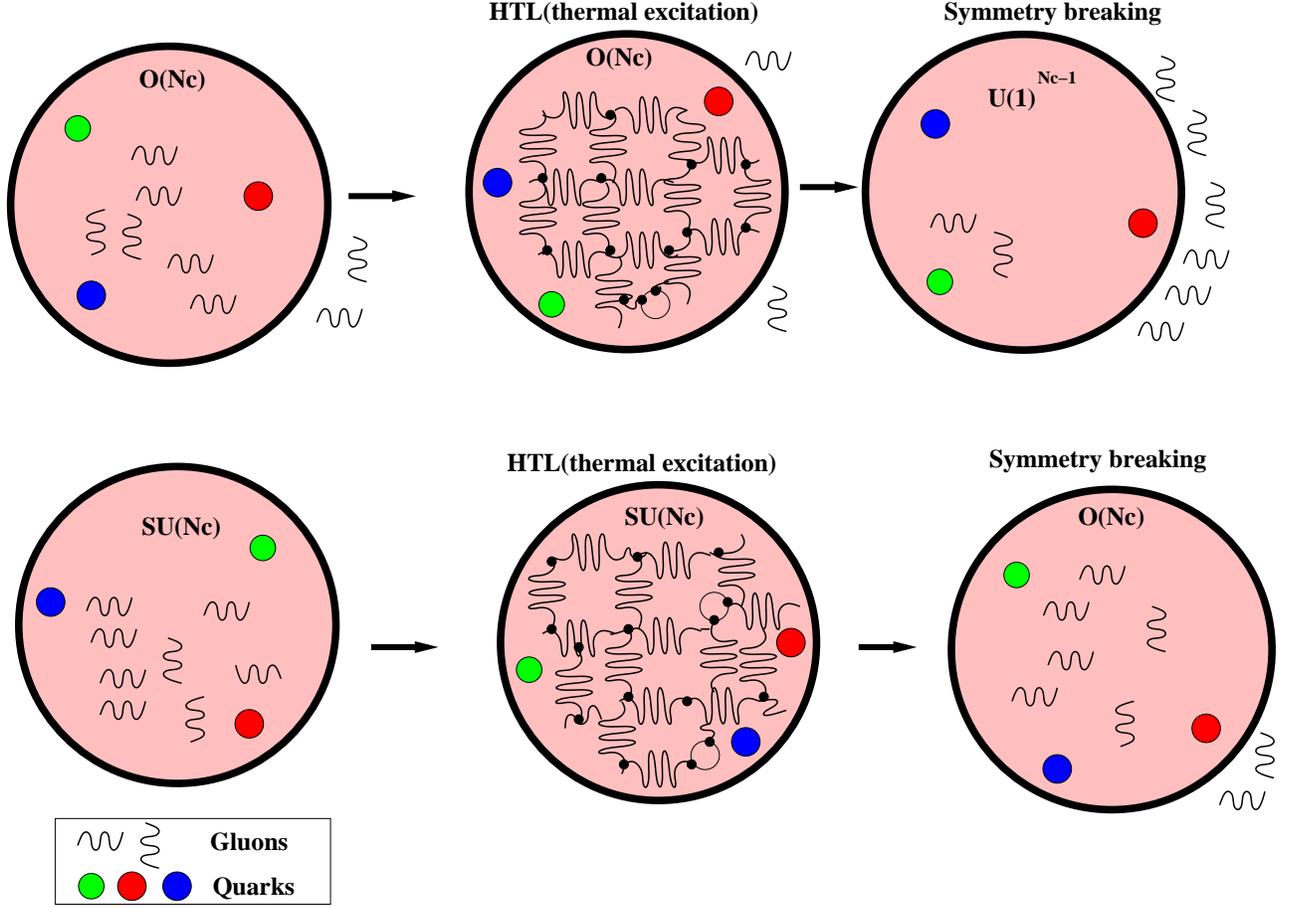}%
\caption{\label{symmetry-sketch1}
(Color online)
The Hagedorn's bag internal symmetry breaking from $SU(N_c)$ to $O(N_c)$ 
and eventually to $U(1)^{N_c-1}$ in the dilute and hot nuclear matter.
The quarks are represented by colored circles 
while the gluons are shown by short wavy lines.
The hard thermal loop is expected to generate degenerate/non-degenerate 
gluon's Debye masses with vanishing/non-vanishing adjoint (mean-) fields 
$\hat{A}^{a}$ depending on the medium.
This mechanism is responsible for the symmetry transmutation
in order to reduce to the constituent quark-gluon self-interactions 
and basically to maintain the quark confinement. 
The liberated gluons acquire finite masses ($m^{2}_{D}\propto g^{2} T^{2}$)
due to the Higgs phenomenon. 
In the present calculation, the liberated gluons are practically 
treated as an ideal gas of massless particles.}
\end{figure}

In order to understand the microscopic mechanism of the multiple 
phase transitions from symmetry group to another, the higher order corrections 
of the hard thermal loops become crucial.
Although these corrections are calculated by perturbation,
the higher order corrections beyond the one-loop correction
are highly non-trivial and also the non-perturbative effect 
may turn to be essential.
The simplest way to understand the symmetry breaking from one group 
to another is to study the symmetry breaking 
from $SU(N_{c})$ to $U(1)^{N_c-1}$.
The program to investigate the local and global symmetry behavior 
at finite temperature and density
such as the symmetry breaking or restoration is a straightforward one.
In fact this program relies basically on the dependence 
of effective mass on the temperature~\cite{Weinberg:1974hy,Dolan:1973qd}.
The symmetry is broken when the effective masses turn to be finite
and non-degenerate ones and their corresponding adjoint mean-fields 
become finite ones when the effect of the higher order terms 
that are beyond the quadratic part in the effective potential 
turn to be non-negligible.
The gluon effective mass in the present case is defined by the gluon's Debye mass.
In order to simplify the discussion to the minimum, 
the effective potential for the adjoint fields with perturbative corrections 
beyond the one loop correction will not be considered.
Fortunately, the one loop correction sheds some information about what
is needed to understand how the symmetry can be broken 
under extreme conditions.
The effective potential for the adjoint fields can be written roughly
in the following way
\begin{eqnarray}
V_{eff}\left(\hat{A}^{a}_{\mu}\right)&\doteq&
\frac{1}{2} { \Pi^{\mu \nu} }^{ a' a } {\hat{A}_{\mu}}^{a'} {\hat{A}_{\nu}}^{a}
+\frac{1}{3!} { {\lambda}_{3}^{\mu\nu\kappa} }^{a b c} 
{\hat{A}_{\mu}}^{a} {\hat{A}_{\nu}}^{b} {\hat{A}_{\kappa}}^{c}
+ \frac{1}{4!} { {\lambda}_{4}^{\mu \mu' \nu \nu'} }^{a a' b b'} 
{\hat{A}_{\mu}}^{a} {\hat{A}_{\mu'}}^{a'} {\hat{A}_{\nu}}^{b} {\hat{A}_{\nu'}}^{b'}
+ \cdots ,
\nonumber\\
&\doteq&
\frac{1}{2} { \Pi^{\mu \nu} }^{ a' a } {\hat{A}_{\mu}}^{a'} {\hat{A}_{\nu}}^{a}
+ \frac{1}{4!} { {\lambda}_{4}^{\mu \mu' \nu \nu'} }^{a a' b b'}
{\hat{A}_{\mu}}^{a} {\hat{A}_{\mu'}}^{a'} {\hat{A}_{\nu}}^{b} {\hat{A}_{\nu'}}^{b'}
+ \cdots ,
\label{potential-eff1}
\end{eqnarray}
where ${ \Pi^{\mu \nu} }^{ a' a }\doteq \left<{\Pi^{\mu \nu}}^{a' a}
\left(p_{0},\vec{p}\right)\right>$ is the gluon polarization tensor
and ${ {\lambda}_{4}^{\mu \mu' \nu \nu'} }^{a a' b b'}$ 
is the quartic gluon coupling tensor. 
The quartic gluon coupling tensor can be calculated from the perturbation 
but also the non-perturbation effect may become signification.
The quartic part in the effective potential is the decisive term 
in determining the process of the symmetry 
transmutation (or breaking)~\cite{Weinberg:1974hy,Dolan:1973qd}.
The gluon polarization tensor in the HTL approximation
is found~\cite{Zakout:2010rt}
\begin{eqnarray}
{\Pi_{\mu\nu}}^{a'a}\left(p_{0},\vec{p}\right)
&=&
\left({\bf m}^{2}_{D}\right)^{a'\,a}\left[
-\delta^{0}_{\mu}\delta^{0}_{\nu}
+p_{0}\int \frac{d\Omega_{k}}{4\pi}
\frac{\hat{k}_{\mu}\hat{k}_{\nu}}
{ p_{0}-\hat{k}\cdot\vec{p} }
\right],
\nonumber\\
&=&
\left({\bf m}^{2}_{D}\right)^{a}\delta^{a'a}\left[
-\delta^{0}_{\mu}\delta^{0}_{\nu}
+p_{0}\int \frac{d\Omega_{k}}{4\pi}
\frac{\hat{k}_{\mu}\hat{k}_{\nu}}
{ p_{0}-\hat{k}\cdot\vec{p} }
\right].
\end{eqnarray}
The information that is needed to understand the Higgs mechanism
and symmetry breaking is decoded implicitly in the gluon Debye mass(es).
The Debye mass is found as follows
\begin{eqnarray}
\left({\bf m}^{2}_{D}\right)^{a'\,a}&=&
\left({\bf m}^{2}_{D}\right)^{a}\,\delta^{a'\,a},
\nonumber\\
&=&
\left[
\left( {\bf m}^{2}_{D(Q)} \right)^{a}
+
\left( {\bf m}^{2}_{D(G)} \right)^{a}
\right]\,\delta^{a'\,a},
\nonumber\\
&=&
\left[
{\bf t}^{a'}_{ij}\,{\bf t}^{a}_{ji}
\left( {\bf m}^{2}_{D(Q)} \right)_{i}
+
{\bf T}^{a'}_{cb}\,{\bf T}^{a}_{bc}
\left( {\bf m}^{2}_{D(G)} \right)^{b}
\right]
\,\delta^{a'\,a},
\label{Debye-mass1}
\end{eqnarray}
where
\begin{eqnarray}
\left( {\bf m}^{2}_{D(Q)} \right)_{i}&=&
2\frac{g^{2}}{\pi^{2}}\sum^{N_f}_{Q=1}\left[
\frac{\pi^{2}}{6}\,T^{2}
+\frac{1}{2}\left(\mu_{Q}+{\mu_{C}}_{i}\right)^{2}
\right],
\end{eqnarray}
while the real gluonic part reads
\begin{eqnarray}
\left( {\bf m}^{2}_{D(G)} \right)^{b=\underbrace{(AB)}}
&\rightarrow&
\Re e\, \left( {\bf m}^{2}_{D(G)} \right)^{b=\underbrace{(AB)}},
\nonumber\\
&=&
\frac{g^{2}}{\pi^{2}}\left[
\frac{\pi^{2}}{3}\,T^{2}
-\frac{1}{2}\left({\mu_{C}}^{b}\right)^{2}\right],
\nonumber\\
&=&
\frac{g^{2}}{\pi^{2}}\left[
\frac{\pi^{2}}{3}\,T^{2}
-\frac{1}{2}\left({\mu_{C}}_{A}-{\mu_{C}}_{B}\right)^{2}\right].
\end{eqnarray}
Eq.(\ref{Debye-mass1}) hints some information about the possible 
symmetry breaking processes.
There is $(N_{c}^2-N_{c})$ non-degenerate masses 
and $(N_{c}-1)$ degenerate masses.
It is naively to conjecture that
$\hat{A}^{a}\propto
{\mu_{C}}^{a}  \equiv
\left({\mu_{C}}_{A}-{\mu_{C}}_{B}\right)$ where $a=\underbrace{(AB)}$.
This means that the $N_{c}^2-N_{c}$ non-degenerate masses
correspond the non-vanishing values  $\hat{A}^{a}$ with $A\neq B$
while the $N_{c}-1$ degenerate masses correspond
$\hat{A}^{a}=0$ with $A=B$.
Hence the $N_{c}^2-N_{c}$ non-degenerate masses
are simply the Higgs masses which are generated
by breaking the symmetry $SU(N_{c})$ while
the $N_{c}-1$ degenerate masses are the elements
of the resultant new symmetry $U(1)^{N_c-1}$. 
These masses in one (Hard-thermal) loop correction are of order 
$m_{D}\sim g\,T$. 
Depending on the higher order interaction
$\left(\hat{A}^{a}_{\mu}\right)^{n>2}$ which appear
in Eq.(\ref{potential-eff1}) and its modification with temperature
and density, it is naturally to surmise multiple phase transitions
such as $SU({N_c})\rightarrow SU({N_c-1})\times U(1)\rightarrow
\cdots \rightarrow U(1)^{N_c-1}$ 
(see for instance Ref.~\cite{Bronoff:1998hr}).
More recent discussions of the Higgs mechanism 
in the semi-quark-gluon plasma can be found 
in Ref.~\cite{Dumitru:2010mj} and the references therein. 
Instead of breaking the symmetry
$SU(N_c)\rightarrow SU(N_c-1)\times U(1)$, the symmetry transmutation 
undergoes multiple symmetry breaking transitions 
to $SU(N_c)\rightarrow O(N_c)\rightarrow SU(N_c-1)\times U(1)$
in order to maintain the quark confinement. 
The intermediate symmetry transmutation $O(N_{c})$ 
smoothes the symmetry breaking from $SU(N_{c})$ to $U(1)^{N_c-1}$
in order to soften the Hagedorn's equation of state.
On the other hand, the microscopic mechanism
in the multi-processes for the formation of symplectic
symmetry $Sp(N_{c})$ is rather more complicated 
because of the correlation between the flavor and color symmetries.
The symmetry reduction undergoes multiple processes, namely,
$SU_{V}(N_{f})\times SU(N_c)\rightarrow 
O(2N_c)\times \cdots\rightarrow Sp(N_c)$.
In the process of forming the symmetry $Sp(N_{c})$,
the color degrees of freedom favor to couple 
the flavor degrees of freedom through 
anti-symmetric channels to form diquark condensates and energy gaps.
The adjoint mean-fields acquire degenerate/non-degenerate 
effective gluon masses depending of the type of the color 
superconductivity that is formed in the medium.
The symmetry breaking leads to massless Nambu-Goldstone bosons
which come from flavor sector and massive gluons which stem from the color sector.
In principle, the energy gaps of the color-superconductivity 
can be included explicitly in the calculation of the canonical ensemble.
The multiple symmetry breaking processes 
from the matter that is dominated by $SU(N_c)$ Hagedorn states 
to another one that is dominated by $U(1)^{N_c-1}$ 
and the corresponding Higgs mechanism
is sketched in Fig.(\ref{symmetry-sketch1}).
In order to simplify the present work, we assume that 
the liberated gluons are approximated 
to an ideal gas of massless particles. 
In this approximation, the gluon effective masses 
which are acquired through the symmetry breaking 
process are neglected and this does not change 
the conclusion of the possible multiple phase transitions.  
%%%%%%%%%%%%%%%%%%%%%%%%%%%%%%%%%%%%%%%%%%%%%
%%%%%%%%%%%%%%%%%%%%%%%%%%%%%%%%%%%%%%%%%%%%%
%%%%%%%%%%%%%%%%%%%%%%%%%%%%%%%%%%%%%%%%
%%%%%%%%%%%%%%%%%%%%%%%%%%%%%%%%%%%%%%%%
%%%%%%%%%%%%%%%%%%%%%%%%%%%%%%%%%%%%%%%%
%%%%%%%%%%%%%%%%%%%%%%%%%%%%%%%%%%%%%%%%

%%%%%%%%%%%%%%%%%%%%%%%%%%%%%%%%%%%%%%%%%%%%%
%%%%%%%%%%%%%%%%%%%%%%%%%%%%%%%%%%%%%%%%%%%%%
%%%%%%%%%%%%%%%%%%%%%%%%%%%%%%%%%%%%%%%%
%%%%%%%%%%%%%%%%%%%%%%%%%%%%%%%%%%%%%%%%
%%%%%%%%%%%%%%%%%%%%%%%%%%%%%%%%%%%%%%%%
%%%%%%%%%%%%%%%%%%%%%%%%%%%%%%%%%%%%%%%%

\section{\label{sect-6b} Conclusion}

The present work is intended to determine the physics behind 
the intermediate processes toward the quark-gluon plasma 
in the $\mu_{B}-T$ phase transition diagram
and the variation of the mass spectral exponent
for the Hagedorn mass spectral density in the medium.
The $\mu_{B}-T$ phase transition diagram 
from the Hadronic phase to the quark-gluon plasma is rich and 
the multiple intermediate transition processes are found essential.
The nuclear matter which is dominated by the unitary Hagedorn states passes 
different phase transitions with different orders depending 
on the nuclear medium's temperature and density.
The dilute nuclear matter 
that is enriched with the unitary Hagedorn states
undergoes phase transition to another 
matter that is dominated by the orthogonal Hagedorn states
when the medium is heated up
while the extreme dense nuclear matter that is dominated 
by the unitary Hagedorn states 
passes  phase transition to another one that 
is dominated by  the symplectic Hagedorn states 
when the medium is further compressed and cooled down. 
It should be noted that 
the Hagedorn states are colorless (by projecting only the color-singlet states) 
regardless of their internal symmetry group.
Furthermore, it is possible to imagine under the assumption of 
certain scenarios that the Hagedorn states  
passes smooth multiple phase transitions 
to  a neutral color gas of metastable colored bags 
at $T\,\approx\,T_{c}$ and $\mu_{B}\,\approx\,0$. 
%%%%%%%%%%%%%%%%%%%%%%%%%%%%%
%%%%%%%%%%%%%%%%%%%%%%%%%%%%%

The tri-critical point in the $\mu_{B}-T$ phase transition 
diagram and the fluid behaviour of the quark-gluon plasma 
at a low baryonic chemical potential $\mu_{B}$ and 
high temperature $T\gg 0$ 
can be interpreted in terms of the modification of 
the Hagedorn mass spectral density due to the modification 
in the quark-gluon bag's internal symmetry.
%%%%%%%%%%%%%%%%%%%%%%%%%%
At low temperature, the nuclear matter is dominated 
by the discrete low-lying mass spectrum (hadrons)
and when the system is heated up 
it undergoes higher order phase transition 
to new nuclear matter that is dominated by the 
continuous high-lying mass spectrum particles 
(i.e. Hagedorns).
When the Hagedorn states freeze out they evaporate 
to the discrete low-lying particles through Gross-Witten phase
transition in an analogous way to the fragmentation  
of the big soap bubble to many tiny bubbles. 
%%%%%%%%%%%%%%%%%%%%%%%%%%

%%%%%%%%%%%%%%%%%%%%%%
The Hagedorn states are determined basically 
by the mass spectrum of the quark and gluon bag 
with a specific internal symmetry. 
The color-singlet constraint which guarantees 
the colorless quark-gluon bags plays 
an essential role in keeping quarks and gluons confined. 
The continuous high-lying hadronic matter is dominated at first 
by the unitary Hagedorn states just above the Gross-Witten point 
(it appears as a line in $\mu_{B}-T$ plane) 
for the phase transition from the discrete low-lying hadronic 
matter to the continuous high-lying one.
The mass spectral exponent $\alpha$ of the continuous 
high-lying mass spectral density
$\rho_{(II)}(m)\,=\,c\,m^{-\alpha}\,e^{b\,m}$
depends on the underlying internal symmetry of the Hagedorn bag.
The composite bag's internal symmetry is modified from 
the colorless unitary state to the colorless orthogonal state 
in the RHIC and/or LHC energy at high temperature. 
On the other hand, 
the quark and gluon bag's internal structure is mutated 
from the colorless unitary state (unitary Hagedorn)
to the colorless symplectic state (symplectic Hagedorn)
under the extreme compressed nuclear matter 
(i.e. when the unitary Hagedorn matter becomes saturated) 
such that one is found in the compact stars.
It is evident that the mass spectral exponent 
$\alpha$ increases with respect to $\mu_{B}$ 
when the system is compressed and cooled down.
This exponent decreases when the system is heated up.
The Hagedorn states  which are expected to be produced 
in the ultra-relativistic heavy ion collisions and beyond 
have a rather low mass spectral exponent.
%%%%%%%%%%%%%%
In the RHIC or LHC energy,
it is expected that the dilute hadronic matter 
which is dominated at first by the unitary Hagedorn states 
at small  $\mu_{B}$
passes Hagedorn phase transition 
to another nuclear matter
that is dominated by the orthogonal Hagedorns
when the medium is heated up to higher temperature.
The resultant orthogonal Hagedorn matter acts 
as the Coulomb liquid.

At small baryonic density, 
the hot nuclear matter which is dominated by 
the colorless orthogonal (Hagedorn) states 
undergoes higher order phase transition 
to the colored $SU(N_c)$ states 
with total color-neutral quark-gluon plasma.
The colored $SU(N_c)$ states 
are colored quark-gluon bags 
(i.e.: color-non-singlet states).
The conserved color charges are adjusted 
by the color chemical potentials.
The colored quark-gluon bags with 
color-non-singlet states are not Hagedorn states.
Aftermath the deconfinement phase transition, 
the medium is enriched by colored quarks and gluons.
In this sense the bag's color-singlet constraint 
is badly broken and the bag becomes a colored one 
where the colored quarks can be liberated or exchanged.
%%%%%%%%%%%%%%%%%%%%%%%%%% 

The phase transition from the hadronic matter which is dominated 
by the colorless Hagedorns to another matter 
that is populated by the colored bags is associated 
with the breaking of the color-singlet constraint.
The breaking of Hagedorn bag's internal symmetry
from the color-singlet state of the unitary symmetry 
to the colored unitary state probably takes place 
through several intermediate processes in the following way:
\begin{eqnarray}
\mbox{colorless}~~ SU(N_c)
&\rightarrow& 
\mbox{colorless}~~ O_{(S)}(N_c)\,+\, \frac{1}{2}N_{c}(N_c+1)\,~\mbox{gluons},
\nonumber\\
&\rightarrow& 
\mbox{colorless}~~ U(1)^{N_c}\,+\, N_c(N_{c}-1)\,~\mbox{gluons},
\nonumber\\
&\rightarrow& 
\mbox{colored}~~ SU(N_c) ~~\left(\mbox{i.e. color-non-singlet states}\right),
\nonumber\\
&\rightarrow& N_c\,~\mbox{quarks}\,+\, (N^2_c-1)\,~\mbox{gluons},
\end{eqnarray}
where the unimodular-like constraint is imposed 
in all the symmetries are involved in the present study.   
%%%%%%%%%%%%%%
In fact, there is $N_{c}^2$ adjoint 
degrees of freedom in the $U(N_{c})$ 
unitary representation and
the unimodular constraint reduces 
the number of degree of freedom by one. 
When the Hagedorn's internal structure 
is mutated from the colorless unitary state 
to the colorless orthogonal state,
the number of (Hagedorn's internal) adjoint gluon degrees 
of freedom is reduced from 
$N_{c}^2$ to $N_c(N_{c}+1)/2$, respectively. 
Therefore, the $N_{c}(N_{c}-1)/2$ 
free colorless gluons escape from the Hagedorn bag 
and emerge in the medium as jets. 
The nuclear matter turns to be dominated 
by the orthogonal Hagedorns
and is enriched by gluonic content. 
The $N_{c}(N_{c}-1)/2$ gluon species may glue 
together and probably emerge as color-neutral glueballs 
or jets.
%%%%%%%%%%%%%
When the temperature $T$ increases, 
the surfaces of colored bags are perturbed and expand 
by the thermal excitations and 
the system undergoes smooth phase transition 
to true deconfined quark-gluon plasma. 
The higher order phase transition is not usually 
associated with the explosive quark-gluon plasma.
Moreover, it seems that the colored bags 
that emerge through smooth phase transition
are rather mechanically stable. 
Hence the preceding scenario leads to smooth 
cross-over phase transition to quark-gluon plasma
at $\mu_{B}\approx 0$ and high $T$ 
and on the left hand side of the tri-critical point.
%%%%%%%%%%%%%%%%%
In the present approximation, the liberated gluons are
considered as an ideal gas of massless particles.
However, the breaking of the Hagedorn's internal symmetry
from the unitary symmetry to the orthogonal one may breaks 
the color global part of the liberated gluons and they acquire
finite mass through the Higgs mechanism. 
This mechanism will suppress the contribution of 
the gluon's partition function in particular 
if the liberated gluons attain large masses. 
%%%%%%%%%%%%%%%%%

On the other hand, when the system is compressed 
and cooled down, the higher order phase transition 
is reduced to the first order one at the moderate nuclear density.  
This takes place as $\mu_{B}$ increases 
and exceeds the tri-critical point 
(i.e. on the right hand side of the tri-critical point).
The system prefers to be dominated by the unitary Hagedorns 
in the stripe above the Gross-Witten point 
and below the quark-gluon plasma. 
When the medium becomes denser and hotter, 
the unitary Hagedorn states will not transmute
to orthogonal Hagedorn states anymore 
but instead they become rather mechanically 
unstable and the system prefers to pass direct 
deconfinement phase transition to quark-gluon plasma. 
Due to  the high thermal excitations, they likely pass 
the first order transition to quark-gluon plasma.
This kind of phase transition breaks badly 
the color-singlet constraint and subsequently 
the colored $SU(N_c)$ symmetry becomes possible.
%%%%%%%%%%%%%%%%%%%%%%
Therefore, the system undergoes first order deconfinement 
phase transition to explosive quark-gluon plasma 
at moderate $\mu_{B}\neq 0$.
Unlike the smooth cross-over phase transition 
which usually takes place 
at small baryonic density (i.e. $\mu_{B}\approx 0$),
the color charges are capable to escape 
and subsequently the color neutrality is violated.  
%%%%%%%%%%%%%%%%%
In this case, the color-singlet constraint is loosed
and the bag's internal symmetry 
is broken violently above the right hand side of the tri-critical point
in the following way:
\begin{eqnarray}
\mbox{colorless: color-singlet}\,SU(N_c)\,\rightarrow\,
\mbox{colored}\,SU(N_c).
\end{eqnarray}
%%%%%%%%%%%%%%%%%

When the system is compressed and cooled down further more 
and becomes extremely dense and saturated 
(i.e. $\mu_{B}$ becomes significantly large), 
the unitary Hagedorn matter 
transmutes to another one that is dominated 
by the symplectic Hagedorn states 
due to the modification of the quark and gluon bag's 
internal structure and the coupling 
of color symmetry with other symmetries 
such as but not limited to the flavor 
symmetry as follows 
\begin{eqnarray}
SU(N_{c})\,\times\,SU_{L}(N_{f})\times SU_{R}(N_{f})
&\rightarrow&
Sp(N)\,\times\,
\left[\mbox{conserved:}\,{U(1)}^{N_{f}-1}\right]. 
\end{eqnarray}
In fact the symplectic symmetry restores additional $N^2-N$ 
adjoint degrees of freedom 
(for instance partially breaking the flavor invariance 
$SU(N_{f})\,\times\,SU(N_{f})$ 
and leaving ${U(1)}^{N_{f}-1}$ invariance) 
besides the original $N_{c}^2-1$ 
adjoint color degrees of freedom.
The phase transition from the unitary Hagedorns 
to the colorless symplectic states 
(i.e. symplectic Hagedorns) may take place 
through more complicated multi-processes 
in the Hagedorn matter.
%%%%%%%%%%%%%%%%%%%%%%%%%%%
Furthermore, the mass spectral exponent $\alpha$ 
will continue to increase when additional symmetries 
are incorporated as the nuclear matter becomes denser. 
Probably, the Hagedorn matter 
with a certain complex internal symmetry 
passes to a new physics regime 
such as forming fluid 
of stable black holes or dark matter.
The deformation of the quark and gluon bag's 
boundary surface besides other corrections 
such as the center of mass correction modify 
the mass spectral exponent $\alpha$ smoothly.
The smooth modification of the mass spectral exponent 
$\alpha$ guarantees the continuity 
of various hadronic phases which are dominated 
by orthogonal, unitary and symplectic etc Hagedorns. 
Moreover, it is worth to mention that 
the symplectic Hagedorn matter is very rich 
due to the color-flavor coupling or color-angular coupling etc. 
The emergence of the color-superconductivity and other related physics 
can be explained in terms of the complex internal structure
such as the symplectic Hagedorn matter.
%%%%%%%%%%%%%%%%%%%%%%%%%%%%%%%%%%%%%%%%%%%%%%%%%%%%%%%%%%%%%
%%%%%%%%%%%%%%%%%%%%%%%%%%%%%%%%%%%%%%%%%%%%%%%%%%%%%%%%%%%%%

The analysis of the chiral phase transition 
at low $\mu_B$ and high $T$
demonstrates that in the case of the mass spectral exponent 
$\alpha\le 7/2$ 
such as the mass spectrum for orthogonal Hagedorn states, 
the chiral restoration phase transition takes 
place far away below the deconfinement phase transition
but at the threshold or above Gross-Witten point.
Therefore, the chiral mass is generated through the multiple
phase transition processes from the orthogonal Hagedorn states 
to the unitary Hagedorn states and subsequently to the low-lying 
mass spectrum states at the Gross-Witten point.

The order and shape of the ($\mu_{B}-T$) phase transition 
diagram depends basically on the internal structure 
of the quark and gluon bags such as but not limited 
to the color, flavor and angular symmetries. 
The Hagedorn's mass spectral exponent $\alpha$ 
is found to depend essentially on the medium. 
Furthermore, the multi-intermediate processes 
in the phase transition diagram from the low-lying hadrons 
to the eventual quark-gluon plasma is found very rich and not trivial. 
The complexity of the multi-process phase transitions increases 
along the temperature axis 
with the rather small baryonic chemical potential 
$\mu_{B}\sim 0$ 
(i.e. dilute nuclear matter).
It is evident that the Hagedorn's internal structure 
is significantly modified with respect 
to both $T$ and $\mu_{B}$. 
The mass spectral exponent seems to increase 
significantly at large baryonic density 
$\mu_{B}\gg 0$ 
and relatively low temperature 
(i.e. the extreme dense nuclear matter 
that is relevant to the compact stars).
The order and shape of the phase transition 
is basically medium dependent. 
It is associated with complicated multi-processes 
of symmetry reconfiguration along the chemical 
potential $\mu_{B}$ and temperature $T$ axes.
Therefore, the QCD phase transition diagram is proved 
to be very rich, tricky and non-trivial one.

%%%%%%%%%%%%%%%%%%%%%%%%%%%%%%%%%%%%
%%%%%%%%%%%%%%%%%%%%%%%%%%%%%%%%%%%%
%%%%%%%%%%%%%%%%%%%%%%%%%%%%%%%%%%%%
\begin{acknowledgments}
The partial support from Alexander von Humboldt foundation 
is acknowledged. 
\end{acknowledgments}
%%%%%%%%%%%%%%%%%%%%%%%%%%%%%%%%%%%%%%%%%%%%%%%%%%%%%%%%%%%%%%%%%%%
%%%%%%%%%%%%%%%%%%%%%%%%%%%%%%%%%%%%%%%%%%%%%%%%%%%%%%%%%%%%%%%%%%%
%%%%%%%%%%%%%%%%%%%%%%%%%%%%%%%%%%%%%%%%%%%%%%%%%%%%%%%%%%%%%%%%%%%%

%\bibliographystyle{plain}
\bibliography{refer1} 

%%%%%%%%%%%%%%%%%%%%%%%%%%%%%%%%%%%
%%%%%%%%%%%%%%%%%%%%%%%%%%%%%%%%%%%
\newpage
%\begin{sidewaystable}[v]
\begin{table}[ht]
\caption{\label{table-1}The mass spectral exponent $\alpha$ 
which appears in $\rho_{(II)}=c\, m^{-\alpha}\, e^{b\,m}$ 
for the color-singlet state  bag (i.e. Hagedorn state) versus 
the bag's internal symmetry that is given by
the $U(1)^{N_c}$, orthogonal $O(N_{c})$, unitary $U(N_{c})$ 
and symplectic $Sp(N)$ ($N=N_c$) 
symmetry groups and they are restricted 
to the unimodular-like constraint. 
The exponent $\alpha$ for the color-non-singlet 
quark-gluon bags 
(i.e. colored $SU(N_c)$) state is included.}
\centering
\begin{tabular}{l cc}
\hline
\hline
Symmetry group & $\alpha (N_c)$ & $\alpha~ (N_c=3)$ \\
[1ex] 
\hline
Non-singlet $SU(N_c)$ state  (not Hagedorn state):~   
& 
$\frac{1}{2}$ & $\frac{1}{2}$
\\
Color-singlet unimodular-like $U(1)^{N_c}$:~ 
& 
$\frac{1}{2} N_c$ & $\frac{3}{2}$ 
\\
Color-singlet unimodular-like orthogonal~ $O_{(S)}(N_c)$:~ 
& 
$\frac{1}{4}\left(N^2_c+N_c\right)$ & $3$ 
\\
Color-singlet unimodular-like unitary $U(N_c)$:~ 
& 
$\frac{1}{2}N^2_c$ & $\frac{9}{2}$ 
\\
Color-singlet unimodular-like symplectic $Sp(N_c)$:~ 
& 
$N^{2}_c-\frac{1}{2}N_c$ & $\frac{15}{2}$ 
\\
\hline
\hline
\end{tabular}
\label{table1}
\end{table}
%\end{sidewaystable}
%%%%%%%%%%%%%%%%%%%%%%%%%%%%%%%%%%%
%%%%%%%%%%%%%%%%%%%%%%%%%%%%%%%%%%%

\newpage 

\newpage
%%%%%%%%%%%%%%%%%%%%

\newpage
%%%%%%%%%%%%%%%%%%%%

\newpage

\newpage

\end{document}